\documentclass[aps,prd,draft,groupedaddress,preprintnumbers,%
              showpacs,showkeys]{revtex4}       
\usepackage[latin1]{inputenc}                    
\usepackage{graphicx}                            
\usepackage{epsfig}
\usepackage{epsf}
\usepackage{latexsym}                            
\usepackage{amsfonts}                            
\usepackage{amssymb}                             
\usepackage{amsmath}                             
\usepackage[mathscr]{eucal}                      
\usepackage{dcolumn}                             
\usepackage{multirow}
\usepackage{bm}                                  
\usepackage{hyperref}                            
\usepackage[usenames]{color}

\graphicspath{{.}{Graphics/}}                    

\newcommand{\be}{\begin{equation}}
\newcommand{\ee}{\end{equation}}
\newcommand{\beq}{\begin{eqnarray}}
\newcommand{\eeq}{\end{eqnarray}}
\newcommand{\bea}{\begin{eqnarray}}
\newcommand{\eea}{\end{eqnarray}}
\newcommand{\tendto}{\mathop{\longrightarrow}}

\usepackage{ifpdf}
\usepackage{graphicx}

\def\eq#1{Eq.~(\ref{#1})}

\def \3{\ss }

\newcommand{\tr}{\operatorname{Tr}}
\newcommand{\re}{\operatorname{Re}}

\newcommand{\beqn}{\begin{eqnarray}}
\newcommand{\eeqn}{\end{eqnarray}}

\def\cyp{a}
\def\cyi{b}
\def\sac{c}
\def\gre{d}
\def\nic{e}
\def\ors{f}

\begin{document}

\begin{titlepage}
  {\vspace{-0.5cm} \normalsize
  \hfill \parbox{60mm}{
DESY 09-160\\
SFB/CPP-09-91
}}\\[10mm]
  \begin{center}
    \begin{LARGE}
      \textbf{Low-lying baryon spectrum with two dynamical twisted
       mass fermions} \\
    \end{LARGE}
  \end{center}

 \vspace{.5cm}

\begin{figure*}[h!]
\begin{center}
\epsfxsize=2.5truecm
\epsfysize=3truecm
 \mbox{\epsfbox{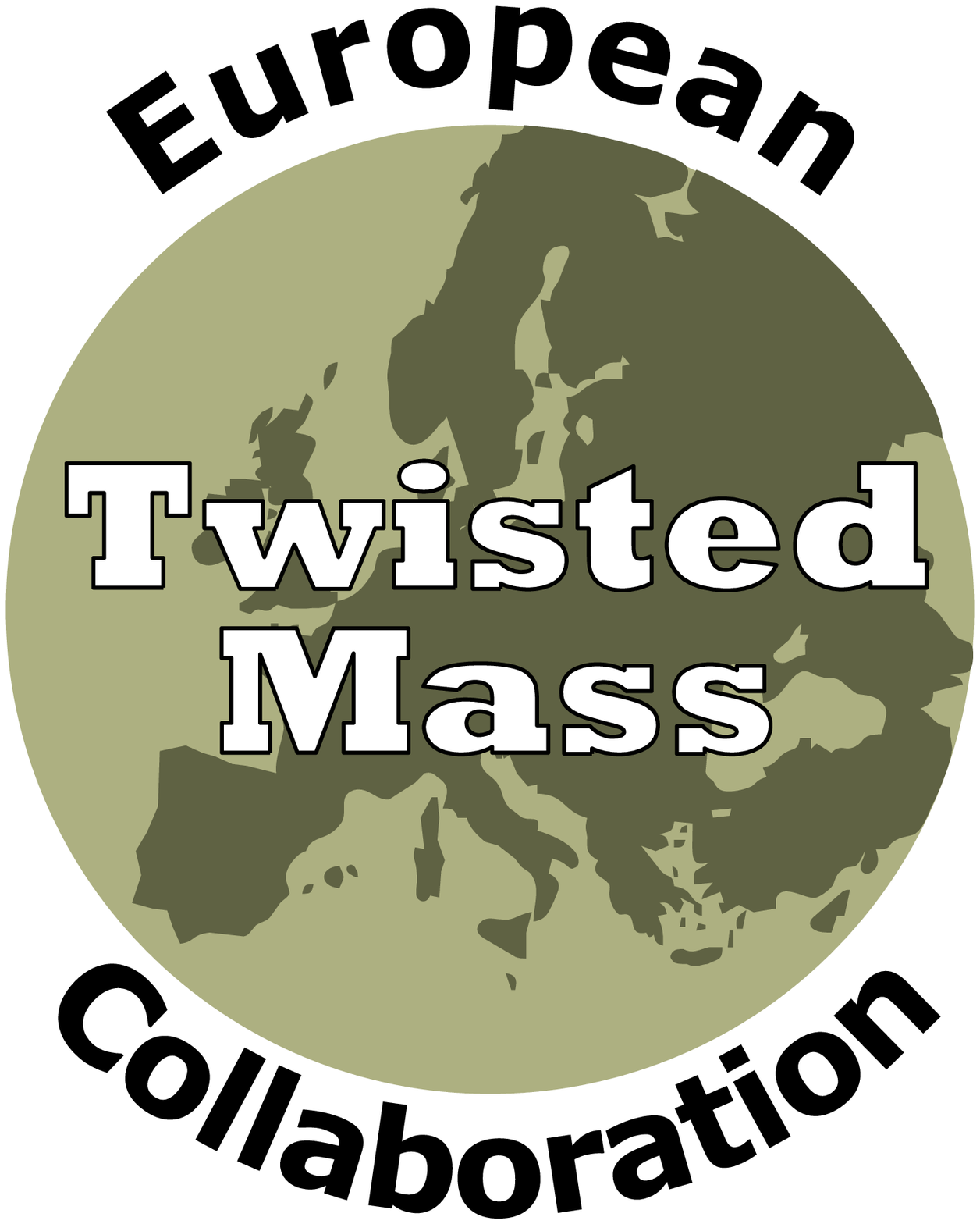}}
    \end{center}
\end{figure*}


 \vspace{-0.8cm}
  \baselineskip 20pt plus 2pt minus 2pt
  \begin{center}
    \textbf{
      C.~Alexandrou$^{(\cyp, \cyi)}$,
      R.~Baron$^{(\sac)}$, 
      J.~Carbonell$^{(\gre)}$, 
      V.~Drach$^{(\gre)}$,
      P.~Guichon$^{(\sac)}$,
      K.~Jansen$^{(\nic)}$,
      T.~Korzec$^{(\cyp)}$ 
      O.~P\`ene$^{(\ors)}$
}
  \end{center}

  \begin{center}
    \begin{footnotesize}
      \noindent 
	
 	$^{(\cyp)}$ Department of Physics, University of Cyprus, P.O. Box 20537,
 	1678 Nicosia, Cyprus\\	
 	$^{(\cyi)}$ Computation-based Science and Technology Research Center, Cyprus Institute,20 Kavafi Str., Nicosia 2121, Cyprus 
	\vspace{0.2cm}
	
	$^{(\sac)}$CEA-Saclay, IRFU/Service de Physique Nucl\'eaire, 91191 Gif-sur-Yvette, France 
	\vspace{0.2cm}

      $^{(\nic)}$ NIC, DESY, Platanenallee 6, D-15738 Zeuthen, Germany\\
      \vspace{0.2cm}

	$^{(\gre)}$ Laboratoire de Physique Subatomique et Cosmologie,
               UJF/CNRS/IN2P3, 53 avenue des Martyrs, 38026 Grenoble, France
	 \vspace{0.2cm}



      $^{(\ors)}$ Laboratoire de Physique Th\'eorique (B\^at.~210), Universit\'e
      de Paris XI,CNRS-UMR8627,  Centre d'Orsay, 91405 Orsay-Cedex, France\\
      \vspace{0.2cm}
      
%
%
      
    \end{footnotesize}
  \end{center}

  \begin{abstract}

The masses of the low lying baryons are evaluated 
using two degenerate flavors of twisted
mass sea quarks corresponding to pseudo scalar masses in the range of about  
270~MeV to 500~MeV.
The strange valence quark mass is tuned to reproduce 
the mass of the kaon in the physical limit.
The tree-level Symanzik improved gauge action is employed.
We use lattices of spatial size 2.1~fm and 2.7~fm at two values of 
the lattice spacing
with $r_0/a=5.22(2)$ and $r_0/a=6.61(3)$.
We check for both finite volume and cut-off effects on the baryon masses. 
We performed a detailed study of the chiral extrapolation of
the octet and decuplet masses using SU(2) $\chi$PT.
The lattice spacings  determined using the nucleon mass at the
physical point are
 consistent with the values extracted using the pion decay constant.
We examine the issue of isospin symmetry breaking for the octet and
decuplet baryons and its dependence on the lattice spacing.
We show that in the continuum limit isospin breaking is consistent with zero, 
as expected.
The baryon masses that we find after taking  the continuum limit and
extrapolating to the physical limit are in good agreement with experiment.

\begin{center}
\today
\end{center}
 \end{abstract}
\pacs{11.15.Ha, 12.38.Gc, 12.38.Aw, 12.38.-t, 14.70.Dj}
\keywords{Octet and Decuplet mass, Lattice QCD}
\maketitle 
\end{titlepage}


\section{Introduction}

In the last couple of years an intense and successful effort in extending
unquenched lattice calculations 
towards realistic values of quark masses, small lattice spacings and large 
volumes has 
been undertaken using a variety of algorithmic techniques and lattice actions.
A review of the salient features of the various discretization  schemes
currently employed can be found in  Ref~\cite{Jansen:2008vs}. 
Of particular relevance to the 
current work are the calculations of the low-lying baryon spectrum
 using two degenerate flavors ($N_f=2$)  
of  light dynamical quarks. Such studies have been carried out   
 by the MILC collaboration
\cite{Bernard:2001av,Aubin:2004fs} using Kogut-Susskind  fermions
and by the European Twisted Mass Collaboration (ETMC)~\cite{Alexandrou:2008tn} for the nucleon ($N$) 
and $\Delta$ baryons using twisted mass fermions. There are also
baryon mass calculations
  using two degenerate flavors of light quarks and
a strange quark with the mass tuned to its physical value  ($N_f=2+1$)
mainly using clover improved Wilson fermions with different levels of
smearing, such as the calculation of  the nucleon mass
by the QCDSF-UKQCD collaboration~\cite{AliKhan:2003cu},
and  the evaluation of the octet and decuplet
spectrum by the PACS-CS~\cite{Aoki:2008sm} and  
BMW~\cite{Durr:2008zz} collaborations.
The LHP Collaboration
 computed the octet and decuplet spectrum using a hybrid
action with domain wall valence 
fermions 
on asqtad improved staggered sea quarks~\cite{WalkerLoud:2008bp}. Preliminary results on
the nucleon mass are also  computed using $N_f=2+1$ domain wall fermions by the RBC-UKQCD
collaboration~\cite{Antonio:2006px,Antonio:2006zz}.

In this work we study the low-lying spectrum of the baryon 
 octet and decuplet
with   twisted mass fermions at maximal twist.
 The light quarks  are  dynamical degrees of freedom
 while
in the strange sector we use an Osterwalder-Seiler 
valence  quark, following the approach employed in the 
study of the pseudo scalar meson decay constants~\cite{Blossier:2007vv,Blossier:2009bx}. 
 The bare strange valence quark mass is taken to be the
same as the one determined in the meson studies tuned by requiring
that the mass of  the kaon  at the physical point matches
its physical value.
Using the ETMC $N_f=2$ configurations~\cite{Boucaud:2007uk,Boucaud:2008xu} we 
calculate the baryon spectrum for pion masses in the range of 270~MeV to 500~MeV
and at two values of the 
lattice spacing  corresponding to $\beta=3.9$ and $\beta=4.05$ 
with $r_0/a=5.22(2)$ and $r_0/a=6.61(3)$, respectively, where $r_0$ is determined from the force between two static quarks. 
Results are also obtained at a third $\beta$-value, namely $\beta=3.8$, which corresponds to $r_0/a=4.46(3)$.
 The latter results are not
taken  into account in the final analysis due to large
autocorrelation effects observed in the Monte Carlo history for quantities 
like the PCAC mass 
 and the plaquette at small sea quark masses.
Data at $\beta=3.8$ are only used as
a consistency check of the continuum extrapolation. 
For the nucleon mass we also performed the calculation at an even finer value
of the lattice spacing corresponding
to  $r_0/a=8.31(5)$ and $\beta=4.2$ to ensure that indeed the
continuum extrapolation using a 
weighted average with results at $\beta=3.9$ and $\beta=4.05$ is valid.
We find that the baryon masses considered here show a very weak
dependence on the lattice spacing and are fully compatible with an ${\cal O}(a^2)$ behaviour with an almost 
vanishing coefficient of the $a^2$ term.
This justifies neglecting the ${\cal O}(a^2)$
term in extrapolating results to the continuum limit.

For a fixed value of the lattice spacing we have used up to five different light quark 
masses and two different volumes.
The corresponding  $m_{\pi}L$ values are in the range 3.3 to 7.4, where
$L$ is the spatial extent of the lattice.
Using these various values of the lattice spacing, quark masses and volumes 
allows us to estimate the volume corrections and  perform a continuum and chiral extrapolation. 
The good precision of our results on the baryon masses allows us to perform 
a study of chiral extrapolations to
the physical point. This study shows that one of 
our main uncertainties in predicting the mass at the physical point is
 caused by the chiral extrapolations.
Another source of systematic error is the partially quenched approximation
that we have used.

An important issue is the restoration of the
explicitly broken isospin symmetry in the continuum limit.
 At finite lattice spacing, baryon masses
display $\mathcal{O}(a^2)$ isospin breaking effects. 
There are, however, theoretical arguments \cite{Frezzotti:2007qv} and numerical evidences
\cite{Dimopoulos:2008sy,Jansen:2008vs} that these isospin breaking effects are 
particularly pronounced
 for the neutral pseudo scalar mass whereas for other quantities studied 
so far by ETMC they are compatible with zero.   
In this paper we will demonstrate that also in the baryon sector these isospin 
breaking
effects are in general small or even compatible with zero. 
For a preliminary account of these
results see Ref.~\cite{Latt08_Vincent}.

The paper is organized as follows:
The details of our lattice setup, namely those concerning 
the twisted mass action, 
the parameters of the simulations and the interpolating fields used, 
are given in Section~II. 
Section~III contains the numerical results of the baryon masses computed
 for different 
lattice volumes, lattice spacings and bare quark masses
as well as the Gell-Mann Okubo relations that are supposed to be 
fulfilled in the exact SU(3) limit.
Lattice artifacts, including finite volume and discretization errors are discussed in Section IV, with special 
emphasis on the $\mathcal{O}(a^2)$ isospin breaking effects inherent
 in the twisted mass formulation
of lattice QCD.
The chiral extrapolations are analyzed in Section~V.
Section~VI  contains a comparison with other existing calculations and
conclusions are finally  drawn in Section~VII.

\section{Lattice setup}
\subsection{The lattice action}

For the gauge fields  we use the  tree-level Symanzik improved
gauge action~\cite{Weisz:1982}, which includes besides the
plaquette term $U^{1\times1}_{x,\mu,\nu}$ also rectangular $(1\times2)$ Wilson 
loops $U^{1\times2}_{x,\mu,\nu}$
\begin{equation}
  \label{eq:Sg}
    S_g =  \frac{\beta}{3}\sum_x\Biggl(  b_0\sum_{\substack{
      \mu,\nu=1\\1\leq\mu<\nu}}^4\left \{1-\re\tr(U^{1\times1}_{x,\mu,\nu})\right \}\Bigr. 
     \Bigl.+
    b_1\sum_{\substack{\mu,\nu=1\\\mu\neq\nu}}^4\left \{1
    -\re\tr(U^{1\times2}_{x,\mu,\nu})\right \}\Biggr)\,  
\end{equation}
with  $b_1=-1/12$ and the
(proper) normalization condition $b_0=1-8b_1$. Note that at $b_1=0$ this
action reduces to the usual Wilson plaquette gauge action.

The fermionic action for two degenerate flavors of quarks
 in twisted mass QCD is given by
\be
S_F= a^4\sum_x  \bar{\chi}(x)\bigl(D_W[U] + m_0 
+ i \mu \gamma_5\tau^3  \bigr ) \chi(x)
\label{S_tm}
\ee
with   $\tau^3$ the Pauli matrix acting in
the isospin space, $\mu$ the bare twisted mass 
and $D_W$ the massless Wilson-Dirac operator given by 
\be
D_W[U] = \frac{1}{2} \gamma_{\mu}(\nabla_{\mu} + \nabla_{\mu}^{*})
-\frac{ar}{2} \nabla_{\mu}
\nabla^*_{\mu} 
\ee
where
\be
\nabla_\mu \psi(x)= \frac{1}{a}\biggl[U_\mu(x)\psi(x+a\hat{\mu})-\psi(x)\biggr]
\hspace*{0.5cm} {\rm and}\hspace*{0.5cm} 
\nabla^*_{\mu}\psi(x)=-\frac{1}{a}\biggl[U_{\mu}^\dagger(x-a\hat{\mu})\psi(x-a\hat{\mu})-\psi(x)\biggr]
\quad .
\ee
Maximally twisted Wilson quarks are obtained by setting the untwisted quark mass $m_0$ to its critical value $m_{\rm cr}$,
 while the twisted
quark mass parameter $\mu$ is kept non-vanishing in order to be away from the chiral limit.
In \eq{S_tm} the quark fields $\chi$
are in the so-called ``twisted basis". The ``physical basis" is obtained for
maximal twist by the simple transformation
\be
\psi(x)=\exp\left(\frac {i\pi} 4\gamma_5\tau^3\right) \chi(x),\qquad
\overline\psi(x)=\overline\chi(x) \exp\left(\frac {i\pi} 4\gamma_5\tau^3\right)
\quad.
 \ee
In terms of the physical fields the action is given by
\be
S_F^{\psi}= a^4\sum_x  \bar{\psi}(x)\left(\frac 12 \gamma_\mu 
[\nabla_\mu+\nabla^*_\mu]+i \gamma_5\tau^3 \left(- 
\frac{ar}{2} \;\nabla_\mu\nabla^*_\mu+ m_{\rm cr}\right ) 
+  \mu \right ) \psi(x)\quad.
\label{S_ph}
\ee
In this paper, unless otherwise stated, the quark fields will be understood as ``physical fields'',
 $\psi$, in particular when we define the baryonic interpolating fields. 

A crucial advantage of the twisted mass formulation is
the fact that, by tuning the bare untwisted quark mass $m_0$ to its critical value
 $m_{\rm cr}$,  physical observables  are automatically 
${\cal O}(a)$ improved. 
In practice, we implement
maximal twist of Wilson quarks by tuning to zero the bare untwisted current
quark mass, commonly called PCAC mass, $m_{\rm PCAC}$, which is proportional to
$m_0 - m_{\rm cr}$ up to ${\cal O}(a)$ corrections. As detailed in Ref.~\cite{ETMClong}, 
$m_{\rm PCAC}$ is conveniently evaluated through
\be
m_{\rm PCAC}=\lim_{t/a >>1}\frac{\sum_{\bf x}\langle \partial_4 \tilde{A}^b_4({\bf x},t) \tilde{P}^b(0) \rangle}
{2\sum_{\bf x} \langle \tilde{P}^b({\bf x},t)\tilde{P}^b(0)\rangle}, \hspace*{1cm} b=1,2 \quad,
\label{PCAC mass}
\ee
 where $\tilde{A}^b_\mu=\bar{\chi}\gamma_\mu \gamma_5 \frac{\tau^b}{2}\chi$ is 
the
axial vector current and
 $\tilde{P}^b=\bar{\chi}\gamma_5 \frac{\tau^b}{2}\chi$ is
the pseudo scalar density in the twisted basis. The large $t/a$ limit
is required in order to isolate the contribution of the
lowest-lying charged pseudo scalar meson state in  the correlators of \eq{PCAC mass}. 
This way of determining $m_{\rm PCAC}$ is equivalent
to imposing on the lattice the validity of the axial Ward identity between the vacuum and the charged  one-pion  zero three-momentum state:
\begin{equation}
\partial_\mu \tilde{A}_\mu^b = 2m_{\rm PCAC} \tilde{P}^b,\;\quad b=1,2 \quad. 
\end{equation}

The value of $m_{\rm cr}$ is determined at each $\beta$ value at the lowest 
twisted mass used in our simulations, a procedure that preserves ${\cal O}(a)$ improvement
and keeps ${\cal O}(a^2)$ small~\cite{Boucaud:2008xu,Frezzotti:2005gi}.
The twisted mass fermionic action breaks parity and isospin at 
non-vanishing lattice spacing, as it is apparent from the form of the Wilson term in 
 Eq.~(\ref{S_ph}).
In particular,  the isospin breaking in physical observables is a 
cut-off effect of ${\cal O}(a^2)$~\cite{Frezzotti:2004}.
To simulate the strange quark in the valence sector several choices are possible. 
We consider a quenched Osterwalder-Seiler fermion \cite{Osterwalder:1977pc} with the 
following action in the twisted basis:

\be
S_s= a^4\sum_x  \bar{\chi_s}(x)\bigl(D_W[U] + m_0 
+ i \mu_s \gamma_5  \bigr ) \chi_s(x) \; .
\label{S_OS}
\ee
 
This is naturally realized in the twisted mass approach by introducing an additional 
doublet of strange quark and keeping only the positive diagonal component of $\tau_3$.
The  $m_0$ value is taken to be equal to the critical mass determined in the light sector, 
thus guaranteeing the $O(a)$  improvement in any observable. 
The reader interested in the advantage of this mixed action in the mesonic sector is referred to the Refs~\cite{Frezzotti:2004wz, AbdelRehim:2006ra, AbdelRehim:2006ve, Blossier:2007vv, Blossier:2009bx}.

\subsection{Simulation details}

The input parameters of the calculation, namely $\beta$, $L/a$ and $a\mu$ 
are summarized in Table~\ref{Table:params}. The corresponding lattice spacing $a$ 
and the pion mass values, spanning a mass range 
from 270~MeV to 500~MeV, are taken 
from Ref.~\cite{Urbach:2007}.  
At $m_{\pi}\approx 300$ MeV we have simulations 
for lattices of spatial size $L=2.1$~fm and $L=2.7$~fm at $\beta=3.9$ 
allowing to investigate finite volume effects. 
Finite lattice spacing effects are investigated using two sets of 
results at $\beta=3.9$  and $\beta=4.05$. The set at $\beta=3.8$ is used only as a 
cross-check and to estimate  cut-off errors. 
These sets of gauge ensembles allow us to estimate all the systematic 
errors in order to have reliable predictions for the baryon spectrum.
\begin{table}[h]
\begin{center}
\begin{tabular}{c|llllll}
\hline\hline
\multicolumn{6}{c}{ $\beta=4.05$, $a=0.0666(6)$~fm from $f_\pi$~\cite{Urbach:2007}, ${r_0/a}=6.61(3)$ }\\
\hline
$32^3\times 64$, $L=2.13$~fm &$a\mu$         & 0.0030     & 0.0060     & 0.0080     & 0.012\\
                               & No. of confs.   &269 &253 &  409 &182\\
                               &$m_\pi$~(GeV) & 0.3070(18) & 0.4236(18) & 0.4884(15) & 0.6881(18) \\
                               &$m_\pi L$     & 3.31       &   4.57     & 5.27       &  7.43      \\ \hline\hline
\multicolumn{6}{c}{$\beta=3.9$, $a=0.0855(6)$~fm,   from $f_\pi$~\cite{Urbach:2007}, ${r_0/a}=5.22(2)$}\\\hline 
$24^3\times 48$, $L=2.05$~fm &$a\mu$         & 0.0030   &    0.0040      &   0.0064     &  0.0085     &   0.010 \\ 
                               &No. of confs & -  &782 &545 & 348 &477 \\ 
                               &$m_\pi$~(GeV) & -  & 0.3131(16) & 0.3903(9) & 0.4470(12) & 0.4839(12)\\
                               &$m_\pi L$     &    & 3.25       & 4.05      & 4.63       & 5.03     \\
$32^3\times 64$, $L=2.74$~fm  &$a\mu$ & 0.003 & 0.004 & & & \\
                               & No. of confs & 659  &232 & & & \\
                               & $m_\pi$~(GeV)& 0.2696(9)   & 0.3082(6) & & &  \\
                               & $m_\pi L$    & 3.74        & 4.28      &&& \\\hline \hline
\multicolumn{6}{c}{ $\beta=3.8$, $a=0.0995(7)$~fm  ${r_0/a}=4.46(3)$}\\\hline
$24^3\times 48$, $L=2.39$~fm &$a\mu$         & 0.0060     & 0.0080     & 0.0110     & 0.0165\\
                               & No. of confs      & 215  & 302 & 248 & 244& \\
                               &$m_\pi$~(GeV) & 0.3667(17) & 0.4128(16) & 0.4799(9) & 0.5855(10) \\
                               &$m_\pi L$     & 4.44       & 5.00       & 5.81      & 7.09      \\ \hline
\end{tabular}
\caption{Input parameters ($\beta,L,\mu$) of our lattice calculation and corresponding lattice spacing ($a$) and pion mass ($m_{\pi}$).}
\label{Table:params}
\end{center}
\vspace*{-.0cm}
\end{table}

\subsection{Tuning of the bare strange quark mass}

In a previous paper from the ETM collaboration \cite{Blossier:2007vv},  pseudo scalar meson 
masses have been computed for different values of the sea and valence quark masses for the 
$\beta=3.9$ gauge configurations. Using the experimental value of 
the mass ratio of the kaon to the pion, $m_K/m_\pi$, the 
bare strange quark mass can be set. 
We use the value of $a\mu_s = 0.0217(22)$ at $\beta= 3.9$ 
 taken from Table~2 of Ref.~\cite{Blossier:2007vv}. 
In a more recent study of the pseudo scalar decay constant of kaons and 
D-mesons \cite{Blossier:2009bx}, the computation was extended to $\beta = 3.8$ 
and $\beta=4.05$. However, this is still a preliminary
 analysis  and an ongoing analysis for the accurate extraction of quark masses
 is still in progress.
One can obtain an estimate of the bare strange quark mass at a given value of 
$\beta$ by taking the results at $\beta=3.9$ as a reference and using  the scaling relation~\cite{private:2008}:
\begin{equation}
 a \mu_s(\beta) = \frac{Z_p(\beta)}{Z_p(\beta=3.9)} \frac{ a(\beta)}{a(\beta=3.9)} a \mu_s(\beta=3.9)\; .
\label{s-mass}
\end{equation}
The values we use for $\beta=3.8$ and $\beta=4.05$ given
in Table~\ref{Table:mus} 
are obtained by applying Eq.~(\ref{s-mass}). 
We use  the value of the renormalization constant $Z_p(\beta)$
found in the preliminary analysis of Ref.~\cite{Dimopoulos:2007qy}
within the RI'-MOM scheme. This value is in agreement with a complementary
analysis given in Ref.~\cite{Cyprus:2009}.

\begin{table}[h!]
\begin{center}
\begin{tabular}{c|llll}
  \hline\hline
	   & $\beta= 3.8$  & $\beta= 3.9$ & $\beta= 4.05$ \\
   $a \mu_s$  & $0.0208(15)(48)$ & $0.0217(22)$ & $0.0166(18)(29)$ \\
\hline\hline
\end{tabular}
\caption{Bare strange quark mass used in the valence sector for different $\beta$ values.}
\label{Table:mus}
\end{center}
\vspace*{-0.8cm}
\end{table}

\subsection{Interpolating fields}

\begin{figure}[h!]
\begin{minipage}{8cm}
\epsfxsize=8truecm
\epsfysize=6truecm
 \mbox{\epsfbox{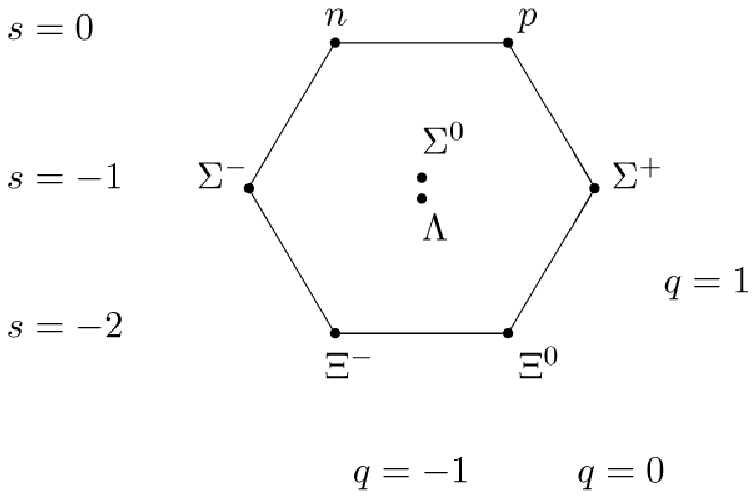}}
\caption{The low lying baryons belonging to the octet representation labeled by the value of $I_3$ and hypercharge.} 
\label{fig:interp. octet}
\end{minipage}
\hfill
\begin{minipage}{8cm}
\epsfxsize=8truecm
\epsfysize=6truecm
 \mbox{\epsfbox{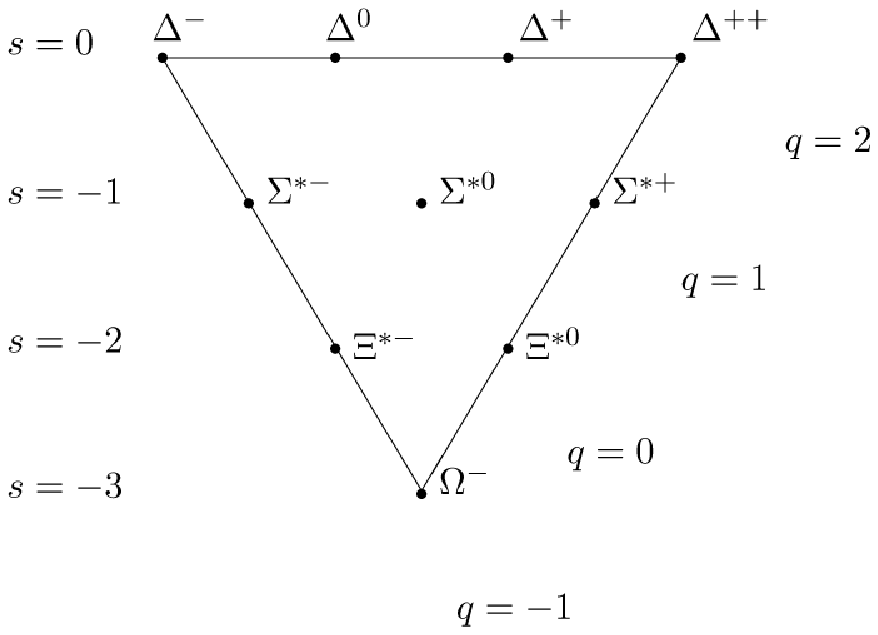}}
\caption{The low lying baryons belonging to the decuplet representation labeled by the value of $I_3$ and hypercharge.}
\label{fig:interp. decuplet}
\end{minipage}
\end{figure}

The low lying baryons belonging to the octet and decuplet representations 
 of $SU(3)$ are given in Figs.~\ref{fig:interp. octet}
 and \ref{fig:interp. decuplet} respectively.
They are classified by giving the isospin, $I$, the third component of the isospin, $I_3$, the strangeness (s), spin and parity.
In order to extract their masses in lattice QCD we evaluate two point correlators. We use interpolating fields to create these states from the vacuum
that have the correct quantum numbers and reduce to the quark model wave functions in the non-relativistic limit.
The interpolating fields used in this work are collected in Tables~\ref{Table:interpolating_octet}~\cite{Ioffe:1981kw, Leinweber:1990dv} and  ~\ref{Table:interpolating_decuplet}~\cite{Ioffe:1981kw, Leinweber:1992hy} for
the octet and decuplet respectively.
\begin{table}[h!]
\begin{center}
\renewcommand{\arraystretch}{1.5}
\begin{tabular}{c|c c c c}
\hline 
 Strangeness  & Baryon  & Interpolating field & $I$  & $I_z$ \\ \hline
\multirow{2}{*}{$s=0$} & $p$ & $~\chi^{p} = \epsilon_{abc}(u_a^T C\gamma_5 d_b) u_c~$ & $1/2$ & $+1/2$\\
 & $n$ & $~\chi^{n} = \epsilon_{abc}(d_a^T C\gamma_5 u_b) d_c~$ &  $1/2$ & $-1/2$\\
\hline
 \multirow{4}{*}{$s=1$} & $\Lambda$ &$~\chi^{\Lambda^8}= \frac{1}{\sqrt{6}}\epsilon_{abc}\big\{2(u_a^T C\gamma_5 d_b )s_c +(u_a^T C\gamma_5 s_b)d_c -(d_a^T C\gamma_5 s_b)u_c\big\}~$ & $0$ & $0$\\
 & $\Sigma^+$ & $~\chi^{\Sigma^+} = \epsilon_{abc}(u_a^T C\gamma_5 s_b )u_c ~$  & $1$ & $+1$\\
 & $\Sigma^0$ & $~\chi^{\Sigma^0} = \frac{1}{\sqrt{2}}\epsilon_{abc}\big\{(u_a^T C\gamma_5 s_b )d_c +(d_a^T C\gamma_5 s_b)u_c \big\}~$ & $1$ & $+0$\\
 & $\Sigma^-$ & $~\chi^{\Sigma^-} = \epsilon_{abc}(d_a^T C\gamma_5 s_b )d_c ~$ & $1$ & $-1$\\
 \hline
\multirow{2}{*}{$s=2$} & $\Xi^0$ & $~\chi^{\Xi^{0}} = \epsilon_{abc}(s_a^T C\gamma_5 u_b) s_c~$ & $1/2$ & $+1/2$ \\
 & $\Xi^-$ & $~\chi^{\Xi^{-}} = \epsilon_{abc}(s_a^T C\gamma_5 d_b) s_c~$& $1/2$ & $-1/2$\\
 \hline \hline
\end{tabular}
\caption{Interpolating fields and quantum numbers for the baryons in the octet representation.}
\label{Table:interpolating_octet}
\end{center}
\end{table}

\begin{table}[h!]
\begin{center}
\renewcommand{\arraystretch}{1.5}
\begin{tabular}{c|c c c c}
\hline\hline
 Strangeness  & Baryon  & Interpolating field & $I$ & $I_z$ \\ \hline
\multirow{4}{*}{$s=0$} &$\Delta^{++}$ & $~\chi^{\Delta^{++}}_{\mu} = \epsilon_{abc}(u_a^T C\gamma_{\mu} u_b) u_c~$ &$3/2$ & $+3/2$\\
 &$\Delta^{+}$   & $~\chi^{\Delta^{+}}_{\mu} =  \frac{1}{\sqrt{3}}\epsilon_{abc}\big\{2 (u_a^T C\gamma_{\mu} d_b) u_c + (u_a^T C\gamma_{\mu} u_b) d_c\big\} ~$ &$3/2$ & $+1/2$ \\
 & $\Delta^{0}$  & $~\chi^{\Delta^{0}}_{\mu} =  \frac{1}{\sqrt{3}}\epsilon_{abc}\big\{2 (d_a^T C\gamma_{\mu} u_b) d_c + (d_a^T C\gamma_{\mu} d_b) u_c\big\}   ~$ &$3/2$ & $-1/2$\\
 & $\Delta^{-}$  & $~\chi^{\Delta^{-}}_{\mu} = \epsilon_{abc}(d_a^T C\gamma_{\mu} d_b) d_c~$ &$3/2$ & $-3/2$ \\
\hline
 \multirow{3}{*}{$s=1$}  & $\Sigma^{\ast +}$ & $~\chi^{\Sigma^{\ast +}}_{\mu}= \frac{1}{\sqrt{3}}\epsilon_{abc}\big\{ (u^{T}_a  C \gamma_{\mu} u_b)s_c+2(s^{T}_a  C \gamma_{\mu}u_b) u_c \big\}  ~$ & $1$ & $+1$\\
 & $\Sigma^{\ast 0}$ & $~\chi^{\Sigma^{\ast 0}}_{\mu} = \sqrt{\frac{2}{3}}\epsilon_{abc}\big\{ (u^{T}_a  C \gamma_{\mu}d_b )s_c + (d^{T}_a  C \gamma_{\mu}s_b) u_c +(s^{T}_a  C \gamma_{\mu}u_b) d_c \big\}~$ &$1$ & $+0$\\
 & $\Sigma^{\ast -}$ & $~\chi^{\Sigma^{\ast -}}_{\mu} = \frac{1}{\sqrt{3}}\epsilon_{abc}\big\{ (d^{T}_a  C \gamma_{\mu}d_b )s_c  +2(s^{T}_a  C \gamma_{\mu}d_b) d_c \big\} ~$ &$1$ & $-1$\\
 \hline
\multirow{2}{*}{$s=2$} & $\Xi^{\ast 0}$ & $~\chi^{\Xi^{\ast 0}}_{\mu} = \epsilon_{abc}(s_a^T C\gamma_\mu u_b) s_c~$  &$1/2$ & $+1/2$\\
 & $\Xi^{\ast -}$  & $~\chi^{\Xi^{\ast -}}_{\mu} = \epsilon_{abc}(s_a^T C\gamma_\mu d_b) s_c~$&$1/2$ & $-1/2$\\
 \hline
\multirow{1}{*}{$s=3$} & $\Omega^{-}$ & $~\chi^{\Omega^{-}}_{\mu} = \epsilon_{abc}(s_a^T C\gamma_{\mu} s_b) s_c~$  &$0$ & $+0$\\
 \hline \hline
\end{tabular}
\caption{Interpolating fields and quantum numbers for baryons in the decuplet representation.}
\label{Table:interpolating_decuplet}
\end{center}
\end{table}
Local interpolating fields  are not optimal for suppressing excited state contributions. We instead apply
 Gaussian smearing to each  quark field,  $q({\bf x},t)$: $q^{\rm smear}({\bf x},t) = \sum_{\bf y} F({\bf x},{\bf y};U(t)) q({\bf y},t)$
using the gauge invariant smearing function  
\be 
F({\bf x},{\bf y};U(t)) = (1+\alpha H)^ n({\bf x},{\bf y};U(t)),
\ee
constructed from the hopping matrix,
\be
H({\bf x},{\bf y};U(t))= \sum_{i=1}^3 \biggl( U_i({\bf x},t)\delta_{{\bf x,y}-i} +  U_i^\dagger({\bf x}-i,t)\delta_{{\bf x,y}+i}\biggr).
\ee
 Furthermore we apply APE smearing to the spatial links  that enter the hopping matrix.
The parameters of the Gaussian and APE smearing are the same as those  used in our previous work devoted to the nucleon and $\Delta$
masses~\cite{Alexandrou:2008tn}.

\subsection{Two-point correlators}

To extract masses in the rest frame we consider two-point correlators defined by 
\bea
C^\pm_X(t,\vec{p}=\vec{0}) = \frac 1 2 {\rm Tr}(1 \pm \gamma_4) \sum_{\bf x_{\rm sink}}
\langle J_X( {\bf x}_{\rm sink}, t_{\rm sink}) \bar J_X({\bf x}_{\rm source}, t_{\rm source})\rangle,\qquad 
t=t_{\rm sink}-t_{\rm source} \quad.
\label{C_X}\eea
Space-time reflection symmetries of the action and the anti-periodic boundary conditions in the temporal direction for the quark fields
imply, for zero three-momentum correlators, that $C_X^+(t) = -C_X^-(T-t)$. So,  In order to decrease errors 
we average correlators in the forward and backward direction and define:
\be
   C_X(t) = C_X^+(t) - C_X^-(T-t) \, .
\ee 
In order to decrease correlation between measurements, we choose the source location  randomly on the whole lattice 
for each configuration.
Masses are extracted from the so called effective mass which is defined by
\be
am_{\rm eff}^X(t)=-\log(C_X(t)/C_X(t-1))= am_X+\log\left(\frac{1+\sum_{i=1}^\infty c_ie^{\Delta_i t}}{1+\sum_{i=1}^\infty c_ie^{\Delta_i (t-1)}}\right)
\tendto_{t\rightarrow \infty}  am_X \quad,
\label{meff}
\ee
where $\Delta_i= m_i-m_X$ is the mass difference of the excited state $i$ with 
respect to the ground state mass $m_X$.

\begin{figure}[h!]
\vspace*{1cm}
\begin{minipage}{8cm}
\epsfxsize=8truecm
\epsfysize=6truecm
 \mbox{\epsfbox{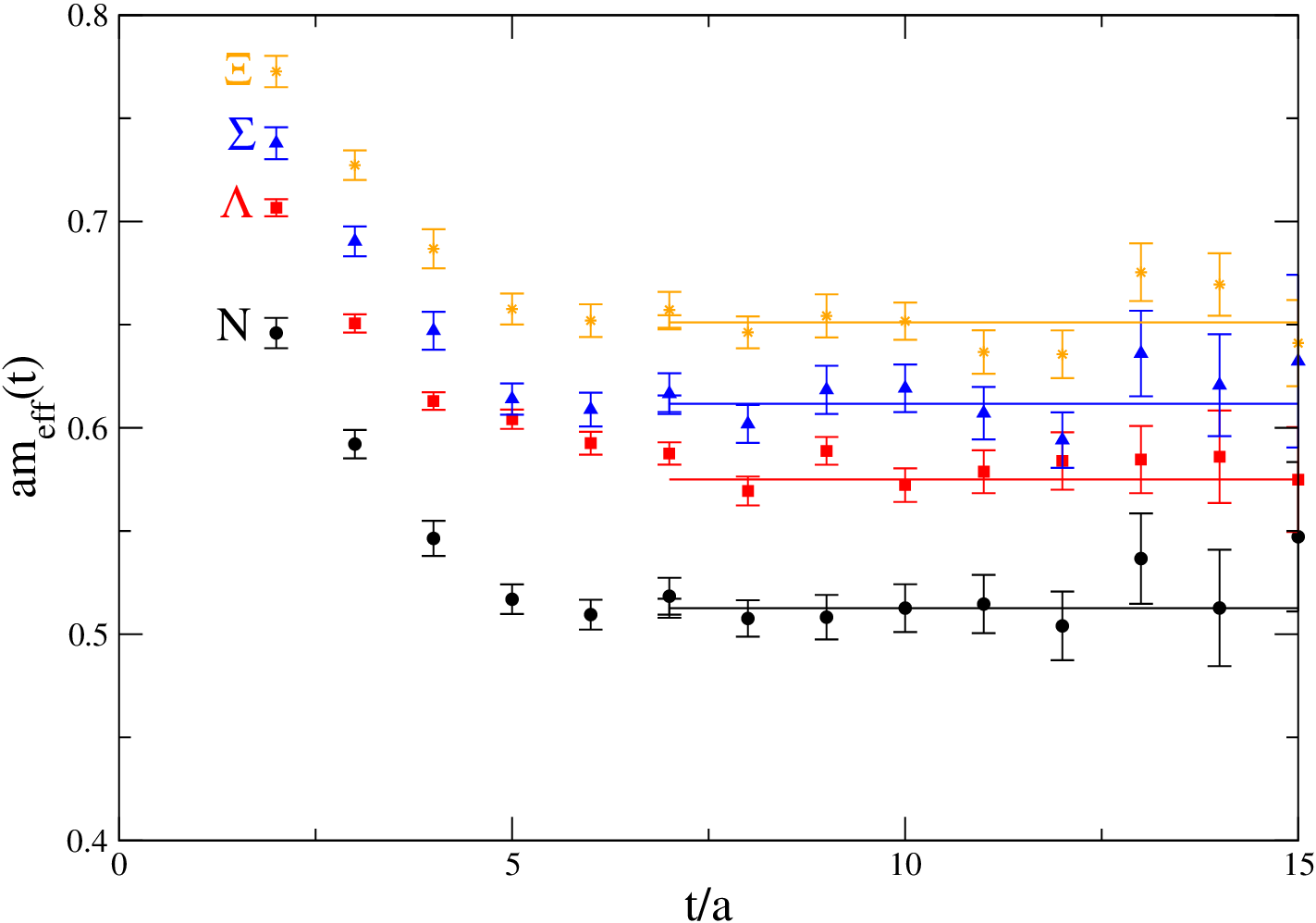}}
\caption{Effective masses of the octet states for $\beta = 3.9$, $a\mu = 0.004$ on a $32^3\times 64$ lattice using  $232$ configurations.} 
\label{fig:meff octet}
\end{minipage}
\hfill
\begin{minipage}{8cm}
\epsfxsize=8truecm
\epsfysize=6truecm
 \mbox{\epsfbox{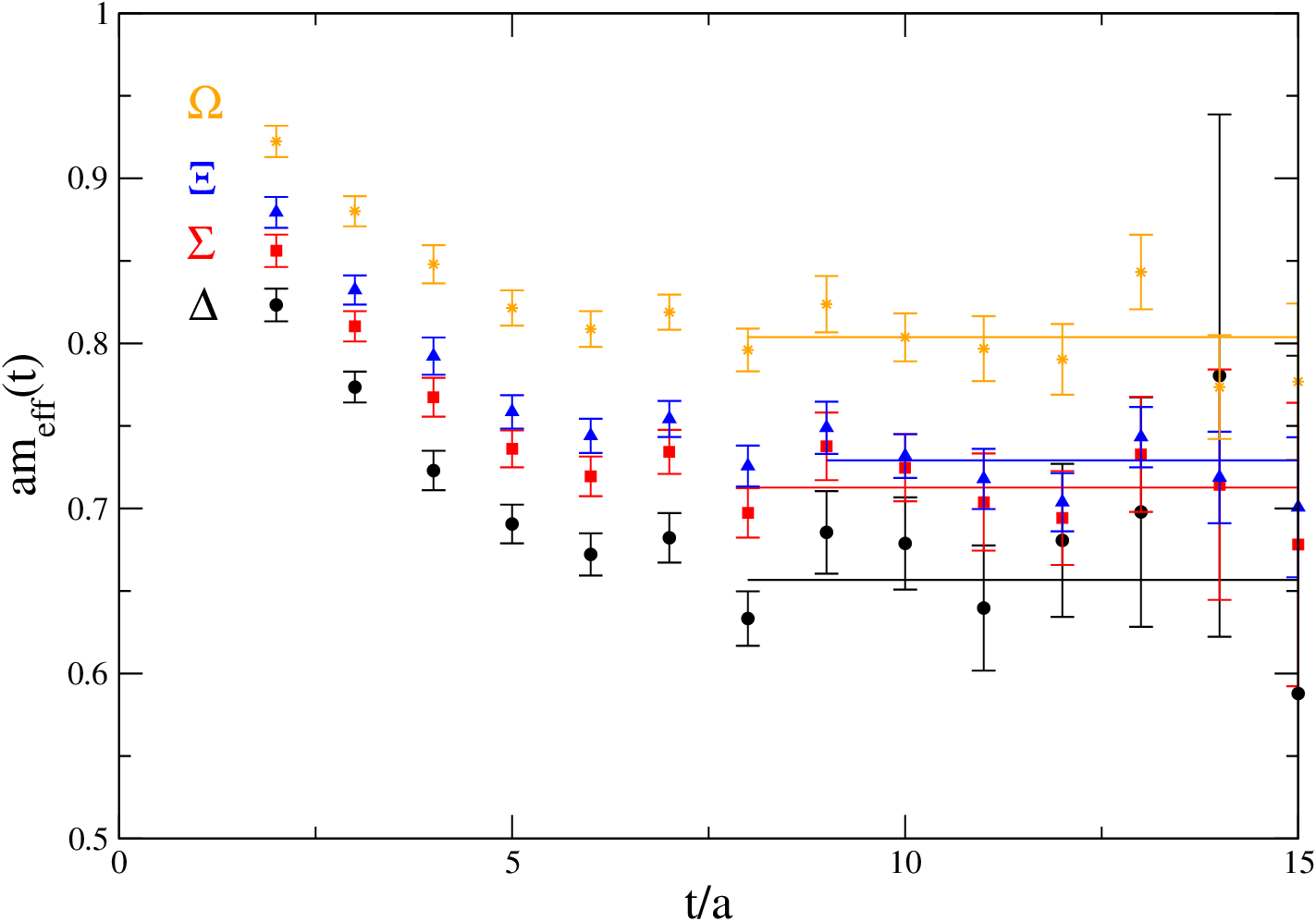}}
\caption{Effective masses of the decuplet states for $\beta = 3.9$, $a\mu = 0.004$ on a $32^3\times 64$ lattice using $232$ configurations.}
\label{fig:meff decuplet}
\end{minipage}
\end{figure}

In Figs.~\ref{fig:meff octet} and \ref{fig:meff decuplet} we show the effective masses of the baryons in the octet 
and decuplet representation respectively.
As can be seen a plateau region can be identified for all baryons. What is
shown in these figures are effective masses extracted from correlators where
smearing is applied both at the sink and source. 
Although local correlators are expected to have the same  value in the
large time limit,
smearing suppresses excited state contributions
yielding a plateau at earlier time separations and  a  better accuracy in the mass extraction.
Our fitting procedure to extract $m_X$ is as follows:
The mass is obtained from the leading term in Eq.~(\ref{meff}), i.e.from a constant fit to $m_X$. A second fit, including the first excited state,
 allows us to estimate the
  systematical error of the previously determined $m_X$
  due to excited states for a given plateau range. The plateau
   range is then chosen such that the
   systematical error on $m_X$ drops below 50\% of its statistical
           error
This criterion is in most of the cases in agreement
 with a $\chi^2/{\rm d.o.f.}< 1$.
In the cases in which this criterion is not satisfied a careful examination of the effective mass is made to ensure that the fit range is in the
plateau region. The results for the masses of the octet and decuplet
at $\beta=3.9$ are are collected in Tables~\ref{tab:masses-octet-3.9} and \ref{tab:masses-decuplet-3.9} respectively. The corresponding
results for the masses at $\beta=4.05$ are given in Table~\ref{tab:masses-octet-4.05} 
and \ref{tab:masses-decuplet-4.05}.
The errors are evaluated using both jackknife and the $\Gamma$-method~\cite{Wolff:2004} to ensure consistency.


\section{Results}

The bulk of the numerical results are presented in this section. Baryon masses are given in lattice units. Our procedure to convert the results to physical
units will be discussed in the next section.

\subsection{Baryon masses}

\begin{figure}[h!]
\begin{minipage}{8cm}
\epsfxsize=8truecm
\epsfysize=6truecm
 \mbox{\epsfbox{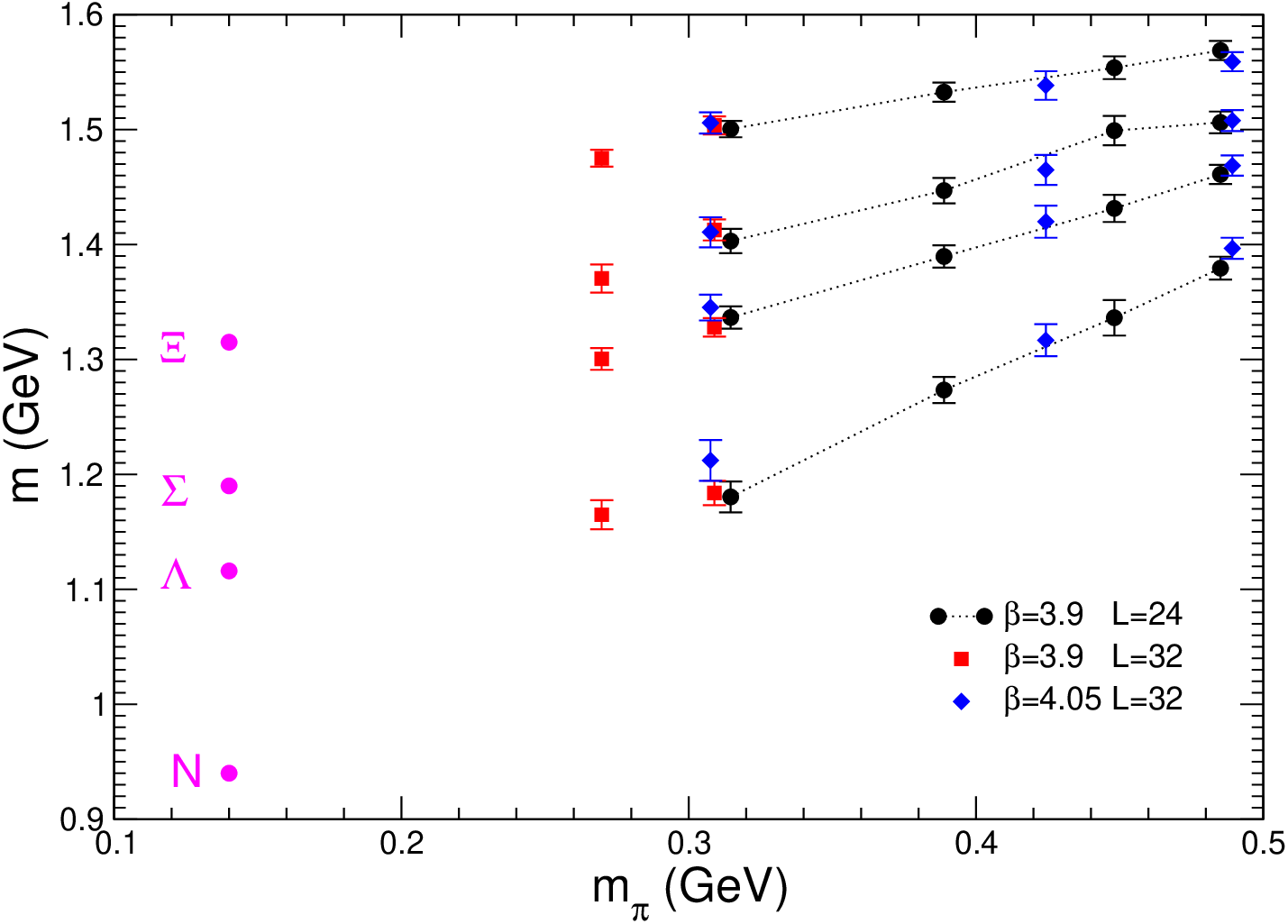}}
\caption{Octet states measured in our different gauge ensembles. Physical points are indicated by their name symbols (magenta). 
The data at $\beta=3.9$, $L/a=24$ are connected by dotted lines to guide the eye.} 
\label{fig:masses_octet}
\end{minipage}
\hfill
\begin{minipage}{8cm}
\epsfxsize=8truecm
\epsfysize=6truecm
 \mbox{\epsfbox{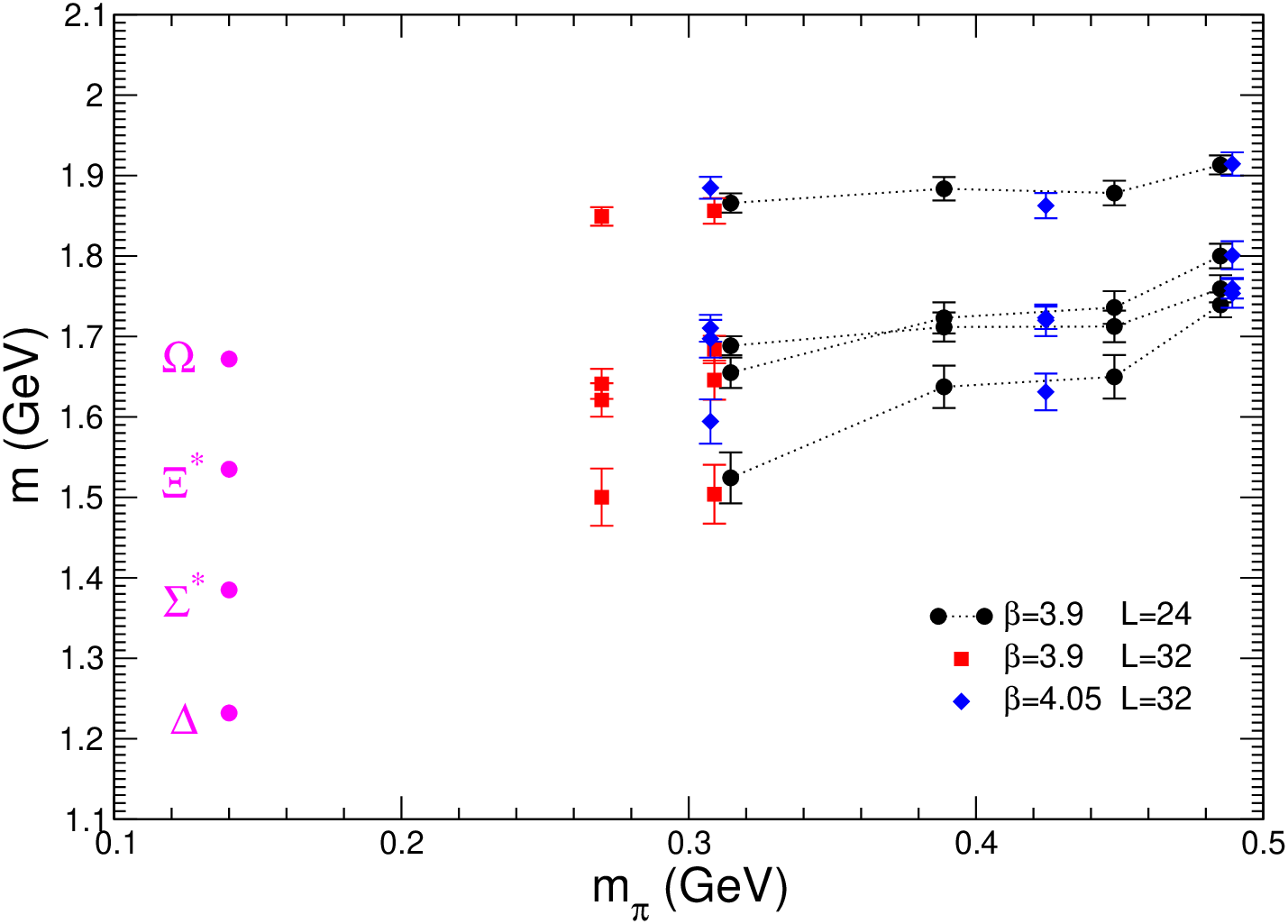}}
\caption{The same as Fig.~\ref{fig:masses_octet} but for the decuplet states. }
\label{fig:masses_decuplet}
\end{minipage}
\end{figure}

In Tables~\ref{tab:masses-octet-3.9} to \ref{tab:masses-decuplet-4.05}
we present the masses of 
the octet and decuplet states with the lattice input parameters given in 
Table~\ref{Table:params}. For the isospin multiplets we have computed separately the 
masses corresponding to each isospin components as well as their averaged value.
These results (averaged values in case of isospin multiplets) are displayed in 
Figs.~\ref{fig:masses_octet} and \ref{fig:masses_decuplet}. The $\beta=3.9$, $L/a=24$ data are 
linked by dotted lines to guide the eye. An inspection of the plots indicates that the lattice artifacts, studied in 
detail in the next section, are small. Notice that the natural order of the $\Sigma^{\ast}$  
and $\Xi^\ast$ states comes out to be correct for $m_\pi\leq 300$ MeV while for larger masses 
this order is inverted.

\begin{table}[h!]
\begin{small}
  \begin{center}
  \begin{tabular}{lcccccccccc}
  \hline\hline
$\Bigl.\Bigr.a\mu$   & stat.    & $am_N$ & $am_{\Lambda}$ & $am_{\Sigma^{\rm Av}}$& $am_{\Sigma^+}$ & $am_{\Sigma^0}$ &$am_{\Sigma^-}$&$am_{\Xi^{\rm Av}}$  &$am_{\Xi^0}$ &$am_{\Xi^-}$  \\
  \hline\hline
  \multicolumn{11}{c}{$24^3\times 48$}\\
  \hline
   $0.0040$ & 782 & $0.5111(58)$ &  $0.5787(42)$ &  $0.6075(46)$ &  $0.6175(66)$ &  $0.6118(48)$ &  $0.5959(52)$ &  $0.6497(31)$ &  $0.6695(42)$ &  $0.6372(31)$  \\
   $0.0064$ & 545 & $0.5514(49)$ &  $0.6017(42)$ &  $0.6265(48)$ &  $0.6487(72)$ &  $0.6278(52)$ &  $0.6131(52)$ &  $0.6636(36)$ &  $0.6876(50)$ &  $0.6500(36)$  \\
   $0.0085$ & 348 & $0.5786(67)$ &  $0.6198(51)$ &  $0.6491(55)$ &  $0.6679(62)$ &  $0.6529(49)$ &  $0.6358(46)$ &  $0.6728(43)$ &  $0.6956(58)$ &  $0.6593(43)$  \\
   $0.0100$ & 477 & $0.5973(43)$ &  $0.6326(36)$ &  $0.6522(41)$ &  $0.6662(56)$ &  $0.6539(43)$ &  $0.6429(44)$ &  $0.6793(36)$ &  $0.6959(49)$ &  $0.6683(32)$  \\
  \hline\hline
  \multicolumn{11}{c}{$32^3 \times 64$}\\
  \hline
   $0.0030$ & 652 & $0.4958(43)$ & $0.5613(33)$  & $0.5891(42)$ & $0.6069(68)$ & $0.5932(50)$ & $0.5775(39)$ & $0.6382(30)$ & $0.6572(44)$ & $0.6275(26)$ \\
   $0.0040$ & 232 & $0.5126(46)$ &  $0.5750(35)$ & $0.6117(40)$ & $0.6281(73)$ & $0.6158(40)$ & $0.5960(47)$ & $0.6511(34)$ & $0.6748(46)$ & $0.6358(32)$  \\
   \hline\hline
  \end{tabular}
  \caption{ Baryon masses  in the octet representation at $\beta =3.9$ in lattice units. }
  \label{tab:masses-octet-3.9}
 \end{center}
 \end{small}
\end{table}

\begin{table}[h!]
\begin{small}
 \begin{center}
 \begin{tabular}{lcccccccccccc}
    \hline\hline
    $\Bigl.\Bigr.a\mu$   & stat.      & $am_{\Delta^{++,-}}$ & $am_{\Delta^{+,0}}$&$am_{\Sigma^{\ast {\rm Av}}}$& $am_{\Sigma^{\ast +}}$ & $am_{\Sigma^{\ast 0}}$ &$am_{\Sigma^{\ast -}}$&$am_{\Xi^{\ast {\rm Av}}}$  &$am_{\Xi^{\ast 0}}$ &$am_{\Xi^{\ast -}}$ &$am_{\Omega}$ \\
    \hline\hline
\multicolumn{12}{c}{$24^3\times 48$}\\
\hline
   $0.0040$ & 782 &  $0.660(14)$  &  $0.670(13)$ &  $0.7166(82)$ &  $0.709(11)$ &  $0.7226(81)$ &  $0.7222(95)$ &  $0.7311(51)$ &  $0.7381(59)$ &  $0.7200(66)$ &  $0.8079(52)$  \\
   $0.0064$ & 545 &  $0.709(11)$  &  $0.711(12)$ &  $0.7461(84)$ &  $0.740(10)$ &  $0.7480(93)$ &  $0.7489(93)$ &  $0.7412(78)$ &  $0.7552(76)$ &  $0.7344(84)$ &  $0.8156(63)$  \\
   $0.0085$ & 348 &  $0.714(12)$  &  $0.733(13)$ &  $0.7517(88)$ &  $0.739(11)$ &  $0.760(11)$ &  $0.7645(98)$ &  $0.7415(85)$ &  $0.7529(81)$ &  $0.7367(84)$ &  $0.8133(66)$  \\
   $0.0100$ & 477 &  $0.7531(67)$  &  $0.7559(75)$ &  $0.7794(66)$ &  $0.7808(62)$ &  $0.7809(64)$ &  $0.7798(69)$ &  $0.7618(73)$ &  $0.7741(64)$ &  $0.7484(74)$ &  $0.8284(51)$  \\
\hline\hline
\multicolumn{12}{c}{$32^3 \times 64$}\\
    \hline
 $0.003$ & 652&  0.6234(139) & 0.6497(133) & 0.6859(96) & 0.6838(93) & 0.6859(106) & 0.7027(101) &  0.7058(50) & 0.7097(58) & 0.7032(53)  & 0.7926(49)\\ 
 $0.004$ & 232 &  $0.651(16)$  &  $0.659(15)$ &  $0.713(10)$ &  $0.705(12)$ &  $0.716(12)$ &  $0.7173(99)$ &  $0.7291(74)$ &  $0.7366(79)$ &  $0.7192(72)$ &  $0.8037(69)$  \\
      \hline\hline
  \end{tabular}
  \caption{Baryon masses  in the  decuplet representation at $\beta =3.9$ in lattice units. }
  \label{tab:masses-decuplet-3.9}
   \end{center}
 \end{small}
\end{table}

\begin{table}[h!]
\begin{small}
  \begin{center}
  \begin{tabular}{lcccccccccc}
  \hline\hline
$\Bigl.\Bigr.a\mu$   & stat.    & $am_N$ & $am_{\Lambda}$ & $am_{\Sigma^{av}}$& $am_{\Sigma^+}$ & $am_{\Sigma^0}$ &$am_{\Sigma^-}$&$am_{\Xi^{av}}$  &$am_{\Xi^0}$ &$am_{\Xi^-}$  \\
  \hline\hline
  \multicolumn{11}{c}{$32^3\times 64$}\\
  \hline
   $0.0030$ & 269 & $0.4091(60)$ &  $0.4540(38)$ &  $0.4761(44)$ &  $0.4885(62)$ &  $0.4774(47)$ &  $0.4651(53)$ &  $0.5082(31)$ &  $0.5177(39)$ &  $0.5007(29)$  \\
   $0.0060$ & 253 & $0.4444(47)$ &  $0.4792(47)$ &  $0.4944(44)$ &  $0.5022(66)$ &  $0.4960(45)$ &  $0.4834(45)$ &  $0.5192(42)$ &  $0.5277(50)$ &  $0.5112(37)$  \\
   $0.0080$ & 409 & $0.4714(31)$ &  $0.4957(30)$ &  $0.5089(31)$ &  $0.5179(41)$ &  $0.5095(32)$ &  $0.5019(31)$ &  $0.5262(28)$ &  $0.5350(34)$ &  $0.5199(25)$  \\
  \end{tabular}
  \caption{ Baryon masses  in the  octet p representation at $\beta =4.05$ in lattice units. }
  \label{tab:masses-octet-4.05}
 \end{center}
 \end{small}
\end{table}

\begin{table}[h!]
\begin{small}
 \begin{center}
 \begin{tabular}{lcccccccccccc}
    \hline\hline
    $\Bigl.\Bigr.a\mu$   & stat.      & $am_{\Delta^{++,-}}$ & $am_{\Delta^{+,0}}$&$am_{\Sigma^{\ast av}}$& $am_{\Sigma^{\ast +}}$ & $am_{\Sigma^{\ast 0}}$ &$am_{\Sigma^{\ast -}}$&$am_{\Xi^{\ast av}}$  &$am_{\Xi^{\ast 0}}$ &$am_{\Xi^{\ast -}}$ &$am_{\Omega}$ \\
    \hline\hline
\multicolumn{12}{c}{$32^3\times 64$}\\
\hline
   $0.0030$ & 269 &  $0.5381(93)$  &  $0.5441(93)$ &  $0.5728(79)$ &  $0.5673(94)$ &  $0.5750(86)$ &  $0.5734(80)$ &  $0.5772(56)$ &  $0.5796(54)$ &  $0.5750(56)$ &  $0.6361(46)$  \\
   $0.0060$ & 253 &  $0.5505(77)$  &  $0.5581(90)$ &  $0.5805(66)$ &  $0.5754(71)$ &  $0.581(11)$ &  $0.5844(68)$ &  $0.5816(47)$ &  $0.5834(50)$ &  $0.5802(48)$ &  $0.6286(53)$  \\
   $0.0080$ & 409 &  $0.5918(60)$  &  $0.5906(63)$ &  $0.6078(59)$ &  $0.6044(68)$ &  $0.5850(74)$ &  $0.6099(57)$ &  $0.5940(43)$ &  $0.6021(50)$ &  $0.5873(43)$ &  $0.6461(49)$  \\
      \hline\hline
  \end{tabular}
  \caption{Baryon masses  in the  decuplet representation at $\beta =4.05$ in lattice units. }
  \label{tab:masses-decuplet-4.05}
   \end{center}
 \end{small}
\end{table}

\newpage
\subsection{Strange quark mass dependence}

\begin{figure}[h!]
\vspace*{1cm}
\begin{minipage}{8cm}
\epsfxsize=8truecm
\epsfysize=6truecm
 \mbox{\epsfbox{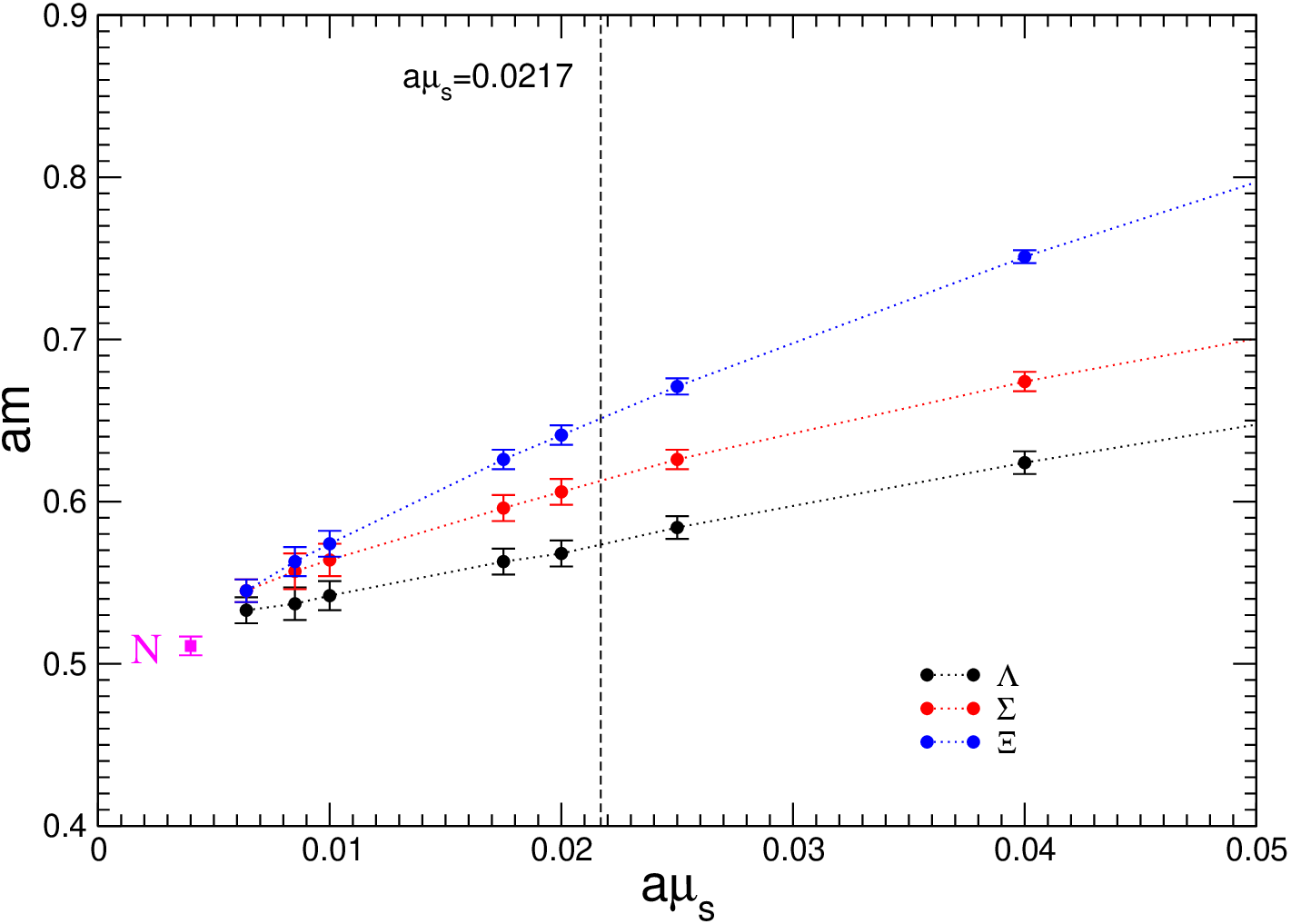}}
\caption{Masses for octet baryons
 at $\beta = 3.9$ and $a\mu = 0.004$ on a lattice of size $24^3\times 48$ as 
a function of $a \mu_s$. The vertical dashed line indicates the value of the tuned bare 
strange quark mass. The dotted lines are to guide the eye.} 
\label{fig:squark_octet}
\end{minipage}
\hfill
\begin{minipage}{8cm}
\epsfxsize=8truecm
\epsfysize=6truecm
 \mbox{\epsfbox{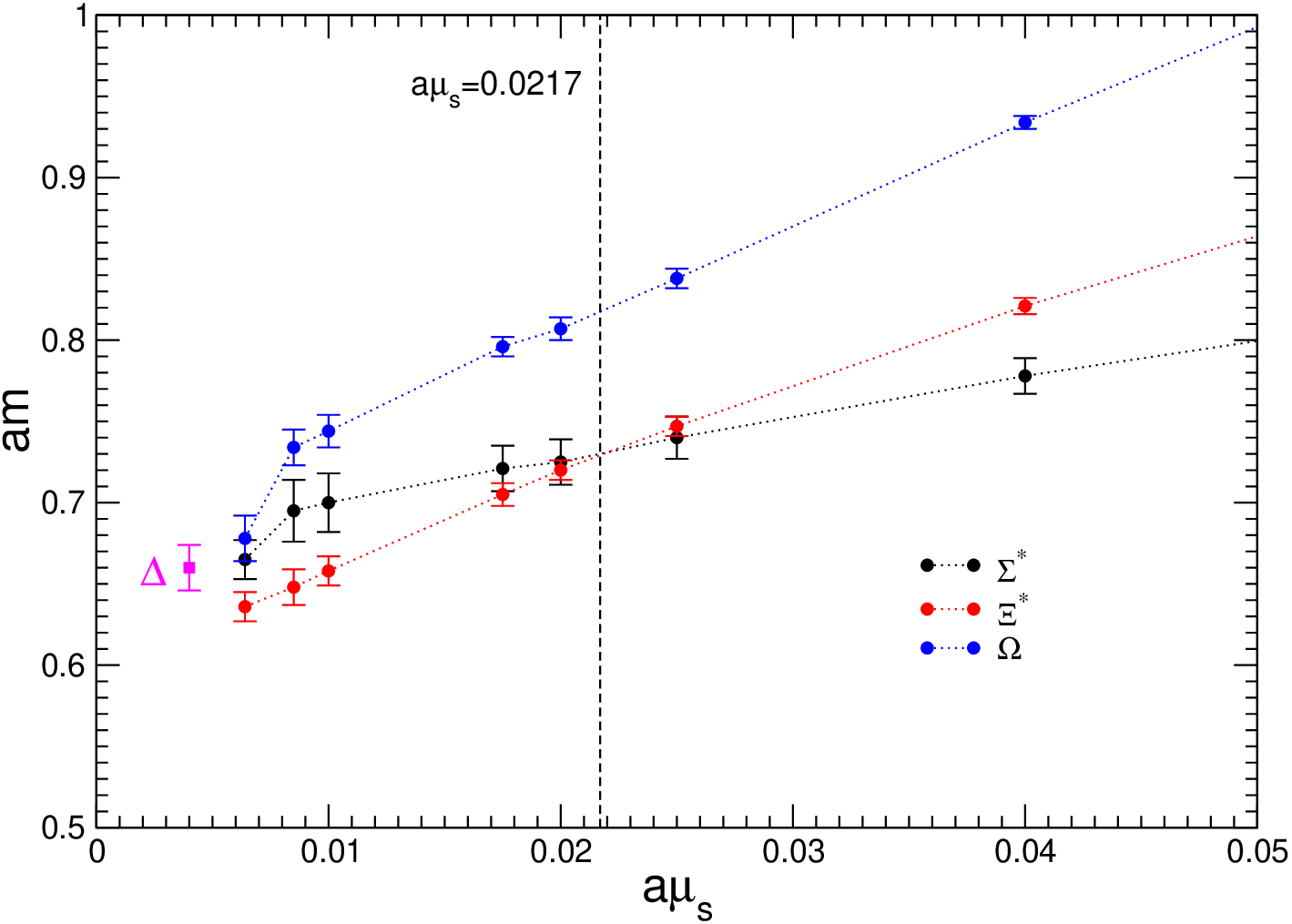}}
\caption{The same as for Fig.~\ref{fig:squark_octet} but 
for the decuplet baryons. }
\label{fig:squark_decuplet}
\end{minipage}
\end{figure}

The dependence of the masses of baryons with strangeness on
 the bare strange quark mass has been investigated at $\beta = 3.9 $ 
for  $a\mu = 0.004 $. 
The results are given in Tables~\ref{tab:squark-octet-3.9} and 
\ref{tab:squark-decuplet-3.9} and displayed in Figs.~\ref{fig:squark_octet} and  
\ref{fig:squark_decuplet}. The vertical dotted line indicates the value of
the tuned bare strange 
quark mass as given in Table \ref{Table:mus}. The $SU(3)$ symmetric
 point  $\mu_s = \mu$ is given by the nucleon and $\Delta$ mass for the octet and decuplet respectively.
As can be seen in the $SU(3)$ limit all the octet and decuplet masses converge to a single point up to
cut-off effects and the fact that we only have $N_f=2$ simulations.
For clarity we only show in Fig.~\ref{fig:squark_octet}  the mass of
$\Lambda$,  $\Sigma^{\rm Av}$ and  $\Xi^{\rm Av}$. They should be degenerate
 with the nucleon in the limit of $\mu_s = \mu$. 
 Indeed, if one computes the nucleon mass with 
the same  statistics with that used for $\Sigma^{\rm Av}$ and $\Xi^{\rm Av}$,
one finds them to be degenerate  within the errors as can be
seen  in Fig.~\ref{fig:squark_octet}. 

The corresponding results for the decuplet-baryons
 are displayed in Fig.~\ref{fig:squark_decuplet}. 
As can be seen, also in the case of the decuplet masses there is
convergence to the  $\Delta$ mass
as predicted in the exact $SU(3)$  limit $\mu_s = \mu$. 

\begin{table}[h!]
\begin{small}
  \begin{center}
  \begin{tabular}{lccccccccc}
  \hline\hline
$\Bigl.\Bigr.a\mu_s$   & stat.     & $am_{\Lambda}$ & $am_{\Sigma^{\rm Av}}$& $am_{\Sigma^+}$ & $am_{\Sigma^0}$ &$am_{\Sigma^-}$&$am_{\Xi^{\rm Av}}$  &$am_{\Xi^0}$ &$am_{\Xi^-}$  \\
  \hline\hline
  \multicolumn{10}{c}{$24^3\times 48$}\\
  \hline
  $0.0064$ & 597 &   $0.533(8)$   &   $0.545(7)$   &   $0.549(12)$   &   $0.563(5)$   &   $0.537(6)$   &   $0.545(7)$   &   $0.560(11)$   &   $0.530(6)$   \\
  $0.0085$ & 316 &   $0.537(10)$  &   $0.557(11)$   &   $0.559(19)$   &   $0.557(12)$   &   $0.554(8)$   &   $0.563(9)$   &   $0.585(14)$   &   $0.549(8)$   \\
  $0.0100$ & 316 &   $0.542(9)$   &   $0.564(10)$   &   $0.567(18)$   &   $0.564(11)$   &   $0.561(7)$   &   $0.574(8)$   &   $0.597(13)$   &   $0.560(7)$   \\
  $0.0175$ & 315 &   $0.563(8)$   &   $0.596(9)$   &   $0.600(14)$   &   $0.593(9)$   &   $0.593(7)$   &   $0.626(6)$   &   $0.644(8)$   &   $0.610(6)$   \\
  $0.0200$ & 308 &   $0.568(8)$   &   $0.606(8)$   &   $0.609(13)$   &   $0.602(9)$   &   $0.603(6)$   &   $0.641(6)$   &   $0.660(8)$   &   $0.625(5)$   \\
  $0.0250$ & 311 &   $0.584(7)$   &   $0.626(6)$   &   $0.627(12)$   &   $0.619(8)$   &   $0.620(6)$   &   $0.671(5)$   &   $0.688(7)$   &   $0.656(5)$   \\
  $0.0400$ & 316 &   $0.624(7)$   &   $0.674(6)$   &   $0.676(10)$   &   $0.667(7)$   &   $0.672(6)$   &   $0.751(4)$   &   $0.764(5)$   &   $0.738(4)$   \\
  $0.0800$ & 314 &   $0.718(7)$   &   $0.780(5)$   &   $0.787(7)$   &   $0.772(7)$   &   $0.776(7)$   &   $0.935(3)$   &   $0.945(4)$   &   $0.926(3)$   \\

   \hline\hline
   \end{tabular}
  \caption{Octet masses for $\beta = 3.9$, $a\mu = 0.004$ on a $24^3\times 48$ lattice as  a function of $a \mu_s$.}
  \label{tab:squark-octet-3.9}
 \end{center}
 \end{small}
\end{table}

\begin{table}[h!]
\begin{small}
 \begin{center}
 \begin{tabular}{lcccccccccc}
    \hline\hline
    $\Bigl.\Bigr.a\mu_s$   & stat.     &$am_{\Sigma^{\ast {\rm Av}}}$& $am_{\Sigma^{\ast +}}$ & $am_{\Sigma^{\ast 0}}$ &$am_{\Sigma^{\ast -}}$&$am_{\Xi^{\ast {\rm Av}}}$  &$am_{\Xi^{\ast 0}}$ &$am_{\Xi^{\ast -}}$ &$am_{\Omega}$ \\
    \hline\hline
\multicolumn{10}{c}{$24^3\times 48$}\\
\hline
  $0.0064$ & 597 &  $0.665(12)$   &   $0.658(18)$   &   $0.669(14)$   &   $0.669(14)$   &   $0.636(9)$   &   $0.645(12)$   &   $0.628(9)$    &   $0.678(14)$    \\  
 $0.0085$ & 316 &  $0.695(19)$   &   $0.719(13)$   &   $0.713(16)$   &   $0.733(9)$   &   $0.648(11)$   &   $0.670(15)$   &   $0.624(10)$    &   $0.734(11)$    \\
 $0.0100$ & 316 &  $0.700(18)$   &   $0.722(12)$   &   $0.715(15)$   &   $0.697(22)$   &   $0.658(9)$   &   $0.680(13)$   &   $0.636(9)$    &   $0.744(10)$    \\   
 $0.0175$ & 315 &   $0.721(14)$   &   $0.729(14)$   &   $0.736(12)$   &   $0.718(19)$   &   $0.705(7)$   &   $0.722(8)$   &   $0.690(7)$    &   $0.796(6)$    \\
 $0.0200$ & 308 &  $0.725(14)$   &   $0.734(13)$   &   $0.741(11)$   &   $0.718(17)$   &   $0.720(6)$   &   $0.734(7)$   &   $0.704(7)$    &   $0.807(7)$    \\ 
 $0.0250$ & 311 &   $0.740(13)$   &   $0.735(16)$   &   $0.753(11)$   &   $0.740(17)$   &   $0.747(6)$   &   $0.759(6)$   &   $0.733(6)$    &   $0.838(6)$    \\
 $0.0400$ & 316 &  $0.778(11)$   &   $0.770(13)$   &   $0.788(9)$   &   $0.778(15)$   &   $0.821(5)$   &   $0.831(5)$   &   $0.811(5)$    &   $0.934(4)$    \\
 $0.0800$ & 314 &  $0.864(7)$   &   $0.854(11)$   &   $0.870(8)$   &   $0.865(9)$   &   $0.993(4)$   &   $0.996(5)$   &   $0.987(4)$    &   $1.169(3)$    \\
      \hline\hline
  \end{tabular}
  \caption{Decuplet masses for $\beta = 3.9$, $a\mu = 0.004$ on a $24^3\times 48$ lattice as  a function of $a \mu_s$}
  \label{tab:squark-decuplet-3.9}
   \end{center}
 \end{small}
\end{table}


The $\mu_s$ dependence of the strange baryon masses
 provides an  estimate of systematic errors 
 due to the uncertainty in the tuning of the strange quark mass. As already explained, the kaon mass
 at the
physical point is used to fix $\mu_s$. This gives $a\mu_s=0.0217(22)$. The $\sim 10$\% uncertainty 
leads to a corresponding error in the strange baryon masses that can be estimated by the variation of
their masses in the vicinity of $\mu_s$. At $\mu=0.004$ we estimate an error that is comparable
to the statistical error.  In what follows we will analyze our results taking into account only statistical
errors. This analysis shows that when the statistical error is given
on the final results of strangeness non-zero baryon masses
 one must bear in mind
that there is a systematic error of about the same 
magnitude  due to the strange quark mass determination.

\subsection{Gell-Mann-Okubo relation}

Assuming a small SU(3) breaking, Okubo derived interesting relations among
baryons masses.
We examine in this section how well 
the Gell-Mann-Okubo (GMO) relations~\cite{Donoghue:1992dd} 
are fulfilled for the baryons masses obtained on our lattices
 at different pion mass values.
As we will discuss in detail in the next Section, 
volume and discretization effects are small, and therefore
it suffices to analyze  the $\beta=3.9$ and $L=24\times 48$ results.
For this study we use the lattice spacing determined from $f_\pi$ to convert to physical units.

\begin{figure}[h!]
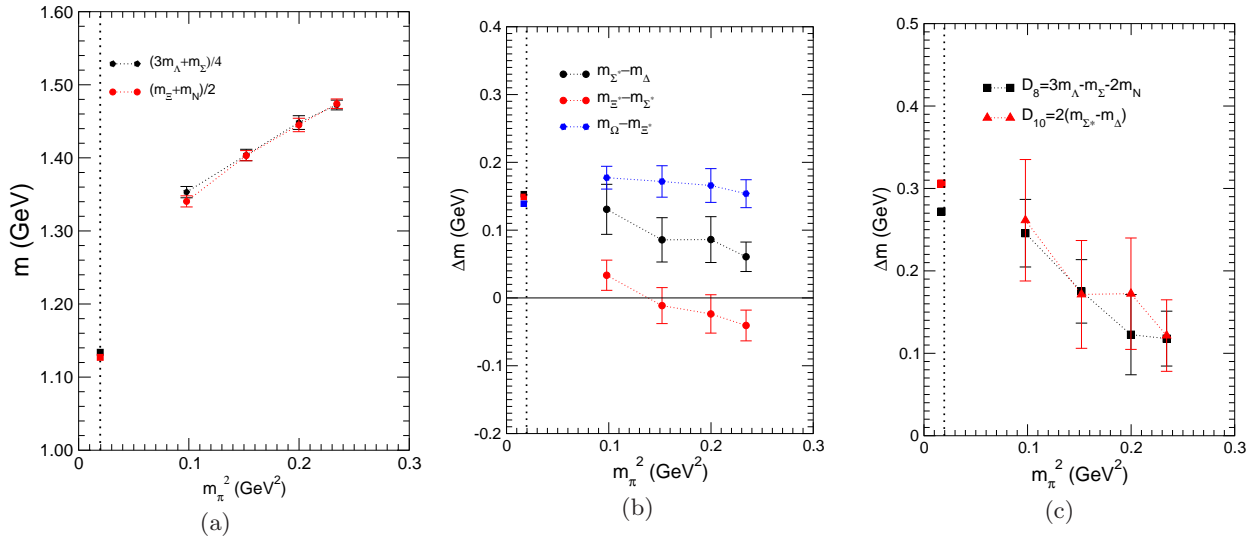

\begin{minipage}{5.5cm}
\mbox{\epsfxsize=5.5cm\epsffile{OKUBO_Octet.eps}} 

(a)

\end{minipage}\hspace{0.cm}
\begin{minipage}{5.5cm}
\mbox{\epsfxsize=5cm\epsffile{OKUBO_Decuplet.eps}}

(b)
\end{minipage}\hspace{0.cm}
\begin{minipage}{5.5cm}
\mbox{\epsfxsize=5cm\epsffile{OKUBO_Octet_Decuplet.eps}}

(c)
\end{minipage}

\caption{Gell Mann Okubo relations for the baryon octet (a) decuplet (b) and mixed 
octet-decuplet (c) as a function of $m_{\pi}^2$. Vertical lines correspond to the 
physical results. Data were obtained from simulations at $\beta=3.9$ and $L/a=24$.
The lattice spacing determined from $f_\pi$ is used to convert to physical units.}
\label{fig:GMO}
\end{figure}

For the $J^{P}=1/2^+$ octet the GMO relation can be written in the form:
\begin{equation}\label{OKUBO_Octet}
\frac{M_{\Xi}+M_N}{2}  = \frac{3M_{\Lambda}+M_{\Sigma}}{4} \; .
\end{equation}
The results are displayed in Fig.~\ref{fig:GMO}
 where the left and right hand side terms
of Eq.~(\ref{OKUBO_Octet}) are separately plotted as a function of $m_{\pi}^2$. The difference
between the two terms 
are compatible with zero at any pion mass. The experimental values, shown by the
squares, are respectively 
254 MeV and 248 MeV.
These results are similar
 to those presented in Ref.~\cite{Beane:2006pt} 
using a mixed action setup with valence domain wall fermions on 
rooted staggered sea fermions.

For the $J^{P}=3/2^+$ decuplet, the GMO relations 
predict equal mass difference among 
two consecutive ($\Delta S=1$) isospin multiplets:
\begin{equation}\label{OKUBO_Decuplet}
 M_{\Sigma^*} - M_{\Delta} =   M_{\Xi^*}- M_{\Sigma^*} = M_{\Omega}-M_{\Xi^*}\; .  
\end{equation}
The results for the decuplet baryons 
are displayed in Fig.~\ref{fig:GMO}. As can be seen, 
the equalities of Eq.~(\ref{OKUBO_Decuplet}) 
are  strongly violated; the three
mass differences of Eq.~(\ref{OKUBO_Decuplet}) are 
spread over about 200 MeV for the range of pion masses 
that have been computed. 
The experimental values for these
mass differences are $153, 149, 139$~MeV, shown in the plot
 by the squares. In the lattice results
the larger deviation comes from $M_{\Xi^*}- M_{\Sigma^*}$,   
while for $M_{\Sigma^*} - M_{\Delta}$ and $M_{\Omega}-M_{\Xi^*}$  
the mass differences are smaller.
The mass difference $M_{\Xi^*}- M_{\Sigma^*}$ is increasing 
as the pion mass decreases.
Unfortunately, with our present statistics it is unclear
whether this increase is sufficient to 
 bring this mass difference in agreement
with experiment but the trend is  definitely in the right direction.

A third relation exists, that connects
the $J^{P}=1/2^+$  octet masses with the $J^{P}=3/2^+$ decuplet masses, 
which reads as:
\begin{equation}\label{OKUBO_8_10}
  3M_{\Lambda}-M_{\Sigma}-2M_N = 2(M_{\Sigma^*}-M_{\Delta}) \; .  
\end{equation}
Experimentally, this relation is fulfilled at the 10\% level yielding 276 MeV 
for the left hand side 
and 305 MeV for the right hand side of Eq.~(\ref{OKUBO_8_10}).
These values are again shown
by the filled squares in Fig.~\ref{fig:GMO}.
The corresponding lattice results are shown in the same figure. 
One can see that, as in the octet case, the relation of Eq.~(\ref{OKUBO_8_10}) is 
satisfied within our statistical uncertainties at each pion mass. It also
 approaches   the experimental results with decreasing pion mass. 

\vspace{+0.8cm}

Fulfillment of the GMO relations is considered a success of $SU(3)$ 
symmetry. Violations of these relations indicate that $SU(3)$ breaking
is not small. Therefore one would expect that these relations
are better satisfied as  we approach the $SU(3)$ limit  $a\mu=a\mu_s=0.0217$, up
 to discretization effects. This corresponds to about $m_{\pi}^2\sim 0.50$~GeV$^2$. For the decuplet mass relation given in Eq.~(\ref{OKUBO_Decuplet}) it is
unclear if this would be indeed satisfied by the lattice data whereas the
other two relations are fulfilled
at all masses.

\section{Systematics}

In order to compare our lattice results collected in 
Tables~\ref{tab:masses-octet-3.9},~\ref{tab:masses-decuplet-3.9},~\ref{tab:masses-octet-4.05} 
and~\ref{tab:masses-decuplet-4.05} to the physical masses we need to check
for finite volume effects, cut-off effects and the extrapolation to the physical
light quark masses. The strange quark was fixed to the physical value using
the kaon mass with the light quarks extrapolated to the physical point
as explained 
in Section~II.B. A check of the effect of this tuning on baryon masses has been discussed 
in Section~III.B. 
In this section we discuss finite volume and
         cutoff-effects, in particular the isospin breaking. 

\subsection{Finite volume effects} 

Finite volume corrections to the nucleon mass in $N_f=2$ 
lattice QCD have been studied in Ref.~\cite{AliKhan:2003cu} within the
$p$ expansion which assumes that finite size effects originate from pions
that propagate around the spatial box. Using
relativistic $SU(2)$ baryon chiral perturbation theory~\cite{Procura:2003ig}
the finite volume corrections to the nucleon mass to ${\cal O}(p^4)$ are:
\be
m_N(\infty)=m_N(L)-\delta m_a(L)-\delta m_b(L)
\ee
where
\beq
\delta m_a(L) &=& \frac{3g_A^2 m_N^0 m_\pi^2}{8\pi^2 f_\pi^2} \int_0^\infty
dx \sum^\prime_{\bf n} K_0\left(L|{\bf n}|\sqrt{(m_N^0)^2 x^2+m_\pi^2(1-x)}\right) \nonumber \\
\delta m_b(L) &=& \frac{3m_\pi^4}{2\pi^2 f_\pi^2} 
 \sum^\prime_{\bf n}\biggl [ (2c_1-c_3) \frac{K_1\left(L|{\bf n}|m_\pi\right)}
{L|{\bf n}|m_\pi}
+c_2\frac{K_2\left(L|{\bf n}|m_\pi\right)}{\left(L|{\bf n}|m_\pi\right)^2} \biggr] \quad.
\label{volume corrections}
\eeq
$K_\nu(x)$ is the modified Bessel function and the sum is over all integer vectors
 ${\bf n}$ excluding ${\bf n}= {\bf 0}$.
The parameters $m_N^0$ and $c_1$ are determined
by fitting first the nucleon mass to the same order~\cite{Steininger98,Bernard:2004,Bernard:2005} given by 
\beq
m_N & = & m_N^0-4 c_1m_\pi^2- \frac{3g_A^2}{16\pi f_\pi^2} m_\pi^3 -4 E_1(\lambda)m_\pi^4 
        +\frac{3 m_\pi^4}{16 \pi^2 f_\pi^2}\biggl[\frac{1}{4} \left (c_2-\frac{2g_A^2}{m_N^0}\right)
- \left(c_2-8c_1+4c_3+\frac{g_A^2}{m_N^0}\right) \log\left (\frac{m_\pi}{\lambda}\right)\biggr]\quad .
\label{HBchi2}
\eea
We take the cut-off scale $\lambda=1$~GeV,  $f_{\pi}=130.70$~MeV and fix the dimension two low energy  constants 
$ c_2=3.2$~GeV$^{-1}$~\cite{Fettes:1998} and 
$c_3=-3.45$~GeV$^{-1}$~\cite{Bernard:2004,Procura:2006}.  These values are consistent with empirical
nucleon-nucleon phase shifts~\cite{Entem:2002sf,Epelbaum:2004}.
The counter-term $E_1$ is taken as an additional fit parameter.
We then use these parameters to estimate the volume corrections
to the nucleon mass. The results are listed 
in Table~\ref{tab:volume corrections}. As can be seen
the corrections for our lattices are, in all cases except one, 
smaller than the statistical
errors. In the analysis that  follows we will use the volume corrected nucleon mass.

\begin{table}[h!]
 \begin{center}
 \begin{tabular}{lcc}
    \hline\hline
    $am_\pi $  &  $am_N(L)$ & $a\delta_a(L)$ + $a\delta_b(L)$
\\
\hline
\multicolumn{3}{c}{$\beta=3.9$ $24^3\times 48$}\\
   0.1362 &  0.5111(58) &  0.0068 \\
   0.1684 &  0.5514(49) &  0.0046  \\
   0.1940 &  0.5786(67) &  0.0026  \\
   0.2100 &  0.5973(43) &  0.0021  \\
\multicolumn{3}{c}{$\beta=3.9$ $32^3\times 64$}\\
   0.1168 &  0.4958(34) &  0.0014 \\
   0.1338 &  0.5126(46) &  0.0011 \\
\multicolumn{3}{c}{$\beta=4.05$ $32^3\times 64$}\\
   0.1038 &  0.4091(60) &  0.0035\\
   0.1432 &  0.4444(47) &  0.0018\\
   0.1651 &  0.4714(31) &  0.0012\\
      \hline\hline
  \end{tabular}
  \caption{Volume correction to the nucleon mass.}
  \label{tab:volume corrections}
   \end{center}
\end{table}

Concerning the other baryons a recent analysis using SU(3) heavy baryon
chiral perturbation theory has shown that
the volume corrections are smaller than for the nucleon~\cite{Ishikawa:2009vc}.
Given that the volume correction found for the nucleon are smaller
than the statistical errors we can safely neglect any volume corrections
for the other baryons computed in this work. This is also corroborated by
our lattice results at $a\mu=0.004$ where simulations at two volumes are used.

\subsection{Isospin breaking}

\begin{figure}[h!]
\vspace*{2cm}
\epsfxsize=10truecm
\epsfysize=12truecm
 \mbox{\epsfbox{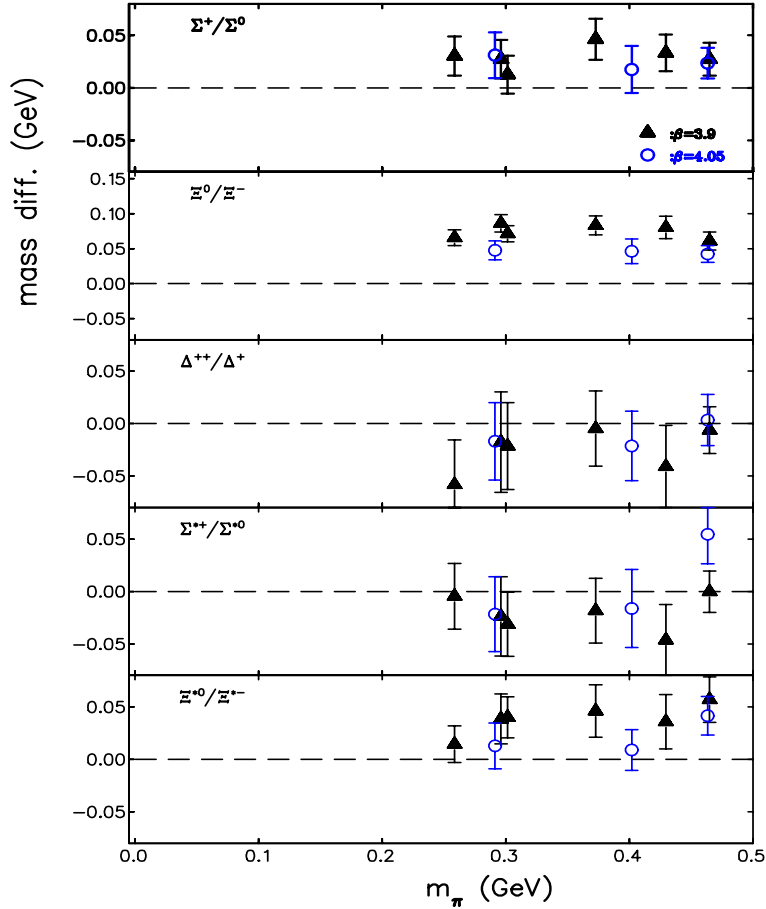}}
\caption{Mass splitting versus the pion mass at $\beta = 3.9$ (filled triangles) and $\beta=4.05$  (open circles) for from top to 
bottom: $\Sigma^+$ and $\Sigma^0$,  $\Xi^0$ and $\Xi^-$,  $\Delta^{++}$ and $\Delta^+$,  $\Sigma^{*+}$ and $\Sigma^{*0}$, and  $\Xi^{*0}$ and $\Xi^{*-}$.}
\label{fig:isospin}
\end{figure}

The twisted mass action breaks isospin explicitly  to ${\cal O}(a^2)$. 
How large this breaking is depends on the size of the ${\cal O}(a^2)$ terms.
It was shown that this cut-off effect is 
particularly large for the neutral 
pion~\cite{Frezzotti:2007qv} but small for other quantities. Indeed
we verified that isospin breaking between the $\Delta^{++,0}$ and $\Delta^{+,-}$
is consistent with zero for lattice spacings below about 
0.1~fm~\cite{Alexandrou:2008tn}. We here address this issue for the octet and
decuplet baryons. We show in Fig.~\ref{fig:isospin} the mass differences
for the $\Sigma$, $\Xi$, $\Delta$, $\Sigma^*$ and $\Xi^*$ charge multiplets as a function of the pion mass
at two values of $\beta$. 

As can be seen, we confirm that for the $\Delta$ system isospin breaking is
consistent with zero. This is also true for the $\Sigma^*$ and $\Xi^*$ 
as well as for the $\Sigma$ at the smaller lattice spacing. In the case
of the $\Xi$ we observe a non-zero splitting that decreases with the 
lattice spacing. If one interpolates the results at the two $\beta$ values
to the same pion mass, as discussed in more detail in the next Section, and makes a linear extrapolation in $a^2$ one
finds that this splitting goes to zero in the continuum limit as expected.
Whereas this confirms that this splitting is a cut-off effect,
to perform a proper analysis one would need results at an additional 
lattice spacing. For the current work we conclude that isospin
splitting at these two lattice spacings
is negligible  for all baryons expect for the $\Xi$ where an isospin 
breaking of about 6\% is observed that vanishes at a rate proportional to $a^2$.

\subsection{Continuum extrapolation}

\begin{figure}[h!]
\begin{minipage}{8cm}
\epsfxsize=8truecm
\epsfysize=10truecm
\mbox{\epsfbox{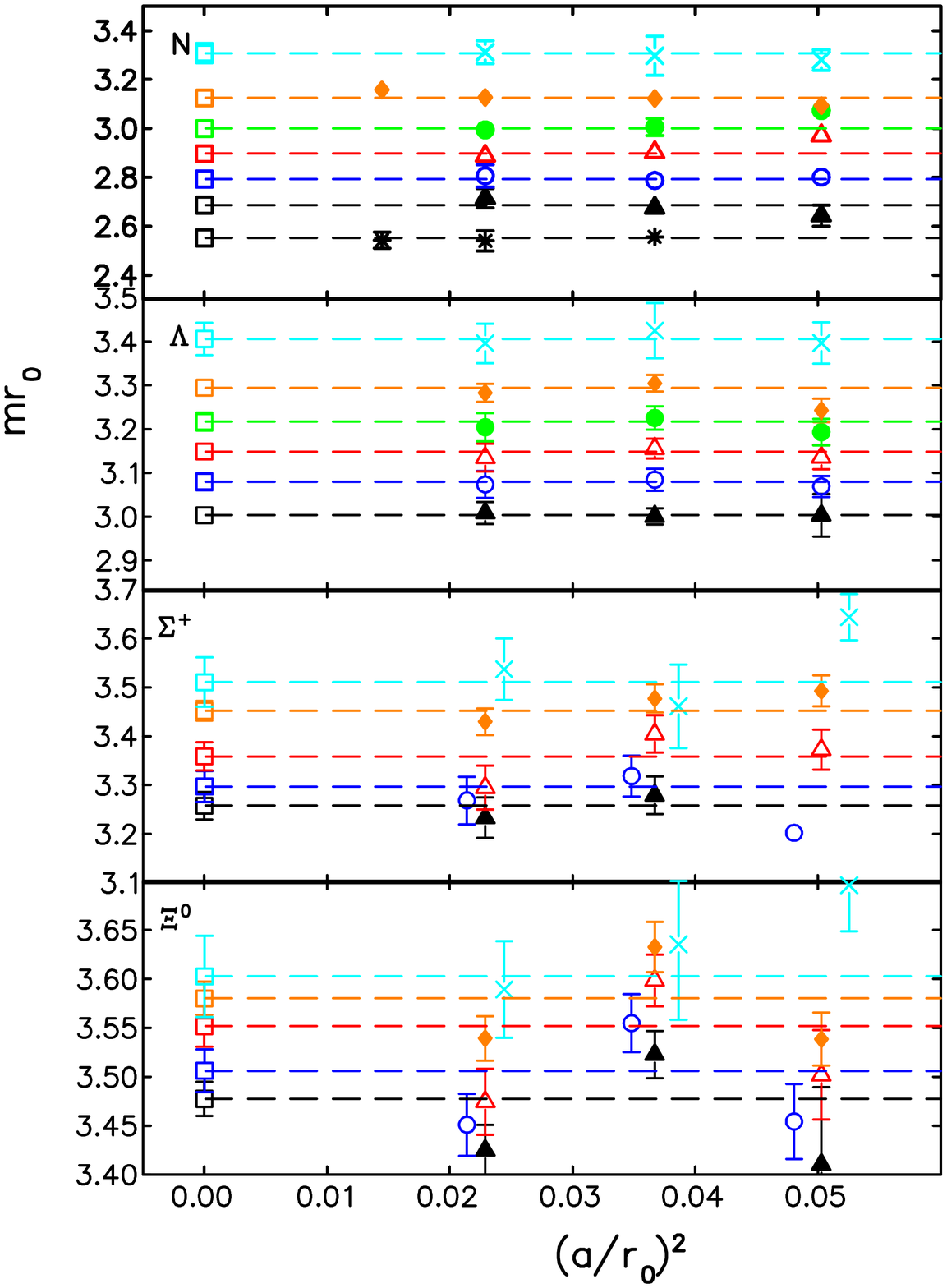}}
\caption{Constant extrapolation to the continuum limit for the octet baryons.
Stars are for $r_0 m_\pi=0.615$,
 filled triangles for  $r_0 m_\pi=0.7$, open circles   for  $r_0 m_\pi=0.8$,
open triangles  for  $r_0 m_\pi=0.9$, filled circles   for  $r_0 m_\pi=1.0$,
rhombii  for  $r_0 m_\pi=1.1$ and crosses for  $r_0 m_\pi=1.25$.
The open squares show the extracted continuum value. For the nucleon we also
show results  at $\beta=4.2$. For the $\Sigma^+$ and $\Xi^0$ we omit the case $r_0 m_\pi=1.0$ and
shift results at  $r_0 m_\pi=1.25$ and $r_0 m_\pi=0.8$ for clarity.   } 
\label{fig:octet continuum const}
\end{minipage}
\hfill
\begin{minipage}{8cm}
\epsfxsize=8truecm
\epsfysize=10truecm
\vspace*{-1.4cm}
\mbox{\epsfbox{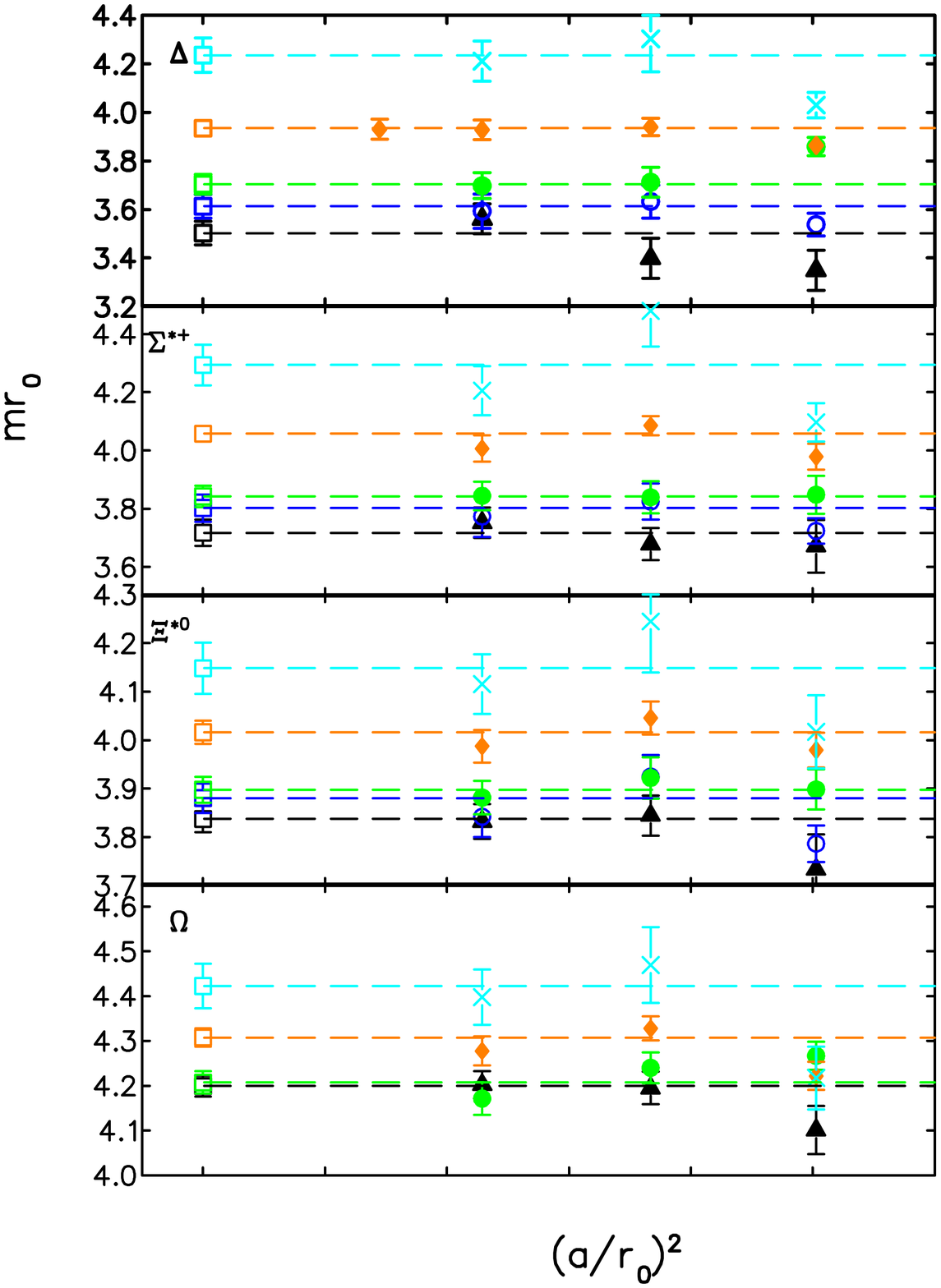}}
\caption{Constant extrapolation to the continuum for the decuplet. The notation
is the same as in Fig.~\ref{fig:octet continuum const}.} 
\label{fig:decuplet continuum const}
\end{minipage}
\end{figure}

In order to assess cut-off effects 
we use results at $\beta=3.9$ and $\beta=4.05$.
The lattice results,
expressed in units of the Sommer scale $r_0$, are interpolated to the same pion mass in units of $r_0$
at each $\beta$-value. 
We give the interpolated results at six values of $m_\pi r_0$ in Tables~\ref{tab:octet-cont-r0}
and \ref{tab:decuplet-cont-r0}. Interpolating linearly or
 with one-loop order chiral 
perturbation theory gives
values consistent within error bars. Given the  size of these
 cut-off effects  a 
weighted average of the baryon masses between these two $\beta$ 
values gives an estimate of the values in the continuum limit. 
It must be stressed that estimating the strange quark mass at $\beta=4.05$ 
 using Eq.~(\ref{s-mass}) may cause residual cut-off effects on the few percentage
level  that are
not taken into account with the continuum extrapolation as performed here.
 The results obtained from the weighted averaging of data at $\beta=3.9$ and $\beta=4.05$ are 
listed in Tables~\ref{tab:octet-cont-r0}
and \ref{tab:decuplet-cont-r0}
and are plotted in Figs.~\ref{fig:octet continuum const} and \ref{fig:decuplet continuum const} .
In the figures we also include results at $\beta=3.8$. If
cut-off effects are small for all $\beta$-values then results
at $\beta=3.8$ should fall onto the same line. As can be seen this is best
fulfilled for the nucleon mass. Furthermore for the nucleon and the $\Delta$
we also show results at a smaller value of the lattice spacing corresponding
to $\beta=4.2$. Essentially, the $a^2$ dependence of the nucleon and $\Delta$
 mass as computed at 
$\beta=3.9$, $4.05$ and $4.2$ is consistent with a constant behaviour,
 verifying
 that for lattice spacings below
$0.1$~fm cut-off effects are indeed small.
For the $\Lambda$ mass results at $\beta=3.8, 3.9$ and $\beta=4.05$ are consistent with a constant. This holds approximately also for the other baryons. Within
the statistical errors one therefore concludes 
that for lattice spacings below
$0.1$~fm cut-off effects are under control.

\begin{table}[h!]
\begin{small}
  \begin{center}
  \begin{tabular}{lccccccccc}
  \hline\hline
$\Bigl.\Bigr.r_0 m_{\pi}$&$r_0 m_N$ &
  $r_0 m_{\Lambda}$ & $r_0 m_{\Sigma^{\rm Av.}}$& $r_0 m_{\Sigma^+}$ &$r_0 m_{\Sigma^0}$ & $r_0 m_{\Sigma^-}$& $r_0 m_{\Xi^{\rm Av.}}$ & $r_0 m_{\Xi^0}$&  $r_0 m_{\Xi^-}$ \\
 \hline\hline
\multicolumn{10}{c}{$\beta=3.9$}\\
   0.60 &  2.571(23)&  2.922(18)&  3.062(22)&  3.156(36)&  3.084(26) & 3.004(21)&  3.324(16) & 3.421(23) &   3.271(14) \\
   0.70 &  2.671(24)&  3.001(18)&  3.193(21)&  3.279(38)&  3.215(21) & 3.111(25)&  3.399(18) & 3.523(24) &   3.319(17) \\
   0.80 &  2.757(30)&  3.085(25)&  3.212(29)&  3.318(42)&  3.219(31) & 3.144(31)&  3.433(21) & 3.555(30) &   3.363(21) \\
   0.90 &  2.880(26)&  3.156(22)&  3.286(25)&  3.404(38)&  3.293(27) & 3.215(27)&  3.472(19) & 3.598(26) &   3.401(19) \\
   1.00 &  2.992(35)&  3.226(27)&  3.381(29)&  3.482(33)&  3.401(26) & 3.310(24)&  3.507(23) & 3.629)30) &   3.436(23)\\ 
   1.10 &  3.111(23)&  3.305(190&  3.405(21)&  3.477(29)&  3.414(22) & 3.358(23)&  3.547(19) & 3.633(26) &   3.491(17)\\ 
\multicolumn{10}{c}{$\beta=4.05$}\\
  0.60 &  2.600(43) &  2.946(29)&  3.107(32)&  3.199(46)&   3.115(34) & 3.034(38)&  3.336(23) & 3.400(29)  &   3.287(22) \\ 
  0.70 &  2.692(40) &  3.008(25)&  3.152(29)&  3.233(41)&   3.161(31) & 3.080(35)&  3.362(21) & 3.425(26)  &   3.313(19)\\ 
  0.80 &  2.788(46) &  3.074(32)&  3.200(34)&  3.269(49)&   3.209(36) & 3.127(40)&  3.391(25) & 3.451(32)  &   3.340(24)\\ 
  0.90 &  2.874(32) &  3.135(32)&  3.242(30)&  3.295(45)&   3.253(31) & 3.165(31)&  3.418(28) & 3.474(34)  &   3.364(25)\\ 
  1.00 &  2.984(33) &  3.205(32)&  3.298(30)&  3.348(45)&   3.308(31) & 3.230(31)&  3.448(29) & 3.504(34)  &   3.397(25)\\ 
  1.10 &  3.119(21) &  3.283(20)&  3.370(21)&  3.430(27)&   3.373(21) & 3.325(21)&  3.481(19) & 3.539(23)  &   3.440(17)\\
\multicolumn{10}{c}{continuum limit}\\
 0.60 &  2.577(20)  &  2.929(15)&  3.077(18) &  3.173(28) & 3.095(21) &  3.011(18) & 3.328(13) & 3.413(18)&  3.275(12) \\ 
 0.70 &  2.676(21)  &  3.003(15)&  3.179(17) &  3.258(28) & 3.198(17) &  3.101(20) & 3.383(13) & 3.477(18)&  3.316(13)\\
 0.80 &  2.766(25)  &  3.080(20)&  3.207(22) &  3.297(32) & 3.215(23) &  3.138(24) & 3.415(16) & 3.506(22)&  3.353(16)\\ 
 0.90 &  2.878(20)  &  3.149(18)&  3.267(19) &  3.359(29) & 3.275(20) &  3.193(20) & 3.456(16) & 3.552(21)&  3.387(15)\\ 
 0.10 &  2.988(24)  &  3.217(21)&  3.342(21) &  3.436(26) & 3.363(20) &  3.280(19) & 3.484(18) & 3.573(23)&  3.418(17)\\ 
 1.10 &  3.116(15)  &  3.295(14)&  3.387(15) &  3.452(20) & 3.392(15) &  3.340(15) & 3.514(13) & 3.580(17)&  3.465(12)\\ 
\hline
\end{tabular}
  \caption{Octet masses computed at reference pion masses in units of $r_0$ and the corresponding 
           continuum limit values.}
  \label{tab:octet-cont-r0}
 \end{center}
 \end{small}
\end{table}

\begin{table}[h!]
\begin{small}
  \begin{center}
  \begin{tabular}{lccccccccc}
  \hline\hline
$\Bigl.\Bigr.r_0 m_{\pi}$& $r_0 m_{\Delta^{{\rm Av.}}}$ & $r_0 m_{\Sigma^{\ast {\rm Av.}}}$ & $r_0 m_{\Sigma^{\ast +}}$ &  $r_0 m_{\Sigma^{\ast 0}}$ &  $r_0 m_{\Sigma^{\ast -}}$ &   $r_0 m_{\Xi^{\ast {\rm Av.}}}$ &   $r_0 m_{\Xi^{\ast 0}}$& $r_0 m_{\Xi^{\ast -}}$& $r_0 m_{\Omega}$ \\
 \hline\hline
\multicolumn{10}{c}{$\beta=3.9$}\\
   0.60 &  3.312(51)&   3.565(51) &  3.558(49) &  3.564(56) &  3.660(53) & 3.671(27) &  3.689(31) &  3.662(28)  & 4.131(26) \\
   0.70 &  3.419(57)&   3.719(55) &  3.679(65) &  3.735(61) &  3.744(52) & 3.805(39) &  3.845(41) &  3.754(38)  & 4.195(36) \\
   0.80 &  3.631(48)&   3.850(50) &  3.824(62) &  3.856(55) &  3.853(55) & 3.856(46) &  3.925(45) &  3.812(49)  & 4.252(37) \\
   0.90 &  3.727(42)&   3.907(44) &  3.871(55) &  3.918(49) &  3.924(49) & 3.873(41) &  3.947(40) &  3.839(44)  & 4.259(33) \\
   1.00 &  3.761(47)&   3.911(46) &  3.839(55) &  3.954(58) &  3.981(52) & 3.862(45) &  3.922(43) &  3.840(44)  & 4.240(35) \\
   1.10 &  3.950(27)&   4.075(35) &  4.085(33) &  4.081(34) &  4.074(36) & 3.981(38) &  4.046(34) &  3.909(39)  & 4.328(27) \\
\multicolumn{10}{c}{$\beta=4.05$} \\
   0.60 & 3.524(47)&  3.769(57) &   3.732(67) &  3.787(65) &  3.766(58) & 3.806(40) &  3.823(39) &  3.789(40)  & 4.221(34) \\
   0.70 & 3.581(43)&  3.789(52) &   3.752(62) &  3.803(57) &  3.795(53) & 3.817(37) &  3.832(36) &  3.802(37)  & 4.202(30) \\
   0.80 & 3.615(50)&  3.808(60) &   3.773(71) &  3.819(70) &  3.822(61) & 3.828(43) &  3.842(42) &  3.816(43)  & 4.183(37) \\
   0.90 & 3.617(40)&  3.804(45) &   3.768(49) &  3.834(76) &  3.829(46) & 3.829(32) &  3.834(34) &  3.825(33)  & 4.141(36) \\
   1.00 & 3.718(41)&  3.876(46) &   3.844(49) &  3.851(76) &  3.901(47) & 3.862(32) &  3.882(35) &  3.847(33)  & 4.171(36) \\
   1.10 & 3.922(40)&  4.028(39) &   4.007(45) &  3.868(49) &  4.042(38) &  3.931(29)&  3.987(33) &  3.885(29)  & 4.277(33)\\
 \multicolumn{10}{c}{continuum limit} \\
$\Bigl.\Bigr.r_0 m_{\pi}$& $r_0 m_{\Delta^{{++}}}$,  $r_0 m_{\Delta^{{+}}}$,& $r_0 m_{\Sigma^{\ast {\rm Av.}}}$ & $r_0 m_{\Sigma^{\ast +}}$ &  $r_0 m_{\Sigma^{\ast 0}}$ &  $r_0 m_{\Sigma^{\ast -}}$ &   $r_0 m_{\Xi^{\ast {\rm Av.}}}$ &   $r_0 m_{\Xi^{\ast 0}}$& $r_0 m_{\Xi^{\ast -}}$& $r_0 m_{\Omega}$ \\
 \hline\hline
   0.60 & 3.439(35)&  3.656(38) &  3.619(40) &  3.660(42) & 3.709(39) &   3.712(22) &  3.741(24) &  3.704(23) & 4.164(21) \\
   0.70 & 3.520(49)&  3.755(38) &  3.717(45) &  3.771(42) & 3.768(37) &   3.811(27) &  3.838(27) &  3.778(26) & 4.199(23) \\
   0.80 & 3.623(35)&  3.833(38) &  3.802(47) &  3.841(43) & 3.839(41) &   3.841(31) &  3.880(30) &  3.814(32) & 4.217(26) \\
   0.90 & 3.668(29)&  3.856(32) &  3.813(36) &  3.893(41) & 3.874(34) &   3.845(25) &  3.882(26) &  3.830(26) & 4.205(24) \\
   1.00 & 3.735(30)&  3.893(32) &  3.842(37) &  3.916(46) & 3.937(35) &   3.862(26) &  3.898(27) &  3.845(26) & 4.207(25) \\
   1.10 & 3.935(20)&  4.054(26) &  4.058(27) &  4.013(28) & 4.059(26) &   3.949(23) &  4.016(24) &  3.893(23) & 4.307(21) \\
\hline
\end{tabular}
  \caption{The same as Table~\ref{tab:octet-cont-r0} but for the decuplet baryons.}
  \label{tab:decuplet-cont-r0}
 \end{center}
 \end{small}
\end{table}

\subsection{Fixing the lattice spacing}

In order to convert to physical units we need to fix the lattice spacing.
The values of the lattice spacing given in Table~\ref{Table:params} were extracted using the pion 
decay constant. Equivalently one can determine the value of 
 $r_0$ by extrapolating the results to the physical point. The value obtained  
is $r_0=0.439(25)$ fm determined in the light meson sector \cite{r0_ETMC_Scaling_Paper} 
where the systematic error is added to the statistical one. Knowing $r_0$ and the ratio $r_0/a$ one
can determine the lattice spacing.

\begin{figure}[h!]
\epsfxsize=8truecm
\epsfysize=6truecm
\mbox{\epsfbox{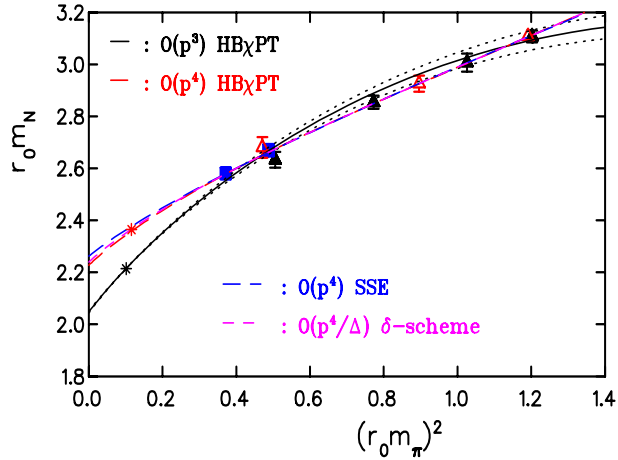}}
\caption{Determination of $r_0^N$ with a simultaneous fit to the
lattice data at $\beta=3.9$ and $\beta=4.05$. The asterisks denote the physical
point determined by the value of $r_0^N$ by  using ${\cal O}( p^3)$
and ${\cal O}(p^4)$ $\chi$PT as described in Ref.~\cite{Alexandrou:2008tn}.}
\label{fig:nucleon r0}
\end{figure}

The nucleon mass can be used to
                    set the scale and this determination
                    seems natural if one is interested
in the study of the  baryon spectrum.
 Our data at $\beta=3.9$ and
$\beta=4.05$ do not show significant lattice spacing effects. Therefore
we can make a combined fit using continuum chiral
perturbation theory results to determine the $r_0^N$. 
Chiral corrections to the nucleon mass are known to 
 ${\cal O}(p^4)$ within several
expansion schemes. We use the same
schemes used in  Ref.~\cite{Alexandrou:2008tn,Alexandrou:2007qq}. The fits
are shown in Fig.~\ref{fig:nucleon r0}.
Using the ${\cal O} (p^3)$ result, which is well established,
 to extrapolate to the physical pion mass we obtain $r_0^N=0.465(6)$~fm.
If we instead  use the data at $\beta=3.9$ and $\beta=4.05$ to perform
the continuum
limit as discussed in the previous subsection and then fit, 
 we find $r_0^N=0.471\pm 0.006(stat.) \pm 0.015(syst.)$~fm. The systematic error is due to the interpolation
to a fixed value of $m_\pi r_0$ and it is obtained by comparing the value of $r_0$ obtained when linear interpolation is used
to the one obtained using ${\cal O}(p^3)$. 
Furthermore, we take the difference in the value of $r_0$ obtained using continuous results and the
value found by fitting the lattice data at $\beta=3.9$ and $\beta=4.05$ to be the systematic error
due to cut-off effects. We therefore take $r_0^N=0.465(6)(14)$~fm,  which
is in agreement with $r_0$ extracted from the value of the pion decay constant,
$f_\pi$, when converting our results at $\beta=3.9$ and $\beta=4.05$ in units of $r_0$.
 Using
the value of $r_0/a=5.22(2)$   and $r_0/a=6.61(3)$ at $\beta=3.9$ and
$\beta=4.05$ we find the for the lattice spacings $a_{\beta=3.9}=0.089(4)$~fm
and $a_{\beta=4.05}=0.070(3)$~fm.
 These values for the lattice spacing extracted using the nucleon mass 
at the physical point are
in agreement with those determined
     from the pion decay constant. This constitutes a nice consistency check
of our lattice formulation. In what follows we will use the lattice spacing extracted from the nucleon mass
using ${\cal O}(p^3)$ heavy baryon chiral perturbation theory 
to convert the rest of the masses into physical units.

\section{Chiral extrapolation}
Given that the cut-off effects are almost negligible in our simulations, 
we apply continuum 
chiral perturbation theory to extrapolate lattice results at 
$\beta=3.9$ and $\beta=4.05$ to the
physical pion mass. In particular, we will use SU(2) chiral perturbation 
theory($\chi$PT)~\cite{Tiburzi:2008bk} for two reasons: The first being that our 
simulations are done for two mass-degenerate dynamical quarks and the second because it
was shown that SU(3) $\chi$PT fails to describe lattice 
data~\cite{Ishikawa:2009vc}. We would like to stress however that the issue of 
the applicability of SU(3) $\chi$PT is not entirely settled and e.g. 
SU(3) fits to lattice results using staggered fermions 
were claimed to be produce reasonable fits~\cite{Frink:2005ru,Frink:2004ic}.

The leading order $SU(2)$ heavy baryon chiral perturbation (HB$\chi$PT) results
are given by
\be
m^{\rm LO}_X(m_\pi) = m_X^{(0)}-4c_X^{(1)} m_\pi^2 \; ,
\label{linear}
\ee
  with two fit-parameters, 
the baryon mass in the chiral limit $m_X^{(0)}$ and $c_X^{(1)}$, 
the latter of which gives the  leading contribution to the $\sigma_X$-term.

The leading one-loop results for the nucleon and the $\Delta$ in HB$\chi$PT  were first derived in Ref.~\cite{Gas88} and successful fits to 
lattice data on the nucleon and $\Delta$ were discussed in our
previous study~\cite{Alexandrou:2008tn}. 
A natural generalization of the  ${\cal O}(p^3)$ results for the nucleon and $\Delta$
 to the rest of the octet
and decuplet baryons~\cite{Nagels:1979xh,Nagels:1978sc} is given by

\beq
m _N(m_\pi)             &=& m^{(0)}_N-4c^{(1)}_N \;m_{\pi}^2 - \frac{3 g_A^2 }{16\pi f_\pi^2} \;m_\pi^3 \nonumber \\
m_\Lambda(m_\pi) &=&m^{(0)}_{\Lambda }-4c^{(1)}_{\Lambda}\; m_\pi^2 - \frac{g_{\Lambda\Sigma}^2 }{16\pi f_\pi^2} \;m_\pi^3\nonumber \\
m_\Sigma(m_\pi)& = & m^{(0)}_{\Sigma }-4c^{(1)}_{\Sigma}m_\pi^2 - \frac{2 g_{ \Sigma \Sigma}^2 + g_{\Lambda\Sigma}^2/3}{16\pi f_\pi^2}\; m_\pi^3   \nonumber \\
m_\Xi(m_\pi)            &=&m^{(0)}_{\Xi }-4c^{(1)}_{\Xi}m_\pi^2 - \frac{3 g_{\Xi \Xi}^2}{16\pi f_\pi^2} \;m_\pi^3 \quad,
\label{LO octet}
\eeq
 for the octet baryons and
\beq
m_\Delta(m_\pi)         &=& m^{(0)}_\Delta           - 4c^{(1)}_\Delta\; m_\pi^2        - \frac{25}{27}\frac{g_{\Delta \Delta}^2}{16\pi f_\pi^2} \;m_\pi^3   \nonumber  \\
m_{\Sigma^*}(m_\pi) &=& m^{(0)}_{\Sigma^* } - 4c^{(1)}_{\Sigma^*}m_\pi^2 - \frac{10}{9}\frac{g_{\Sigma^* \Sigma^*}^2 }{16\pi f_\pi^2}\;m_\pi^3 \nonumber \\
m_{\Xi^*}(m_\pi)         &=& m^{(0)}_{\Xi^*}          - 4c^{(1)} _{\Xi^*} m_\pi^2           - \frac{5}{3}\frac{g_{\Xi^* \Xi^*}^2 }{16\pi f_\pi^2}\;m_\pi^3 \nonumber \\
m_\Omega(m_\pi)      &=& m^{(0)}_\Omega      - 4c^{(1)}_\Omega m_\pi^2 \; ,    \label{LO decuplet}
\eeq
for the decuplet baryons.

In addition we consider  a cubic term of the following form 
\be m_X(m_\pi)=m_X^{(0)}-4c_X^{(1)} m_\pi^2+c_X^{(2)} m_\pi^3
\label{cubic}
\ee
 treating $c_X^{(2)}$ as an additional fit parameter.

The next to leading order SU(2) $\chi$PT results~\cite{Tiburzi:2008bk} for the octet are given by
\beq
m^{NLO}_N(m_\pi)              &=& m^{LO} _N(m_\pi)      - \frac{3 g_A^2 }{16\pi f_\pi^2} \;m_\pi^3            -  \frac{8 g_{N\Delta}^2}{3(4\pi f_\pi)^2} \; {\cal F}(m_\pi,\Delta_{N\Delta}, \lambda) \nonumber \\
m^{NLO}_\Lambda(m_\pi) &=& m^{LO}_{\Lambda}(m_\pi)  - \frac{g^2_{\Lambda\Sigma}}{(4\pi f_\pi)^2} \; {\cal F}(m_\pi,\Delta_{\Lambda \Sigma},\lambda)
                                                   - \frac{ 4g^2_{\Lambda\Sigma^*} } {(4\pi f_\pi)^2} \; {\cal F}(m_\pi,\Delta_{\Lambda \Sigma^*},\lambda) \nonumber \\
m^{NLO}_\Sigma(m_\pi)    &=& m^{LO}_{\Sigma}(m_\pi)  - \frac{2 g_{ \Sigma \Sigma}^2}{16\pi f_\pi^2}\; m_\pi^3  -\frac{g^2_{\Lambda\Sigma}}{3(4\pi f_\pi)^2} \; {\cal F}(m_\pi,-\Delta_{\Lambda \Sigma},\lambda)
                                                   - \frac{4g^2_{\Lambda\Sigma^*}}{3(4\pi f_\pi)^2} \; {\cal F}(m_\pi,\Delta_{\Sigma\Sigma^*},\lambda) \nonumber \\
m^{NLO}_\Xi(m_\pi)            &=& m^{LO}_{\Xi}(m_\pi) - \frac{3 g_{\Xi \Xi}^2}{16\pi f_\pi^2} \;m_\pi^3-\frac{2g_{\Xi^*\Xi}^2}{(4\pi f_\pi)^2} \; {\cal F}(m_\pi,\Delta_{\Xi\Xi^*},\lambda)    \label{NLO octet}
\eeq
and for the decuplet baryons:
\beq
m^{NLO}_\Delta(m_\pi)         &=& m^{LO} _\Delta(m_\pi)   - \frac{25}{27}\frac{g_{\Delta \Delta}^2}{16\pi f_\pi^2} \;m_\pi^3  -\frac{2 g_{\Delta N}^2}{3 (4\pi f_\pi)^2}) \; {\cal F}(m_\pi,-\Delta_{N\Delta},\lambda) \nonumber \\
m^{NLO}_{\Sigma^*}(m_\pi) &=& m^{LO} _{\Sigma^* }(m_\pi)  - \frac{10}{9}\frac{g_{\Sigma^* \Sigma^*}^2 }{16\pi f_\pi^2}\;m_\pi^3  - \frac{2}{3(4\pi f_\pi)^2} \left[g_{\Sigma^*\Sigma} ^2
\; {\cal F}(m_\pi,-\Delta_{\Sigma\Sigma^*,\lambda}) + g_{\Lambda\Sigma^*}^2 \; {\cal F}(m_\pi,-\Delta_{\Lambda\Sigma^*,\lambda}) \right] \nonumber \\ 
m^{NLO}_{\Xi^*} (m_\pi)       &=& m^{LO} _{\Xi^*}(m_\pi)  - \frac{5}{3}\frac{g_{\Xi^* \Xi^*}^2 }{16\pi f_\pi^2}\;m_\pi^3- \frac{g_{\Xi^* \Xi}^2}{(4\pi f_\pi)^2}  \; {\cal F}(m_\pi,-\Delta_{\Xi\Xi^*,\lambda}) \nonumber \\
m^{NLO}_\Omega(m_\pi)     &=& m^{LO} _\Omega(m_\pi)
\label{NLO decuplet}
\eeq
with the non analytic function~\cite{Tiburzi:2005na}
\be\label{F}
{\cal F}(m,\Delta,\lambda) =(m^2-\Delta^2)\sqrt{\Delta^2-m^2+i\epsilon}
\;\log\left(\frac{\Delta-\sqrt{\Delta^2-m^2+i\epsilon}}{\Delta+\sqrt{\Delta^2-m^2+i\epsilon}}\right)
-\frac{3}{2}\Delta m^2\log\left(\frac{m^2}{\lambda^2}\right)-\Delta^3\log\left(\frac{4\Delta^2}{m^2}\right) \quad
\ee
depending on the threshold parameter $\Delta_{XY}= m^{(0)}_{Y}-m^{(0)}_X$ and on the 
scale $\lambda$ of chiral perturbation theory, fixed to
  $\lambda=1$~GeV.
For $\Delta>0$ the real part of the function ${\cal F}(m,\Delta,\lambda)$ has the property
\be
{\cal F}(m,-\Delta,\lambda) = \left \{\begin{array}{ll} -{\cal F}(m,\Delta,\lambda)& m<\Delta \\
-{\cal F}(m,\Delta,\lambda)+2\pi\left(m^2-\Delta^2\right)^{3/2} & m>\Delta\\
\end{array}\right.
\label{F symm}
\ee 
which corrects a typo in the sign of the
second term in Ref.~\cite{WalkerLoud:2008bp}.
In our fits, the nucleon axial charge $g_A$ and pion decay constant $f_{\pi}$ are fixed to their experimental values
(we use the convention such that  $f_{\pi}=130.70$ MeV). The remaining pion-baryon axial coupling constants are taken from  SU(3) relations~\cite{Tiburzi:2008bk}:
\be
\begin{array}{lllll}
{\rm Octet:}          &\quad g_{A} = D+F,                          &\quad g_{\Sigma\Sigma}       = 2F,                                      &\quad g_{\Xi\Xi}=D-F,                                         &\quad g_{\Lambda\Sigma}=2D \\
{\rm Decuplet:}   &\quad g_{\Delta \Delta} = {\cal H}, &\quad g_{\Sigma^*\Sigma^*} = \frac{2}{3} {\cal H},          &\quad  g_{\pi\Xi^*\Xi^*}=\frac{1}{3} {\cal H}    & \\
{\rm Transition:} &\quad g_{\Delta N} = {\cal C},         &\quad g_{\Sigma^*\Sigma}   = \frac{1}{\sqrt{3}} {\cal C}, &\quad g_{\Xi^*\Xi}=\frac{1}{\sqrt{3}} {\cal C},  &\quad g_{\Lambda\Sigma^*}= -\frac{1}{\sqrt{2}} {\cal C}
\end{array}
\label{ACC}
\ee

As can be seen, in the  octet case, and once $g_A$ is fixed, the axial
coupling constants depend on the single 
parameter written as $\alpha={D\over D+F}$. Its value is poorly known. 
It can be taken  either from the quark model ($\alpha=3/5$),  from the phenomenology of
semi-leptonic decays  or from hyperon-nucleon scattering. 
We take $2D=1.47$ or  $\alpha=0.58$ as given in Ref.~\cite{Tiburzi:2008bk} .
The decuplet coupling constants depend on a single parameter for which
we again take the value ${\cal H}=2.2$ from Ref.~\cite{Tiburzi:2008bk}.
This value is  not far from that predicted by SU(6) symmetry, ${\cal H}= {9\over5} g_A =  2.29$ used in our previous work~\cite{Alexandrou:2008tn}
resulting in the 
same cubic term for the nucleon and $\Delta$.
For fixing the octet-decuplet transition couplings we take the value   ${\cal C}=1.48$ from  Ref.~\cite{Tiburzi:2005na} .

With the coupling constants fixed in this way, the LO, the one-loop as well as the NLO fits are left with the two independent fit parameters $m_X^{(0)}$ and $c_X^{(1)}$. 
All mass parameters $m_X^{(0)}$ are treated independently
unlike what is done in Ref.~\cite{Tiburzi:2008bk} where a universal mass
parameter was used for all barons with the same strangeness.

A noticeable result of this expansion is the absence of a cubic term in the   
expression for the $\Lambda$ and $\Omega$ masses given in Eqs.~(\ref{NLO octet}) 
and (\ref{NLO decuplet}).
In the case of $\Omega$, it follows from the absence of light  valence quarks.
However the absence of a cubic term in the NLO expression  of  $\Lambda$,
 although a consequence of  $\chi PT$,  is nevertheless a questionable result, since it relies on the assumption that $m_{\pi} \ll M_{\Sigma}-M_{\Lambda}$.
In the limit $\Delta \to 0$  the non analytic function  (\ref{F}) becomes
$${\cal F}(m_{\pi},\Delta\to 0,\lambda)  = \pi m_{\pi}^3 \;  \quad,$$
which generates  a cubic term for the $\Lambda$ and slightly  modifies the one for $\Sigma$.
The corresponding expressions are given by
\begin{eqnarray}
m_\Lambda(m_\pi) &=&m^{(0)}_{\Lambda }-4c^{(1)}_{\Lambda}\; m_\pi^2 \nonumber  - \frac{g_{ \Lambda \Sigma}^2}{16\pi f_\pi^2}\; m_\pi^3          \label{ILO_Lambda}  \; ,    \\
m_\Sigma(m_\pi)    &=&m^{(0)}_{\Sigma }-4c^{(1)}_{\Sigma}m_\pi^2 - \frac{2 g^2_{ \Sigma \Sigma}+g^2_{ \Lambda \Sigma}/3 }{16\pi f_\pi^2}\; m_\pi^3   \label{ILO_Sigma} \; ,
\end{eqnarray}
in agreement with the results of Eq.~(\ref{LO octet}).

\begin{figure}[h!]
\begin{minipage}{8cm}
\vspace*{1cm}
\epsfxsize=8truecm
\epsfysize=6truecm
\mbox{\epsfbox{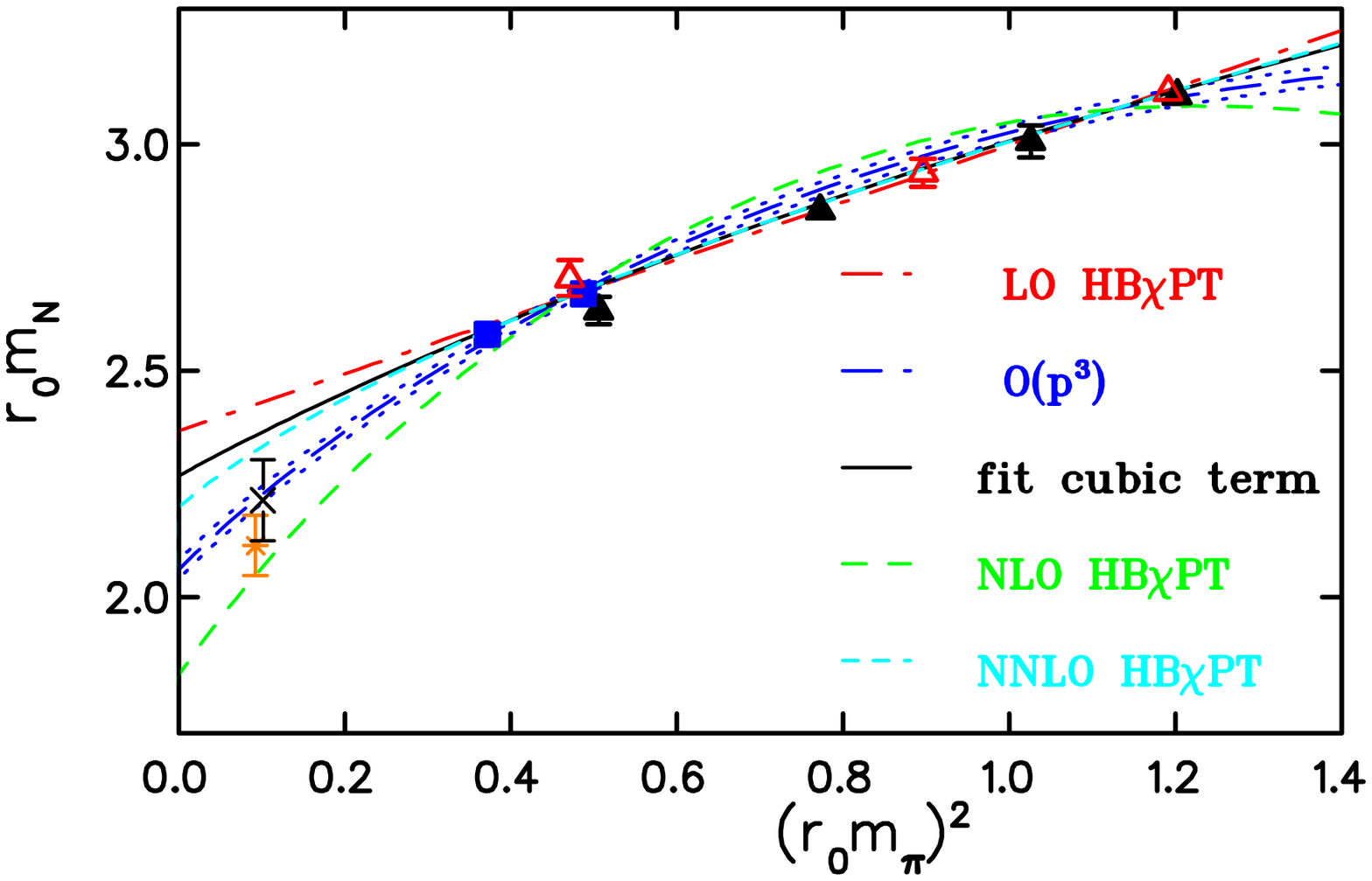}}
\caption{Chiral extrapolation of the nucleon mass in units of $r_0$ .
 Filled triangles and squares
are results at $\beta=3.9$ on  $24^3\times 48$ and
$32^3\times 64$ lattice sizes respectively. Open triangles are
results at $\beta=4.05$. We show chiral extrapolations 
linear in $m_\pi^2$ as in Eq.~(\ref{linear}),  
to ${\cal O}(p^3)$ as in Eq.~(\ref{LO octet}),  fit in the  cubic term as in
Eq.~(\ref{cubic}),
  NLO and NNLO in SU(2) chiral perturbation theory as 
in Eqs.~(\ref{NLO octet},\ref{NNLO octet}), respectively.
We include an error band only for the ${\cal O}(p^3)$ fit for clarity.
The physical point shown by the asterisk uses the value of $r_0$ extracted from $f_\pi$,
whereas the cross  uses $r^N_0$ determined from the nucleon mass.}
\label{fig:nucleon chiral}
\end{minipage}
\hfill
\begin{minipage}{8cm}
\epsfxsize=8truecm
\epsfysize=6truecm
\vspace*{-2.2cm}
\mbox{\epsfbox{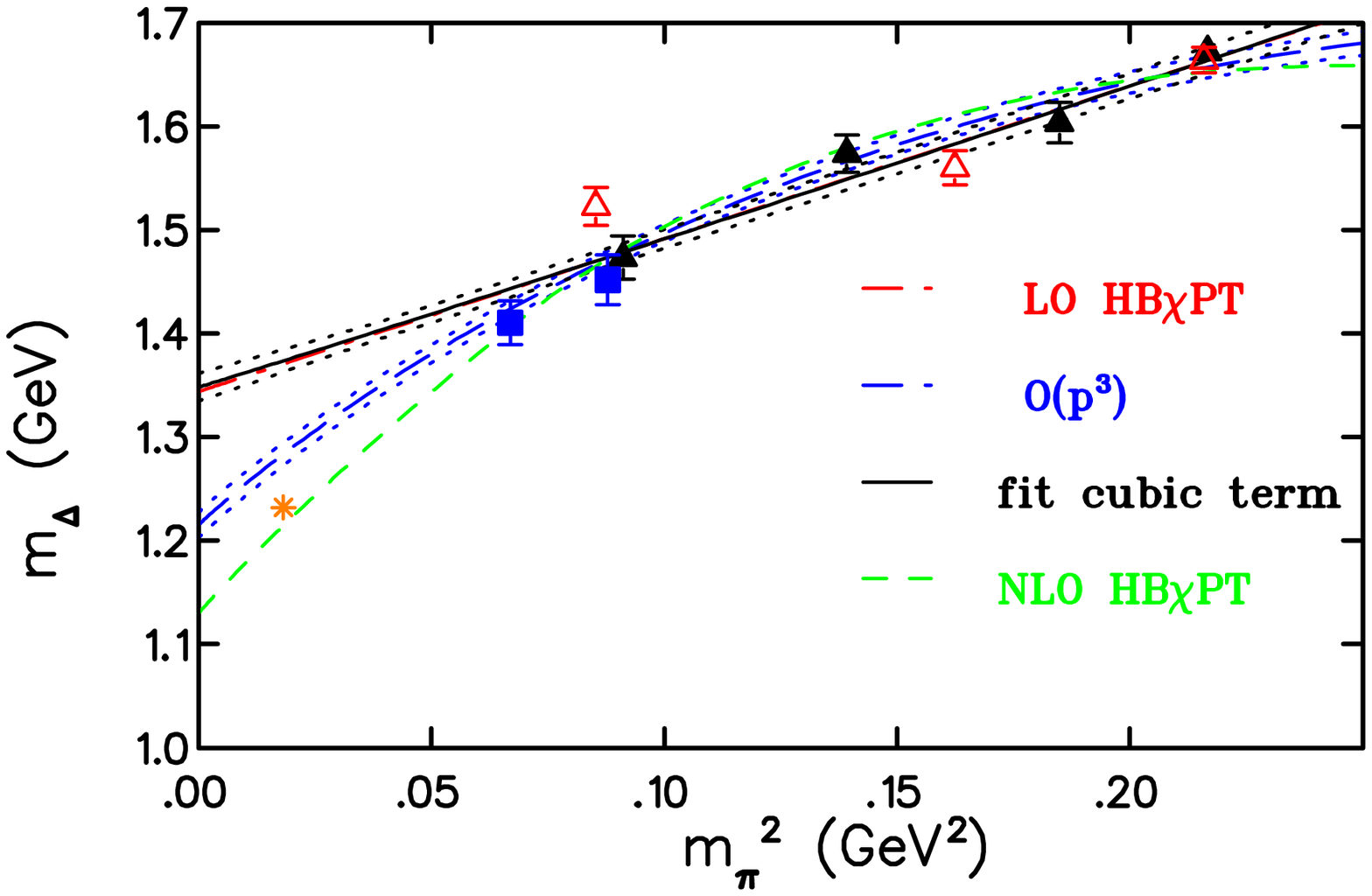}}
\caption{Chiral extrapolation of the $\Delta$ mass.
The notation is the same as that in Fig.~\ref{fig:nucleon chiral} but in physical units. Here we also include an error band for the cubic fit.}
\label{fig:Delta chiral}
\end{minipage}
\end{figure}

\begin{figure}[h!]
\begin{minipage}{8cm}
\epsfxsize=8truecm
\epsfysize=6truecm
\mbox{\epsfbox{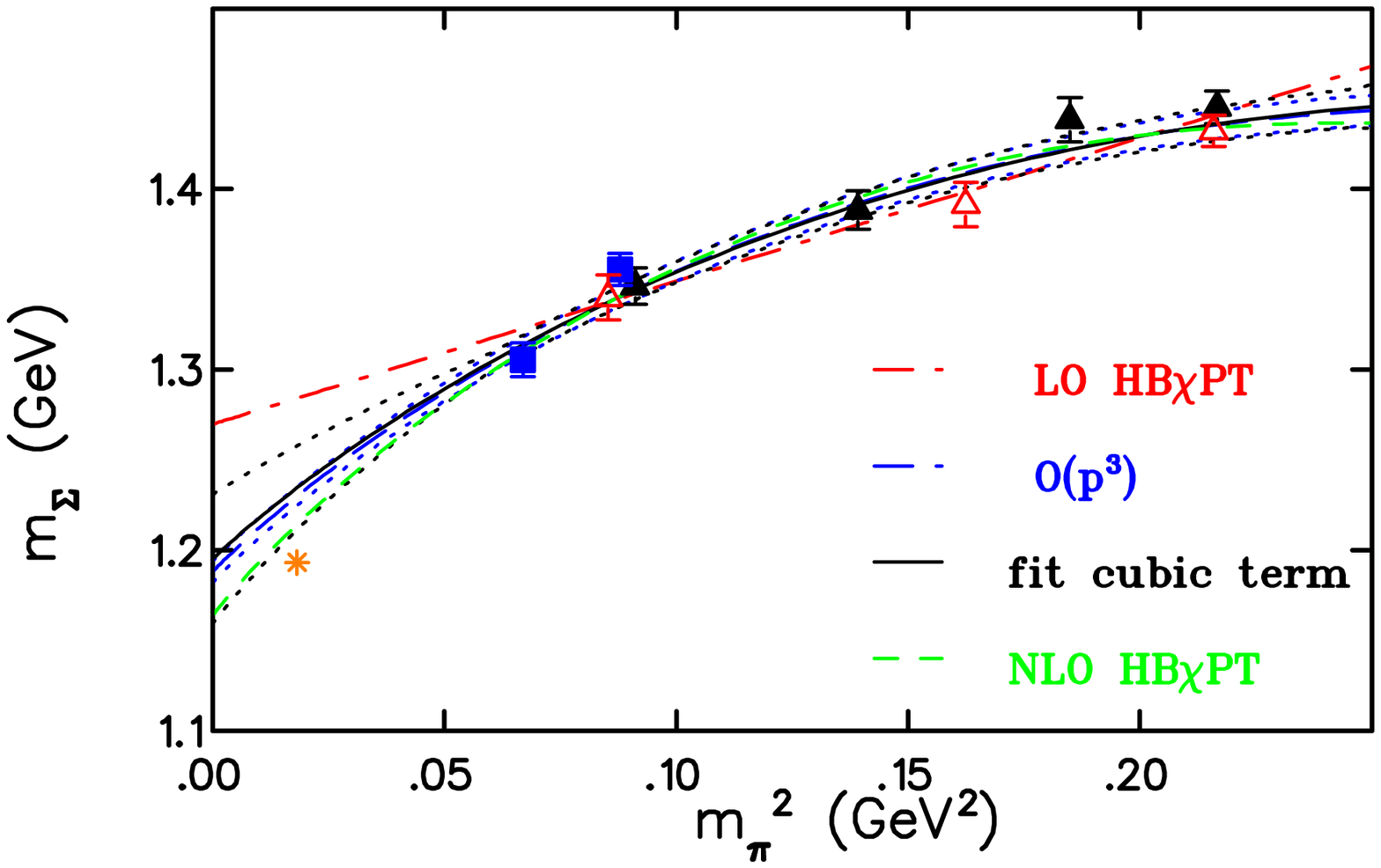}}
\caption{Chiral extrapolation of the $\Sigma$ mass in physical units. We show chiral extrapolations 
linear in $m_\pi^2$, using Eq.~\ref{cubic},
 ${\cal O}(p^3)$, NLO and NNLO in SU(2) chiral perturbation theory given in Eqs.~(\ref{LO octet},
\ref{NLO octet})
respectively. The rest of the notation is the same as that in Fig.~\ref{fig:Delta chiral}.}
\label{fig:Sigma chiral}
\end{minipage}
\hfill
\begin{minipage}{8cm}
\epsfxsize=8truecm
\epsfysize=6truecm
\vspace*{-1.2cm}
\mbox{\epsfbox{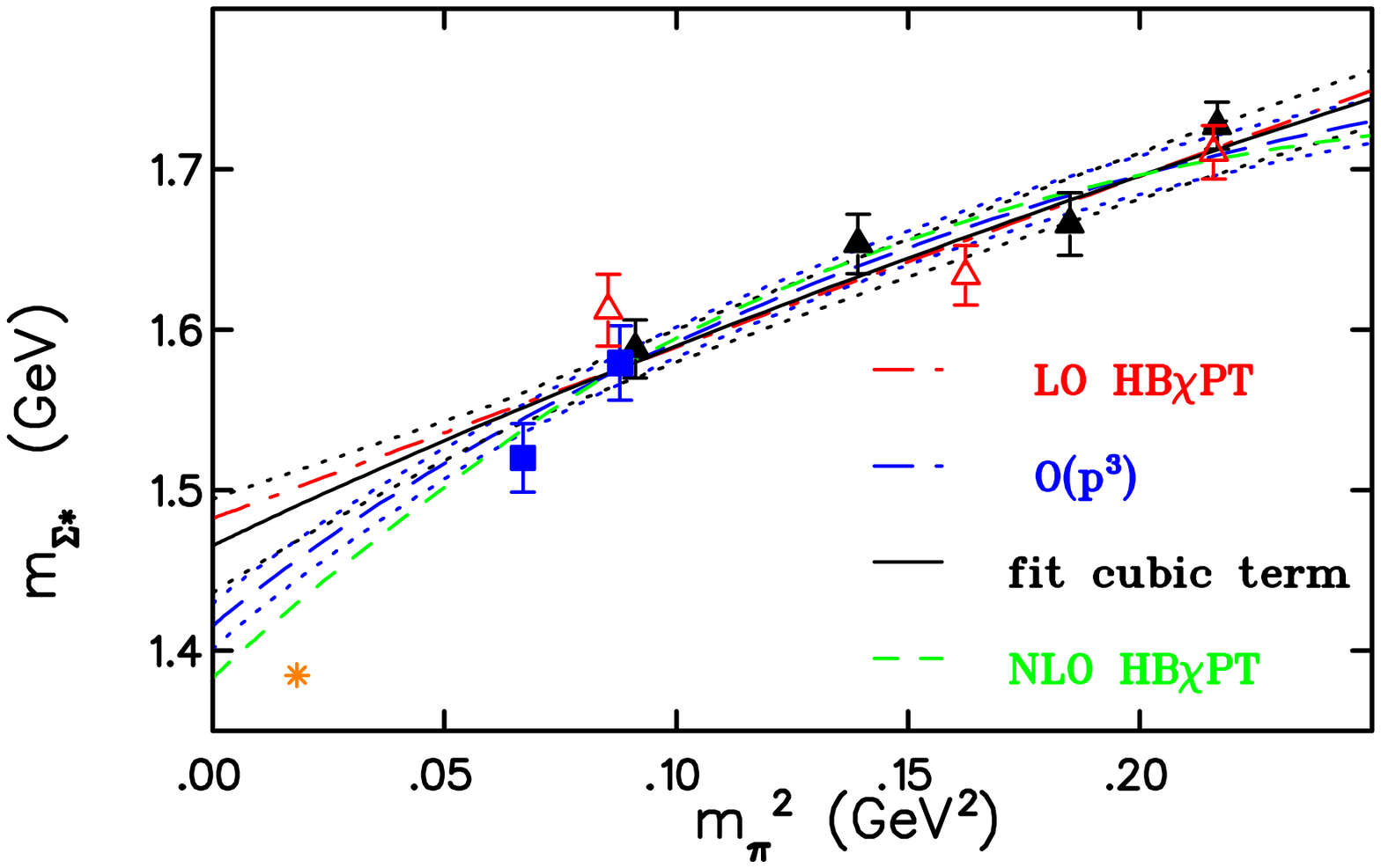}}
\caption{Chiral extrapolation of the $\Sigma^*$ mass.
The notation is the same as that in Fig.~\ref{fig:Sigma chiral}.}
\label{fig:Sigmastar chiral}
\end{minipage}
\end{figure}

\begin{figure}[h!]
\begin{minipage}{8cm}
\epsfxsize=8truecm
\epsfysize=6truecm
\mbox{\epsfbox{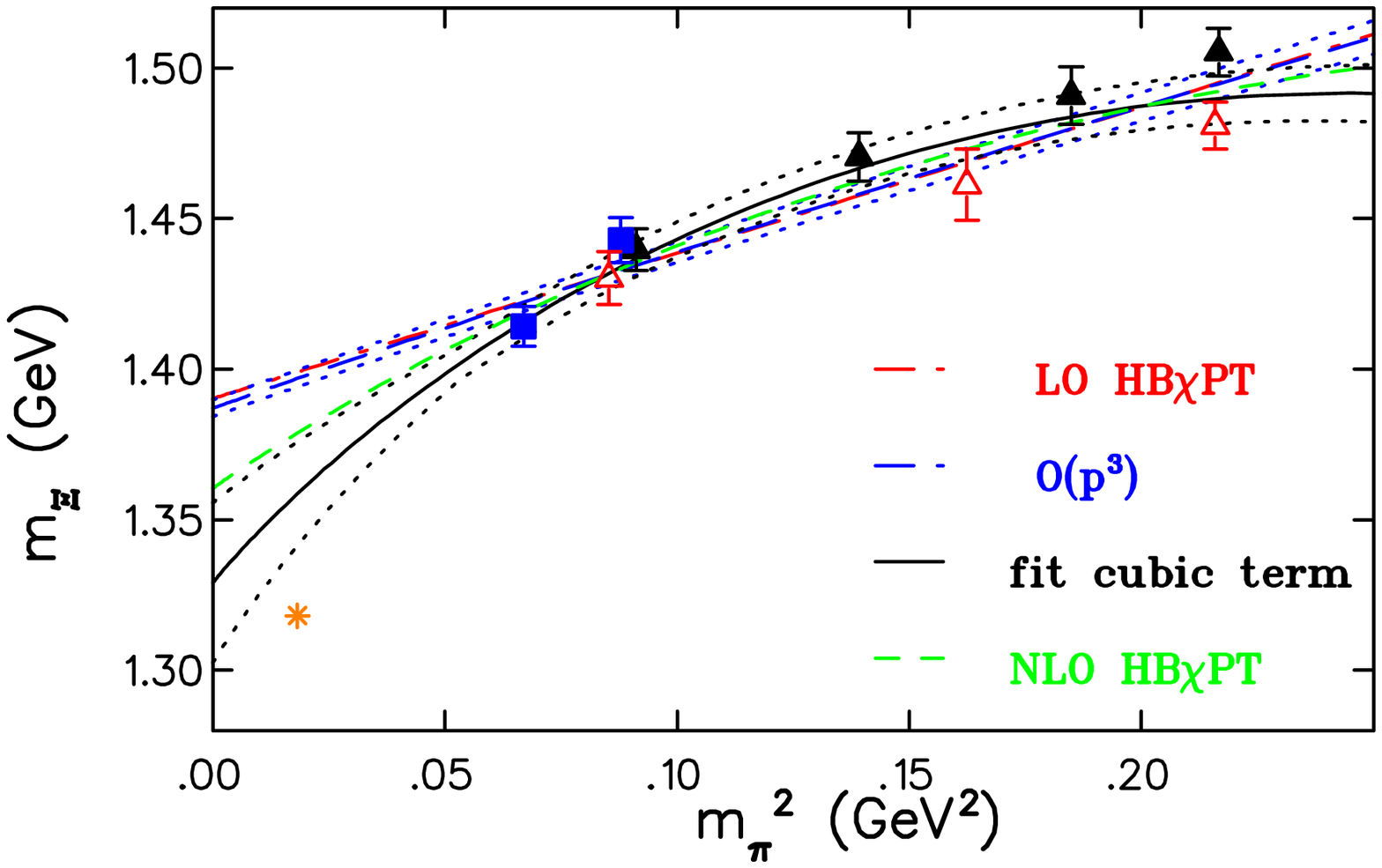}}
\caption{Chiral extrapolation of the $\Xi$ mass.The notation is the same as that in Fig.~\ref{fig:Sigma chiral}.}
\label{fig:Xi chiral}
\end{minipage}
\hfill
\begin{minipage}{8cm}
\epsfxsize=8truecm
\epsfysize=6truecm
\mbox{\epsfbox{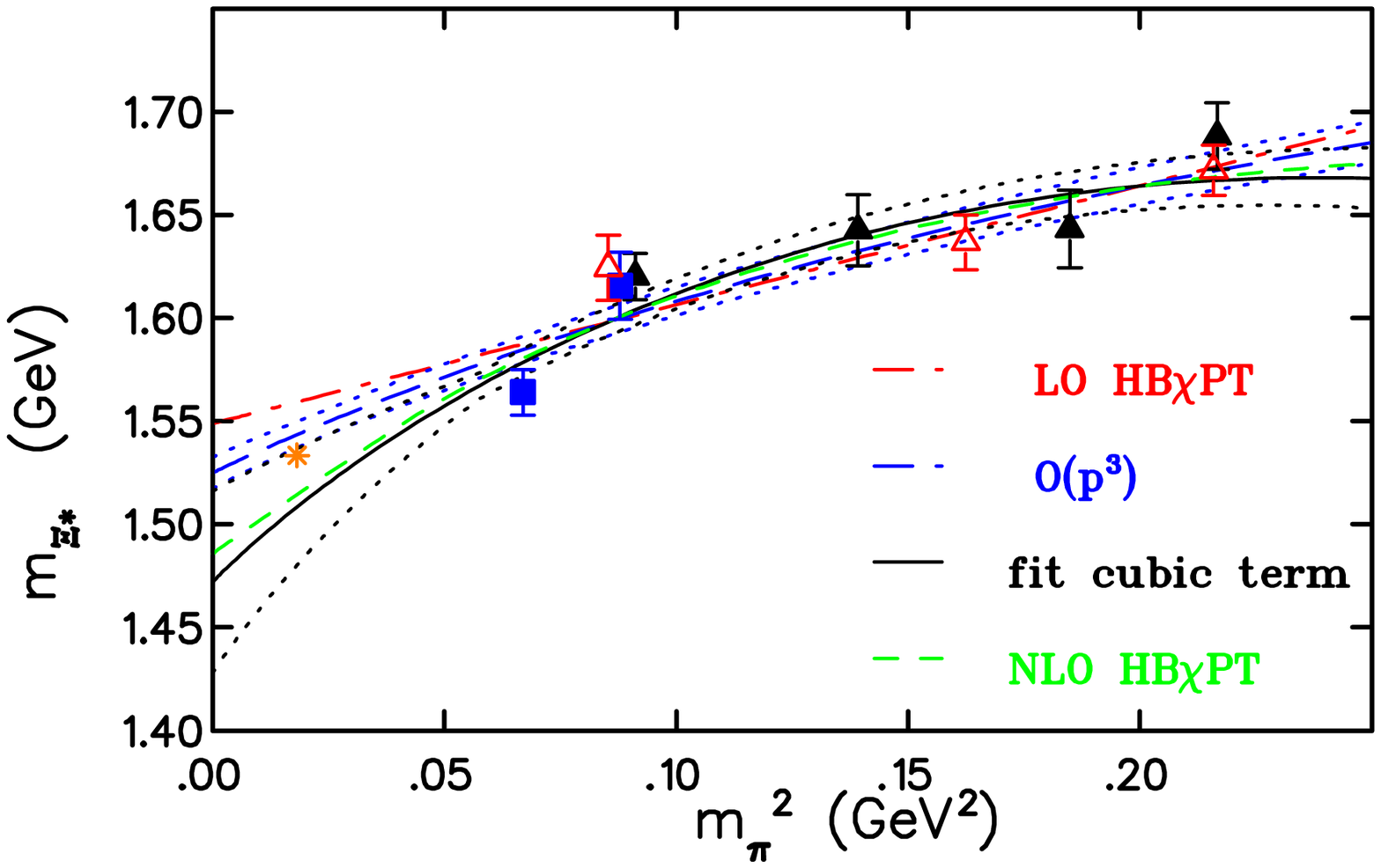}}
\caption{Chiral extrapolation of the $\Xi^*$ mass.
The notation is the same as that in Fig.~\ref{fig:Sigma chiral}.}
\label{fig:Xistar chiral}
\end{minipage}
\end{figure}

\begin{figure}[h!]
\begin{minipage}{8cm}
\epsfxsize=8truecm
\epsfysize=6truecm
\mbox{\epsfbox{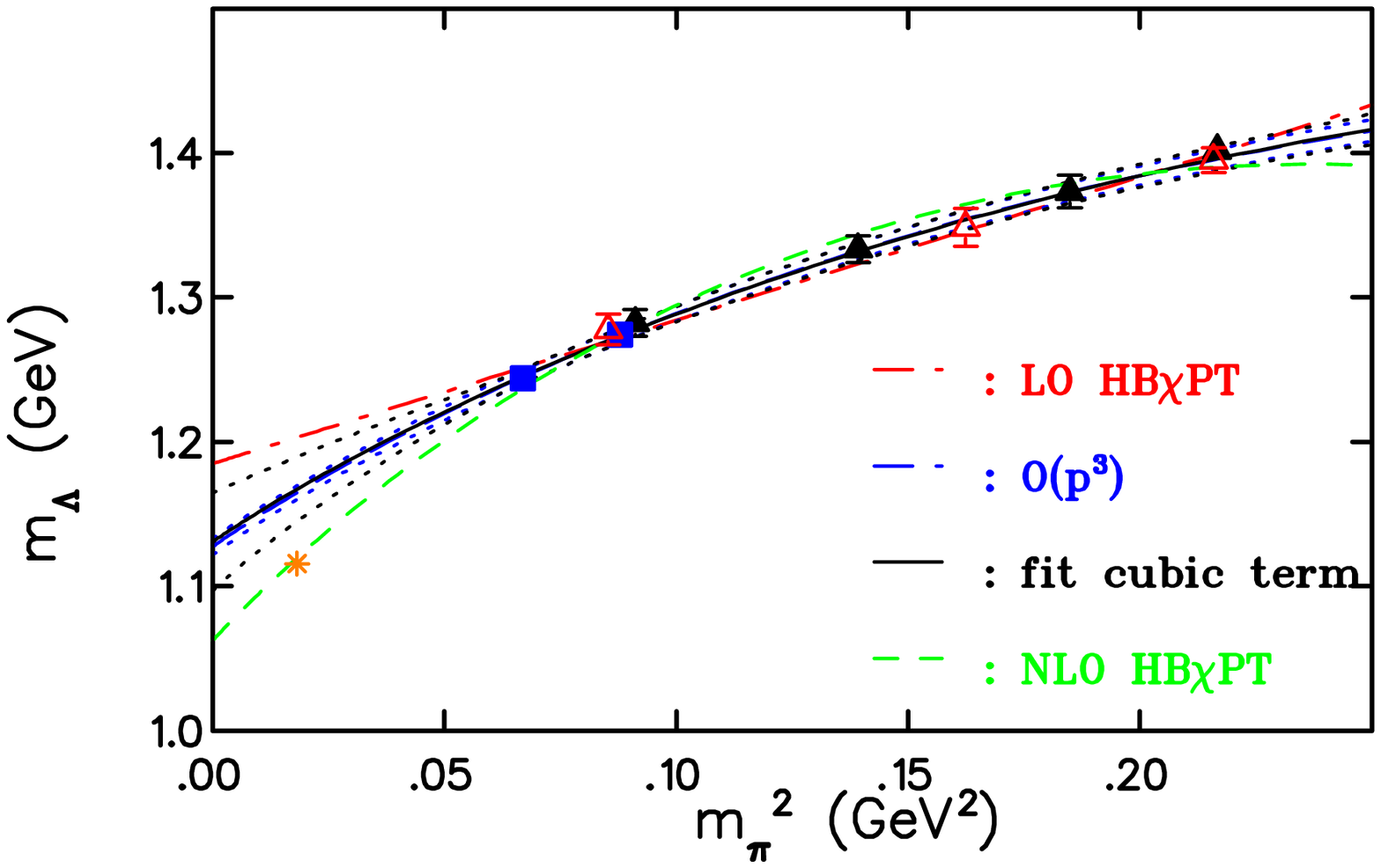}}
\caption{Chiral extrapolation of the $\Lambda$ mass.The notation is the same as that in Fig.~\ref{fig:Sigma chiral}.}
\label{fig:Lambda chiral}
\end{minipage}
\hfill
\begin{minipage}{8cm}
\epsfxsize=8truecm
\epsfysize=6truecm
\mbox{\epsfbox{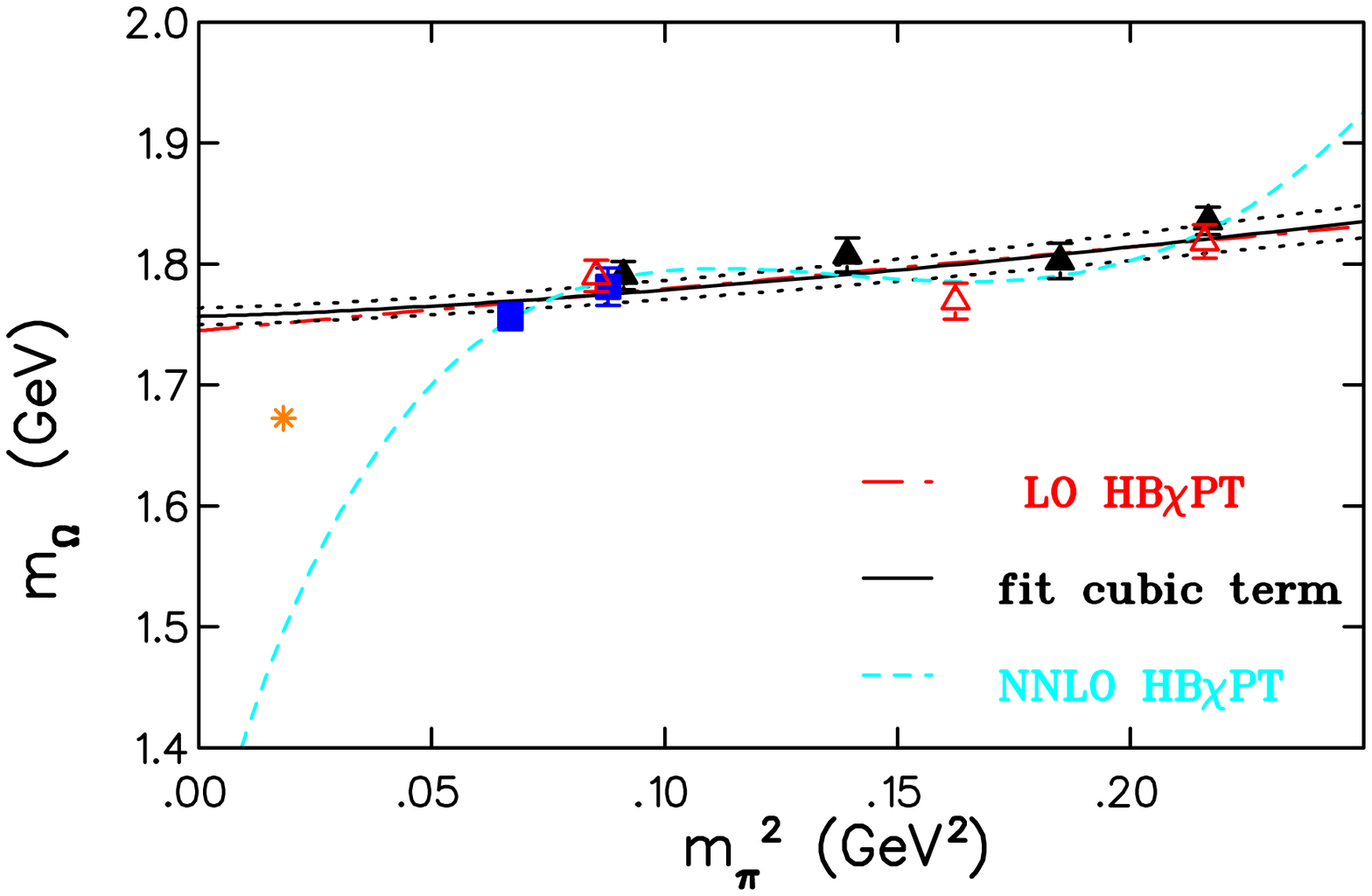}}
\caption{Chiral extrapolation of the $\Omega$ mass. We show chiral extrapolations 
linear in $m_\pi^2$, using Eq.~\ref{cubic},
 and NNLO in SU(2) chiral perturbation theory given in Eq.~(\ref{NNLO Omega}).
The rest of the notation is the same as that in Fig.~\ref{fig:nucleon chiral}.}
\label{fig:Omega chiral}
\end{minipage}
\end{figure}

The expressions for the strange baryon masses to NNLO in $\chi$PT 
given in Ref.~\cite{Tiburzi:2008bk} involve in general more unknown low energy constants
and only if we perform a constrained fit we have enough data to extract them. 
We found however no real advantage in using constrained fits, 
which generally gave larger $\chi/{\rm d.o.f.}$ 
and did not improve the prediction of  
the mass at the physical point as compared to the unconstrained fits. 
For the nucleon, $\Delta$ and  $\Omega$ masses, unconstrained
fits can still be performed with four fit parameters~\cite{WalkerLoud:2008bp}, 
namely $m^{(0)}$, $c^{(1)}$, $\alpha$ and $\beta$ appearing in the expressions of 
NNLO $\chi$PT given below.

\beq
m^{NNLO}_N(m_\pi)&=& m^{NLO}_N(m_\pi)+m_\pi^4\biggl[\beta_N+\frac{16g_{\Delta N}^2 c_N^{(1)}}{(4\pi f_\pi)^2}
-\frac{9g_{\Delta N}^2}{4m_N^{(0)}(4\pi f_\pi)^2}-\frac{45 g_A^2}{32 4m_N^{(0)}(4\pi f_\pi)^2} \biggr] \nonumber \\
&+& \frac{16g_{\Delta N}^2c_N^{(1)}}{(4\pi f_\pi)^2} m_\pi^2 \;{\cal J}(m_\pi,\Delta,\lambda) +\frac{m_\pi^4}{(4\pi f_\pi)^2}\log\left(\frac{m_\pi^2}{\lambda^2}\right) \biggl[{12c_N^{(1)} }
-\frac{3\alpha_N}{4\pi f_\pi} -\frac{27 g_A^2}{16 m_N^{(0)}} - \frac{5g_{\Delta N}}{2m_N{(0)}} \biggr]   \label{NNLO octet}
\eeq
\beq
m^{NNLO}_\Delta(m_\pi)&=& m^{NLO}_\Delta(m_\pi)  + \frac{12 c_\Delta^{(1)}}{(4\pi f_\pi)^2} \;m_\pi^4\log\left(\frac{m_\pi^2}{\lambda^2}\right)
-\frac{25 g_{\Delta \Delta}^2}{48 (m_\Delta^{(0)}+\Delta_{\Delta N})(4\pi f_\pi)^2} \; m_\pi^4\left(\log\left(\frac{m_\pi^2}{\lambda^2}\right)+\frac{19}{10}\right) \nonumber\\
&-&\frac{5 g_{\Delta N}^2}{8(m_\Delta^{(0)}+\Delta_{\Delta N})(4\pi f_\pi)^2}\; m_\pi^4\left(\log\left(\frac{m_\pi^2}{\lambda^2}\right)-\frac{1}{10}\right)
+\frac{4 c_\Delta^{(1)} g_{\Delta N}^2}{(4\pi f_\pi)^2}\; m_\pi^2 \;{\cal J}(m_\pi,-\Delta_{\Delta N},\lambda) \nonumber \\
&+& \beta_\Delta m_\pi^4  + \frac{ \alpha_\Delta}{(4\pi f_\pi)^3}\; m_\pi^4 \log\left(\frac{m_\pi^2}{\lambda^2}\right)   \label{NNLO Delta}\\
m^{NNLO}_\Omega(m_\pi)&=& m^{NLO}_\Omega(m_\pi)  +  \frac{m_{\pi}^4}{(4\pi f_{\pi})^3} \left[    \alpha_{\Omega} \log\left({m_{\pi}^2\over\lambda^2}\right) + \beta_{\Omega}  \right]
\label{NNLO Omega}
\eeq

where 
\be
{\cal J}(m,\Delta,\lambda)=m^4 \log\left(\frac{m^2}{\lambda^2}\right)+ 2\Delta \sqrt{\Delta^2-m^2+i\epsilon}
\; \log\left(\frac{\Delta-\sqrt{\Delta^2-m^2+i\epsilon}}{\Delta+\sqrt{\Delta^2-m^2+i\epsilon}}\right) +2 \Delta^2\log\left(\frac{4\Delta^2}{m^2}\right) \quad.
\ee
and the real part of ${\cal J}$ satisfies
\be
{\cal J}(m,-\Delta,\lambda) = \left\{ \begin{array}{ll}{\cal J}(m,\Delta,\lambda) & m<\Delta \\
         {\cal J}(m,\Delta,\lambda)-2\pi\Delta\left(m^2-\Delta^2\right)^{1/2} & m>\Delta \\
\end{array}\right.
\label{J symm}
\ee

Using the above Ans\"atze the chiral extrapolations  of lattice results at 
 $\beta=3.9$ and $\beta=4.05$ given in Tables~V-VII are performed.
In Fig.~\ref{fig:nucleon chiral} we show the fits for the nucleon 
in units of $r_0$.
The nucleon mass at the physical point has
been expressed in units of $r_0$  using the value determined from  nucleon mass as well as from the pion
decay constant. As can be seen these values are consistent. The ${\cal O}(p^3)$ being the one used to determine
the scale passes through the physical point. The other curves show the dependence on the chiral Ansatz used.
It comes as no surprise that the $NLO$ result does badly for the nucleon underestimating
the mass at the physical point whereas the $NNLO$ fits over-correct
and yield a higher mass. Lattice results
at $\beta=3.9$ and 4.05 expressed in units of $r_0$ fall on a universal
curve confirming that finite cut-off effects are small. Therefore 
we corroborate the conclusion that we can
use continuum chiral perturbation theory to extrapolate lattice results at   $\beta=3.9$ and $\beta=4.05$.
For the chiral extrapolation of the other baryons we use the scale 
determined from the nucleon mass to convert to physical units.
 
We show in Figs.~\ref{fig:Sigma chiral}, \ref{fig:Xi chiral}, \ref{fig:Lambda chiral} the
chiral extrapolations for the octet baryon masses and in 
Figs. \ref{fig:Delta chiral}, \ref{fig:Sigmastar chiral}, 
\ref{fig:Xistar chiral}, \ref{fig:Omega chiral} the corresponding fits for the decuplet baryons given in physical units. 
We emphasize that the physical point is not included in these fits. 

The LO expression describes well the lattice results but leads to extrapolated 
values inconsistent with the experimental point.
The ${\cal O}(p^3)$ $HB\chi PT$ expansion given in
 Eqs.~(\ref{LO octet}) and (\ref{LO decuplet})  
with two fit parameters $m^{(0)}$ and $c^{(1)}$ 
provides a  good description of lattice data and the results extrapolated to the physical point 
are in general in agreement with experiments.
The NLO leads to a clear improvement in the case of the $\Lambda$ and $\Xi$ masses, 
whereas for the rest of the baryons the improvement is marginal.
Apart from the preceding remarks, there is no clear advantage in using higher order fits, 
especially NNLO, which even turns out to be numerically unstable for the case of the 
$\Delta$ and $\Omega$ masses.
Therefore our main conclusion is that the ${\cal O}(p^3)$ HB$\chi$PT 
 provides a reasonable description for the nucleon and $\Delta$ masses whereas NLO $SU(2)$ for the the
strange baryon masses,
yielding values  at the physical point that are
in agreement with experiment.

\begin{table}[h!]
  \begin{center}
  \begin{tabular}{lcccccccc}
  \hline\hline
 & $\sigma_N$ & $\sigma_\Lambda$  &   $\sigma_{\Sigma^{\rm Av.}}$ & $\sigma_{\Xi_{\rm Av.}}$ & $\sigma_{\Delta^{{\rm Av.}}}$ & $\sigma_{\Sigma^{\ast {\rm Av.}}}$  &   $\sigma_{\Xi^{\ast {\rm Av.}}}$ & $\sigma_{\Omega}$ \\
 \hline\hline
{${\cal O}(p^3)$} & 64.2(8) & 34.7(9) & 37.1(8) & 9.7(1.1) & 62.2(1.1) & 38.0(1.7)& 17.3(1.3) & 6.3(1.3) \\
fit with cubic term & 33.4(6.9) &33.7 (9.8) &35.6(10.6) & 26.3 (7.7) & 25.7 (2.5) & 24.2 (6.4) & 32.4 (13.4) & 2.9(1.4) \\
$NLO$ &  92.5(7) & 52.8(8) &43.3(9) &17.2(1.0) & 79.5(1.0)&44.1(1.7) & 27.9(1.3) &  6.3(1.3) \\
\hline
\end{tabular}
  \caption{$\sigma$-term in MeV using the fit parameters determined from ${\cal O}(p^3)$ $\chi$PT, using a cubic fit Eq.~(\ref{cubic}) and NLO. We used the
scale from the nucleon
 mass to convert to physical units.} 
  \label{tab:sigma term}
\end{center}
\end{table}

\begin{figure}[h!]
\begin{minipage}{8cm}
\epsfxsize=7.5truecm
\epsfysize=12truecm
\mbox{\epsfbox{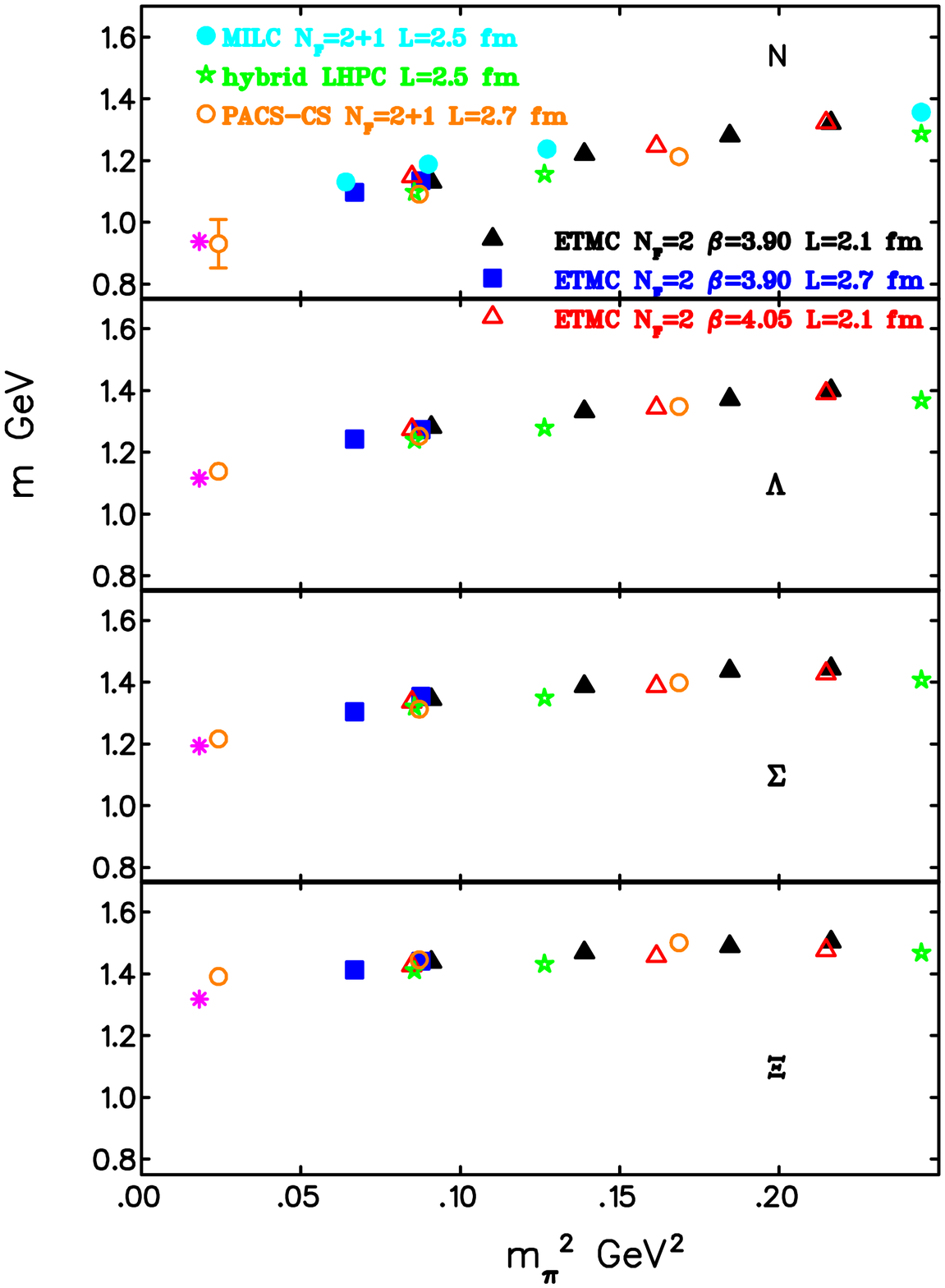}}
\caption{\label{fig:octet} Comparison of masses for the low lying octet 
baryons. Results from this work are shown by the 
filled (black) triangles for $L=2.1$~fm and (blue) squares for $L=2.7$~fm with $
a=0.089$~fm and with the open (red) triangles  for $L=2.1$~fm and $a=0.070$~fm. 
Results with the hybrid action (LHPC) are shown with the (green) asterisks for $a=0.124
$~fm and results using $N_f=2+1$ 
Clover fermions (PACS-CS) with
the open (orange) circles and $a=0.0907$~fm. 
For the nucleon we also show results
using $N_f=2+1$ asqtad improved staggered fermions (MILC) denoted by the filled (light blue) circles.
The physical masses are shown by the (purple) star. }
\end{minipage}
\hfill
\begin{minipage}{8cm}
\epsfxsize=7.5truecm
\epsfysize=12truecm \vspace*{-3cm}
\mbox{\epsfbox{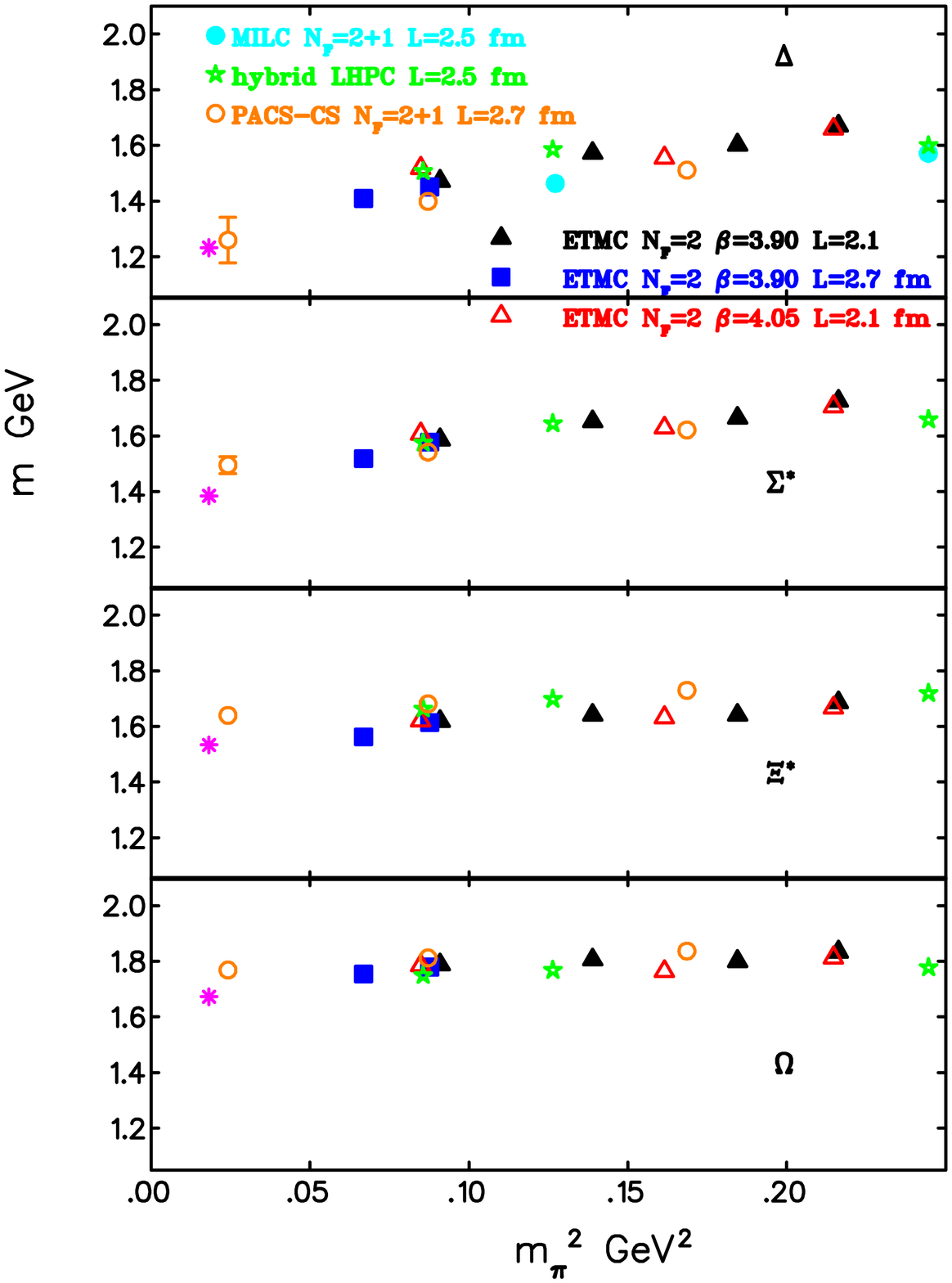}}
\caption{\label{fig:decuplet} Comparison of masses for the low lying decuplet
 baryons.The notation is the same as that of Fig.~\ref{fig:octet}. }
\end{minipage}
\end{figure}

We use the relation
$m_\pi^2\sim \mu$ to evaluate the nucleon $\sigma$-term by computing 
$m_\pi^2\frac{dM_N}{dm_\pi^2}$. Using our ${\cal O}(p^3)$ fit   we find
$\sigma_N=64.2(9)$~MeV in agreement with the value given in Ref.~\cite{Alexandrou:2008tn}.
 Generalizing the relation 
we can evaluate the corresponding $\sigma$-terms for the other
hadrons. We list in Table~\ref{tab:sigma term}
the values we obtain using the nucleon mass to set the scale.
As can be seen, the value extracted depends on the fitting Ansatz. In
the most interesting case of the nucleon the result of  ${\cal O}(p^3)$ 
changes by
two standard deviations if the coefficient of the cubic term in $m_\pi$
is fitted.  In the case of the $\Lambda$ fitting the cubic term gives the
same value as that obtained to ${\cal O}(p^3)$,  compatible 
to that of the nucleon. This is
another indication of the argument presented above in favor for the
presence of a cubic term in $m_\pi$ of comparable size as that of the nucleon. In fact
the main conclusion of this exercise is that  allowing the 
coefficient of the cubic term to be determined from the data produces a
$\sigma$-term that for all baryons except the $\Omega$ is comparable within error to the value of nucleon $\sigma$-term. Comparing to the results of NLO we can see that for the nucleon
this fit produces too much curvature as already observed for instance in 
Ref.~\cite{WalkerLoud:2008bp}. For the other baryons a reasonable value is obtained 
depending on the quality of
the fits as seen in the figures.

\section{Comparison with other lattice results and with experiment}
In this section we show a comparison of recent lattice
results on the baryon masses from various collaborations in Figs.~\ref{fig:octet} and \ref{fig:decuplet}.
For our results we use the lattice spacing determined from the nucleon mass to convert physical units. Results from
the other collaborations are converted to physical units using the lattice spacing that they provide.
The level of agreement  of lattice QCD 
results using a variety of fermion discretization schemes
before taking the continuum limit or other lattice artifacts into account 
is quite impressive. Small discrepancies seen mainly in
the decuplet masses can be attributed  to lattice artifacts.
 In
particular results using asqtad improved staggered fermions may suffer the most 
from discretization errors. The MILC collaboration has simulations on finer lattices
and an update on the masses is  expected in the the near future. We note 
that results very close to the physical point obtained
using Clover fermions from the PACS-CS collaboration~\cite{Aoki:2008sm}
may have large finite volume effects due to the fact
that $m_\pi L<3.5$ in this simulation.

\begin{figure}[h!]
\mbox{\epsfxsize=10.cm\epsfysize=10.cm\epsffile{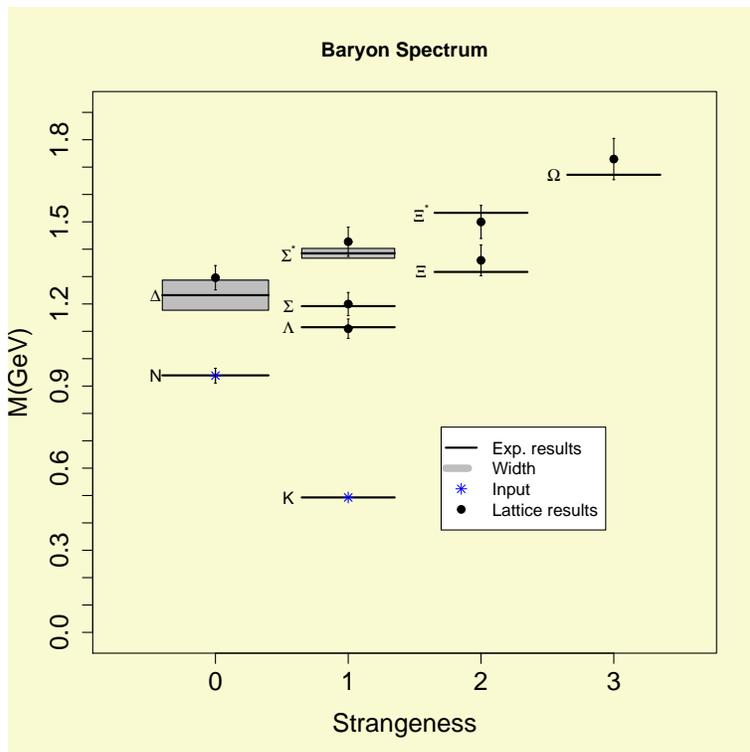}}
\caption{The  octet and decuplet spectrum.
Data are shown in the continuum and at the physical point. 
For the nucleon and $\Delta$  using ${\cal O}(p^3)$ and for the rest
 using NLO SU(2) HB$\chi$PT
for the chiral extrapolation.}
\label{fig:spectrum}
\end{figure}

 Finally we show our continuum results 
 in Fig.~\ref{fig:spectrum}. We take the
continuum limit using results at $\beta=3.9$ and $\beta=4.05$ after
interpolating at a given value of $r_0m_\pi$. 
The continuum values used 
 are collected in Table XIII and VIX for the octet and decuplet respectively.
Residual cut-off effects that may result
from using   Eq.~(\ref{s-mass}) to estimate $\mu_s$ at $\beta=4.05$ are not included in the  systematic errors.  
 For the nucleon and the $\Delta$ we use
${\cal O}(p^3)$ to extrapolate to the physical point as done in our previous
work~\cite{Alexandrou:2008tn}. For the strangeness non-zero baryons we use
NLO SU(2)
 HB$\chi$PT to  
extrapolate to the physical point. In the statistical error we have added
the error arising from the uncertainty in $r_0^N$, i.e. the difference in
the resulting masses  when we use $r_0^N=0.471$~fm and $r_0^N\pm 0.021$~fm to
set the scale.
As can be seen, our results compare well with 
experiment within the estimated uncertainties.

\section{Summary and Conclusions}

The focus of this work is
 the computation of the low-lying baryon masses
using twisted mass fermions at maximal twist.
It is in line with ongoing 
efforts by other lattice collaborations worldwide 
to predict the baryon spectrum from first principles.
Comparison of lattice results with experiment is regarded 
 as an important benchmark for  
lattice QCD and justifies the use of
 different lattice actions, each with  different systematic errors. 
For example, the twisted mass action with only one parameter to tune, 
provides automatic ${\cal O}(a)$ improvement. However
it restores isospin symmetry only in 
the continuum limit. 
We have examined the consequences of isospin breaking
  on the baryon masses and found them
to be either small or compatible
with zero. On our finer lattice at  $\beta=4.05$  the maximal
isospin violation is obtained in the octet  only
in the case of the Cascade where we find that $m_{\Xi^0}-m_{\Xi^-}\sim 50$~MeV.
 Finite volume corrections are estimated in the case of the nucleon and found
to be smaller than statistical errors.
The continuum extrapolation using results at $\beta=3.9$ and $\beta=4.05$
are verified using a finer lattice at $\beta=4.2$ in the case of the nucleon
supporting the analysis carried out.  Therefore this study shows that both
cut-off effects and finite volume corrections are small and continuum
results can be extracted using lattice data at  $\beta=3.9$ and $4.05$. 

An investigation of the Gell-Mann-Okubo relations has been carried out. For the 
octet baryon masses we find that these relations are  
  satisfied at all pion  masses even at a non-vanishing lattice spacing.
For the decuplet baryon masses we see deviations and it will be
 interesting to study these relations
at finer values of the lattice spacing and smaller quark masses.

Comparison of the  results at given lattice spacings
with those of other collaborations
 reveals consistency among groups using improved actions
with lattice spacing being smaller than 0.1~fm.
This is a non-trivial consistency  check of the several 
lattice formulations directly on lattice data without without
          the necessity of any extrapolations.
This level of agreement
 among different lattice actions, is a welcome outcome of
the collective effort of several collaborations.
The final continuum results at the physical limit shown in Fig.~\ref{fig:spectrum}
are in excellent agreement with
experiment. The largest uncertainty in the final value comes
 from systematic errors
in setting the scale, which are an order of magnitude larger 
than statistical errors.

Besides the masses we have extracted from the various fits the
$\sigma$-term. To ${\cal O}(p^3)$ $\chi$PT  we find a value of $64(1)$~MeV 
for the nucleon
$\sigma$-term. Allowing the coefficient of the cubic term in $m_\pi$ to
be determined from the data yields a smaller value of $39(12)$~MeV albeit with
a large statistical error. Fitting with the latter Ansatz produces for all
baryons expect the $\Omega$ a value of the $\sigma$-term compatible 
with  that of the nucleon. Clearly this is a result that requires further study. 
In particular results at smaller pion masses will help to better
determine the curvature of fits.

The next step for the ETM collaboration is to perform the analysis
using a dynamical strange quark. Within the twisted mass formalism this
is accomplished by simulating a non-degenerate doublet. 
Such $N_F=2+1+1$ simulations are already available at two values of the lattice spacing~\cite{Baron:2008xa}  comparable
to the ones studied in this work. 
This future study will provide a nice comparison to the
present work and enable  us to gauge unquenching effects in the
strange quark sector.

\section*{Acknowledgments}
We would like to thank all members of ETMC for a
very constructive and enjoyable collaboration and the many fruitful 
discussions that took place during the development of this work.

This work was performed using HPC resources from GENCI at IDRIS,
Grant 2009-2009052271. We have also largely benefited from
computer and storage resources in the CCIN2P3 (Lyon).
 Computer time for this project was also 
made available to us by the
John von Neumann-Institute for Computing on the JUMP and 
 Jugene systems at the research center
in J\"ulich and the Stella system at the
Donald Smits Center for Information Technology in Groningen.
 We thank the staff members for their kind and sustained support. 
This work has been supported in part by  the DFG
Sonder\-for\-schungs\-be\-reich/ Trans\-region SFB/TR9.
This work was partly supported by funding received  from the
 Cyprus Research Promotion Foundation under contracts EPYAN/0506/08,
KY-$\Gamma$/0907/11/ and TECHNOLOGY/$\Theta$E$\Pi$I$\Sigma$/0308(BE)/17. 
\bibliography{strange_baryons_V3}

\begin{thebibliography}{54}
\expandafter\ifx\csname natexlab\endcsname\relax\def\natexlab#1{#1}\fi
\expandafter\ifx\csname bibnamefont\endcsname\relax
  \def\bibnamefont#1{#1}\fi
\expandafter\ifx\csname bibfnamefont\endcsname\relax
  \def\bibfnamefont#1{#1}\fi
\expandafter\ifx\csname citenamefont\endcsname\relax
  \def\citenamefont#1{#1}\fi
\expandafter\ifx\csname url\endcsname\relax
  \def\url#1{\texttt{#1}}\fi
\expandafter\ifx\csname urlprefix\endcsname\relax\def\urlprefix{URL }\fi
\providecommand{\bibinfo}[2]{#2}
\providecommand{\eprint}[2][]{\url{#2}}

\bibitem[{\citenamefont{Jansen}(2008)}]{Jansen:2008vs}
\bibinfo{author}{\bibfnamefont{K.}~\bibnamefont{Jansen}}
  (\bibinfo{year}{2008}), \eprint{0810.5634}.

\bibitem[{\citenamefont{Bernard et~al.}(2001)}]{Bernard:2001av}
\bibinfo{author}{\bibfnamefont{C.~W.} \bibnamefont{Bernard}}
  \bibnamefont{et~al.}, \bibinfo{journal}{Phys. Rev.}
  \textbf{\bibinfo{volume}{D64}}, \bibinfo{pages}{054506}
  (\bibinfo{year}{2001}), \eprint{hep-lat/0104002}.

\bibitem[{\citenamefont{Aubin et~al.}(2004)}]{Aubin:2004fs}
\bibinfo{author}{\bibfnamefont{C.}~\bibnamefont{Aubin}} \bibnamefont{et~al.}
  (\bibinfo{collaboration}{MILC}), \bibinfo{journal}{Phys. Rev.}
  \textbf{\bibinfo{volume}{D70}}, \bibinfo{pages}{114501}
  (\bibinfo{year}{2004}), \eprint{hep-lat/0407028}.

\bibitem[{\citenamefont{Alexandrou et~al.}(2008)}]{Alexandrou:2008tn}
\bibinfo{author}{\bibfnamefont{C.}~\bibnamefont{Alexandrou}}
  \bibnamefont{et~al.} (\bibinfo{collaboration}{European Twisted Mass}),
  \bibinfo{journal}{Phys. Rev.} \textbf{\bibinfo{volume}{D78}},
  \bibinfo{pages}{014509} (\bibinfo{year}{2008}), \eprint{0803.3190}.

\bibitem[{\citenamefont{Ali~Khan et~al.}(2004)}]{AliKhan:2003cu}
\bibinfo{author}{\bibfnamefont{A.}~\bibnamefont{Ali~Khan}} \bibnamefont{et~al.}
  (\bibinfo{collaboration}{QCDSF-UKQCD}), \bibinfo{journal}{Nucl. Phys.}
  \textbf{\bibinfo{volume}{B689}}, \bibinfo{pages}{175} (\bibinfo{year}{2004}),
  \eprint{hep-lat/0312030}.

\bibitem[{\citenamefont{Aoki et~al.}(2009)}]{Aoki:2008sm}
\bibinfo{author}{\bibfnamefont{S.}~\bibnamefont{Aoki}} \bibnamefont{et~al.}
  (\bibinfo{collaboration}{PACS-CS}), \bibinfo{journal}{Phys. Rev. D}
  \textbf{\bibinfo{volume}{79}}, \bibinfo{pages}{034503}
  (\bibinfo{year}{2009}), \eprint{0807.1661}.

\bibitem[{\citenamefont{Durr et~al.}(2008)}]{Durr:2008zz}
\bibinfo{author}{\bibfnamefont{S.}~\bibnamefont{Durr}} \bibnamefont{et~al.},
  \bibinfo{journal}{Science} \textbf{\bibinfo{volume}{322}},
  \bibinfo{pages}{1224} (\bibinfo{year}{2008}).

\bibitem[{\citenamefont{Walker-Loud et~al.}(2009)}]{WalkerLoud:2008bp}
\bibinfo{author}{\bibfnamefont{A.}~\bibnamefont{Walker-Loud}}
  \bibnamefont{et~al.}, \bibinfo{journal}{Phys.Rev. D}
  \textbf{\bibinfo{volume}{79}}, \bibinfo{pages}{054502}
  (\bibinfo{year}{2009}), \eprint{0806.4549}.

\bibitem[{\citenamefont{Antonio et~al.}(2006{\natexlab{a}})}]{Antonio:2006px}
\bibinfo{author}{\bibfnamefont{D.~J.} \bibnamefont{Antonio}}
  \bibnamefont{et~al.} (\bibinfo{collaboration}{RBC and UKQCD}),
  \bibinfo{journal}{PoS} \textbf{\bibinfo{volume}{LAT2006}},
  \bibinfo{pages}{189} (\bibinfo{year}{2006}{\natexlab{a}}).

\bibitem[{\citenamefont{Antonio et~al.}(2006{\natexlab{b}})}]{Antonio:2006zz}
\bibinfo{author}{\bibfnamefont{D.~J.} \bibnamefont{Antonio}}
  \bibnamefont{et~al.} (\bibinfo{collaboration}{RBC and UKQCD}),
  \bibinfo{journal}{PoS} \textbf{\bibinfo{volume}{LAT2006}},
  \bibinfo{pages}{189} (\bibinfo{year}{2006}{\natexlab{b}}).

\bibitem[{\citenamefont{Blossier et~al.}(2008)}]{Blossier:2007vv}
\bibinfo{author}{\bibfnamefont{B.}~\bibnamefont{Blossier}} \bibnamefont{et~al.}
  (\bibinfo{collaboration}{European Twisted Mass}), \bibinfo{journal}{JHEP}
  \textbf{\bibinfo{volume}{04}}, \bibinfo{pages}{020} (\bibinfo{year}{2008}),
  \eprint{0709.4574}.

\bibitem[{\citenamefont{Blossier et~al.}(2009)}]{Blossier:2009bx}
\bibinfo{author}{\bibfnamefont{B.}~\bibnamefont{Blossier}} \bibnamefont{et~al.}
  (\bibinfo{year}{2009}), \eprint{0904.0954}.

\bibitem[{\citenamefont{Boucaud et~al.}(2007)}]{Boucaud:2007uk}
\bibinfo{author}{\bibfnamefont{P.}~\bibnamefont{Boucaud}} \bibnamefont{et~al.}
  (\bibinfo{collaboration}{ETM}), \bibinfo{journal}{Phys. Lett.}
  \textbf{\bibinfo{volume}{B650}}, \bibinfo{pages}{304} (\bibinfo{year}{2007}),
  \eprint{hep-lat/0701012}.

\bibitem[{\citenamefont{Boucaud et~al.}(2008{\natexlab{a}})}]{Boucaud:2008xu}
\bibinfo{author}{\bibfnamefont{P.}~\bibnamefont{Boucaud}} \bibnamefont{et~al.}
  (\bibinfo{collaboration}{ETM}), \bibinfo{journal}{Comput. Phys. Commun.}
  \textbf{\bibinfo{volume}{179}}, \bibinfo{pages}{695}
  (\bibinfo{year}{2008}{\natexlab{a}}), \eprint{0803.0224}.

\bibitem[{\citenamefont{Frezzotti and Rossi}(2007)}]{Frezzotti:2007qv}
\bibinfo{author}{\bibfnamefont{R.}~\bibnamefont{Frezzotti}} \bibnamefont{and}
  \bibinfo{author}{\bibfnamefont{G.}~\bibnamefont{Rossi}},
  \bibinfo{journal}{PoS} \textbf{\bibinfo{volume}{LAT2007}},
  \bibinfo{pages}{277} (\bibinfo{year}{2007}), \eprint{0710.2492}.

\bibitem[{\citenamefont{Dimopoulos et~al.}(2008)}]{Dimopoulos:2008sy}
\bibinfo{author}{\bibfnamefont{P.}~\bibnamefont{Dimopoulos}}
  \bibnamefont{et~al.} (\bibinfo{collaboration}{ETM}) (\bibinfo{year}{2008}),
  \eprint{0810.2873}.

\bibitem[{\citenamefont{Drach et~al.}(2008)}]{Latt08_Vincent}
\bibinfo{author}{\bibfnamefont{V.}~\bibnamefont{Drach}} \bibnamefont{et~al.}
  (\bibinfo{collaboration}{ETM Collaboration}), \bibinfo{journal}{PoS}
  \textbf{\bibinfo{volume}{LAT2008}}, \bibinfo{pages}{123}
  (\bibinfo{year}{2008}).

\bibitem[{\citenamefont{Weisz}(1983)}]{Weisz:1982}
\bibinfo{author}{\bibfnamefont{P.}~\bibnamefont{Weisz}},
  \bibinfo{journal}{Nucl. Phys.} \textbf{\bibinfo{volume}{B212}},
  \bibinfo{pages}{1} (\bibinfo{year}{1983}).

\bibitem[{\citenamefont{Boucaud et~al.}(2008{\natexlab{b}})}]{ETMClong}
\bibinfo{author}{\bibfnamefont{P.}~\bibnamefont{Boucaud}} \bibnamefont{et~al.}
  (\bibinfo{year}{2008}{\natexlab{b}}), \eprint{arXiv:0803.0224 [hep-lat]}.

\bibitem[{\citenamefont{Frezzotti et~al.}(2006)\citenamefont{Frezzotti,
  Martinelli, Papinutto, and Rossi}}]{Frezzotti:2005gi}
\bibinfo{author}{\bibfnamefont{R.}~\bibnamefont{Frezzotti}},
  \bibinfo{author}{\bibfnamefont{G.}~\bibnamefont{Martinelli}},
  \bibinfo{author}{\bibfnamefont{M.}~\bibnamefont{Papinutto}},
  \bibnamefont{and} \bibinfo{author}{\bibfnamefont{G.}~\bibnamefont{Rossi}},
  \bibinfo{journal}{JHEP} \textbf{\bibinfo{volume}{0604}}, \bibinfo{pages}{038}
  (\bibinfo{year}{2006}), \eprint{arXiv:hep-lat/0503034]}.

\bibitem[{\citenamefont{Frezzotti and
  Rossi}(2004{\natexlab{a}})}]{Frezzotti:2004}
\bibinfo{author}{\bibfnamefont{R.}~\bibnamefont{Frezzotti}} \bibnamefont{and}
  \bibinfo{author}{\bibfnamefont{G.}~\bibnamefont{Rossi}},
  \bibinfo{journal}{JHEP} \textbf{\bibinfo{volume}{0408}}, \bibinfo{pages}{007}
  (\bibinfo{year}{2004}{\natexlab{a}}).

\bibitem[{\citenamefont{Osterwalder and Seiler}(1978)}]{Osterwalder:1977pc}
\bibinfo{author}{\bibfnamefont{K.}~\bibnamefont{Osterwalder}} \bibnamefont{and}
  \bibinfo{author}{\bibfnamefont{E.}~\bibnamefont{Seiler}},
  \bibinfo{journal}{Ann. Phys.} \textbf{\bibinfo{volume}{110}},
  \bibinfo{pages}{440} (\bibinfo{year}{1978}).

\bibitem[{\citenamefont{Frezzotti and
  Rossi}(2004{\natexlab{b}})}]{Frezzotti:2004wz}
\bibinfo{author}{\bibfnamefont{R.}~\bibnamefont{Frezzotti}} \bibnamefont{and}
  \bibinfo{author}{\bibfnamefont{G.~C.} \bibnamefont{Rossi}},
  \bibinfo{journal}{JHEP} \textbf{\bibinfo{volume}{10}}, \bibinfo{pages}{070}
  (\bibinfo{year}{2004}{\natexlab{b}}), \eprint{hep-lat/0407002}.

\bibitem[{\citenamefont{Abdel-Rehim et~al.}(2007)\citenamefont{Abdel-Rehim,
  Lewis, Woloshyn, and Wu}}]{AbdelRehim:2006ra}
\bibinfo{author}{\bibfnamefont{A.~M.} \bibnamefont{Abdel-Rehim}},
  \bibinfo{author}{\bibfnamefont{R.}~\bibnamefont{Lewis}},
  \bibinfo{author}{\bibfnamefont{R.~M.} \bibnamefont{Woloshyn}},
  \bibnamefont{and} \bibinfo{author}{\bibfnamefont{J.~M.~S.} \bibnamefont{Wu}},
  \bibinfo{journal}{Eur. Phys. J.} \textbf{\bibinfo{volume}{A31}},
  \bibinfo{pages}{773} (\bibinfo{year}{2007}), \eprint{hep-lat/0610090}.

\bibitem[{\citenamefont{Abdel-Rehim et~al.}(2006)\citenamefont{Abdel-Rehim,
  Lewis, Woloshyn, and Wu}}]{AbdelRehim:2006ve}
\bibinfo{author}{\bibfnamefont{A.~M.} \bibnamefont{Abdel-Rehim}},
  \bibinfo{author}{\bibfnamefont{R.}~\bibnamefont{Lewis}},
  \bibinfo{author}{\bibfnamefont{R.~M.} \bibnamefont{Woloshyn}},
  \bibnamefont{and} \bibinfo{author}{\bibfnamefont{J.~M.~S.} \bibnamefont{Wu}},
  \bibinfo{journal}{Phys. Rev.} \textbf{\bibinfo{volume}{D74}},
  \bibinfo{pages}{014507} (\bibinfo{year}{2006}), \eprint{hep-lat/0601036}.

\bibitem[{\citenamefont{Urbach}(2007)}]{Urbach:2007}
\bibinfo{author}{\bibfnamefont{C.}~\bibnamefont{Urbach}},
  \bibinfo{journal}{PoS} \textbf{\bibinfo{volume}{LAT2007}},
  \bibinfo{pages}{022} (\bibinfo{year}{2007}).

\bibitem[{\citenamefont{Lubicz and Tarantino}(2008)}]{private:2008}
\bibinfo{author}{\bibfnamefont{V.}~\bibnamefont{Lubicz}} \bibnamefont{and}
  \bibinfo{author}{\bibfnamefont{C.}~\bibnamefont{Tarantino}},
  \bibinfo{howpublished}{private communication} (\bibinfo{year}{2008}).

\bibitem[{\citenamefont{Dimopoulos et~al.}(2007)\citenamefont{Dimopoulos,
  Frezzotti, Herdoiza, Urbach, and Wenger}}]{Dimopoulos:2007qy}
\bibinfo{author}{\bibfnamefont{P.}~\bibnamefont{Dimopoulos}},
  \bibinfo{author}{\bibfnamefont{R.}~\bibnamefont{Frezzotti}},
  \bibinfo{author}{\bibfnamefont{G.}~\bibnamefont{Herdoiza}},
  \bibinfo{author}{\bibfnamefont{C.}~\bibnamefont{Urbach}}, \bibnamefont{and}
  \bibinfo{author}{\bibfnamefont{U.}~\bibnamefont{Wenger}}
  (\bibinfo{collaboration}{ETM Collaboration}), \bibinfo{journal}{PoS}
  \textbf{\bibinfo{volume}{LAT2007}} (\bibinfo{year}{2007}),
  \eprint{arXiv:0710.2498 [hep-lat]}.

\bibitem[{\citenamefont{Alexandrou et~al.}()\citenamefont{Alexandrou,
  Constantinou, and Korzec}}]{Cyprus:2009}
\bibinfo{author}{\bibfnamefont{C.}~\bibnamefont{Alexandrou}},
  \bibinfo{author}{\bibfnamefont{M.}~\bibnamefont{Constantinou}},
  \bibnamefont{and} \bibinfo{author}{\bibfnamefont{T.}~\bibnamefont{Korzec}},
  \bibinfo{note}{private communication; in preparation}.

\bibitem[{\citenamefont{Ioffe}(1981)}]{Ioffe:1981kw}
\bibinfo{author}{\bibfnamefont{B.~L.} \bibnamefont{Ioffe}},
  \bibinfo{journal}{Nucl. Phys.} \textbf{\bibinfo{volume}{B188}},
  \bibinfo{pages}{317} (\bibinfo{year}{1981}).

\bibitem[{\citenamefont{Leinweber et~al.}(1991)\citenamefont{Leinweber,
  Woloshyn, and Draper}}]{Leinweber:1990dv}
\bibinfo{author}{\bibfnamefont{D.~B.} \bibnamefont{Leinweber}},
  \bibinfo{author}{\bibfnamefont{R.~M.} \bibnamefont{Woloshyn}},
  \bibnamefont{and} \bibinfo{author}{\bibfnamefont{T.}~\bibnamefont{Draper}},
  \bibinfo{journal}{Phys. Rev.} \textbf{\bibinfo{volume}{D43}},
  \bibinfo{pages}{1659} (\bibinfo{year}{1991}).

\bibitem[{\citenamefont{Leinweber et~al.}(1992)\citenamefont{Leinweber, Draper,
  and Woloshyn}}]{Leinweber:1992hy}
\bibinfo{author}{\bibfnamefont{D.~B.} \bibnamefont{Leinweber}},
  \bibinfo{author}{\bibfnamefont{T.}~\bibnamefont{Draper}}, \bibnamefont{and}
  \bibinfo{author}{\bibfnamefont{R.~M.} \bibnamefont{Woloshyn}},
  \bibinfo{journal}{Phys. Rev.} \textbf{\bibinfo{volume}{D46}},
  \bibinfo{pages}{3067} (\bibinfo{year}{1992}), \eprint{hep-lat/9208025}.

\bibitem[{\citenamefont{Wolff}(2004)}]{Wolff:2004}
\bibinfo{author}{\bibfnamefont{U.}~\bibnamefont{Wolff}}
  (\bibinfo{collaboration}{ALPHA}), \bibinfo{journal}{Comput. Phys. Commun.}
  \textbf{\bibinfo{volume}{156}}, \bibinfo{pages}{143} (\bibinfo{year}{2004}),
  \eprint{hep-lat/0306017}.

\bibitem[{\citenamefont{Donoghue et~al.}(1992)\citenamefont{Donoghue, Golowich,
  and Holstein}}]{Donoghue:1992dd}
\bibinfo{author}{\bibfnamefont{J.~F.} \bibnamefont{Donoghue}},
  \bibinfo{author}{\bibfnamefont{E.}~\bibnamefont{Golowich}}, \bibnamefont{and}
  \bibinfo{author}{\bibfnamefont{B.~R.} \bibnamefont{Holstein}},
  \bibinfo{journal}{Camb. Monogr. Part. Phys. Nucl. Phys. Cosmol.}
  \textbf{\bibinfo{volume}{2}}, \bibinfo{pages}{1} (\bibinfo{year}{1992}).

\bibitem[{\citenamefont{Beane et~al.}(2007)\citenamefont{Beane, Orginos, and
  Savage}}]{Beane:2006pt}
\bibinfo{author}{\bibfnamefont{S.~R.} \bibnamefont{Beane}},
  \bibinfo{author}{\bibfnamefont{K.}~\bibnamefont{Orginos}}, \bibnamefont{and}
  \bibinfo{author}{\bibfnamefont{M.~J.} \bibnamefont{Savage}},
  \bibinfo{journal}{Phys. Lett.} \textbf{\bibinfo{volume}{B654}},
  \bibinfo{pages}{20} (\bibinfo{year}{2007}), \eprint{hep-lat/0604013}.

\bibitem[{\citenamefont{Procura et~al.}(2004)\citenamefont{Procura, Hemmert,
  and Weise}}]{Procura:2003ig}
\bibinfo{author}{\bibfnamefont{M.}~\bibnamefont{Procura}},
  \bibinfo{author}{\bibfnamefont{T.~R.} \bibnamefont{Hemmert}},
  \bibnamefont{and} \bibinfo{author}{\bibfnamefont{W.}~\bibnamefont{Weise}},
  \bibinfo{journal}{Phys. Rev.} \textbf{\bibinfo{volume}{D69}},
  \bibinfo{pages}{034505} (\bibinfo{year}{2004}), \eprint{hep-lat/0309020}.

\bibitem[{\citenamefont{Steininger et~al.}(1998)\citenamefont{Steininger,
  Meissner, and Fettes}}]{Steininger98}
\bibinfo{author}{\bibfnamefont{S.}~\bibnamefont{Steininger}},
  \bibinfo{author}{\bibfnamefont{U.-G.} \bibnamefont{Meissner}},
  \bibnamefont{and} \bibinfo{author}{\bibfnamefont{N.}~\bibnamefont{Fettes}},
  \bibinfo{journal}{JHEP} \textbf{\bibinfo{volume}{9809}}, \bibinfo{pages}{008}
  (\bibinfo{year}{1998}).

\bibitem[{\citenamefont{Bernard et~al.}(2004)\citenamefont{Bernard, Hemmert,
  and Meissner}}]{Bernard:2004}
\bibinfo{author}{\bibfnamefont{V.}~\bibnamefont{Bernard}},
  \bibinfo{author}{\bibfnamefont{T.}~\bibnamefont{Hemmert}}, \bibnamefont{and}
  \bibinfo{author}{\bibfnamefont{U.-G.} \bibnamefont{Meissner}},
  \bibinfo{journal}{Nucl. Phys.} \textbf{\bibinfo{volume}{A732}},
  \bibinfo{pages}{149} (\bibinfo{year}{2004}).

\bibitem[{\citenamefont{Bernard et~al.}(2005)\citenamefont{Bernard, Hemmert,
  and Meissner}}]{Bernard:2005}
\bibinfo{author}{\bibfnamefont{V.}~\bibnamefont{Bernard}},
  \bibinfo{author}{\bibfnamefont{T.}~\bibnamefont{Hemmert}}, \bibnamefont{and}
  \bibinfo{author}{\bibfnamefont{U.-G.} \bibnamefont{Meissner}},
  \bibinfo{journal}{Phys. Lett.} \textbf{\bibinfo{volume}{B622}},
  \bibinfo{pages}{141} (\bibinfo{year}{2005}).

\bibitem[{\citenamefont{Fettes et~al.}(1998)\citenamefont{Fettes, Meissner, and
  Steininger}}]{Fettes:1998}
\bibinfo{author}{\bibfnamefont{N.}~\bibnamefont{Fettes}},
  \bibinfo{author}{\bibfnamefont{U.-G.} \bibnamefont{Meissner}},
  \bibnamefont{and}
  \bibinfo{author}{\bibfnamefont{S.}~\bibnamefont{Steininger}},
  \bibinfo{journal}{Nucl. Phys.} \textbf{\bibinfo{volume}{A640}},
  \bibinfo{pages}{199} (\bibinfo{year}{1998}), \eprint{hep-ph/9803266}.

\bibitem[{\citenamefont{Procura et~al.}(2006)}]{Procura:2006}
\bibinfo{author}{\bibfnamefont{M.}~\bibnamefont{Procura}} \bibnamefont{et~al.},
  \bibinfo{journal}{Phys. Rev. D} \textbf{\bibinfo{volume}{73}},
  \bibinfo{pages}{114510} (\bibinfo{year}{2006}).

\bibitem[{\citenamefont{Entem and Machleidt}(2002)}]{Entem:2002sf}
\bibinfo{author}{\bibfnamefont{D.~R.} \bibnamefont{Entem}} \bibnamefont{and}
  \bibinfo{author}{\bibfnamefont{R.}~\bibnamefont{Machleidt}},
  \bibinfo{journal}{Phys. Rev.} \textbf{\bibinfo{volume}{C66}},
  \bibinfo{pages}{014002} (\bibinfo{year}{2002}), \eprint{nucl-th/0202039}.

\bibitem[{\citenamefont{Epelbaum et~al.}(2005)\citenamefont{Epelbaum, Glockle,
  and Meissner}}]{Epelbaum:2004}
\bibinfo{author}{\bibfnamefont{E.}~\bibnamefont{Epelbaum}},
  \bibinfo{author}{\bibfnamefont{W.}~\bibnamefont{Glockle}}, \bibnamefont{and}
  \bibinfo{author}{\bibfnamefont{U.-G.} \bibnamefont{Meissner}},
  \bibinfo{journal}{Nucl. Phys.} \textbf{\bibinfo{volume}{A747}},
  \bibinfo{pages}{362} (\bibinfo{year}{2005}), \eprint{nucl-th/0405048}.

\bibitem[{\citenamefont{Ishikawa et~al.}(2009)}]{Ishikawa:2009vc}
\bibinfo{author}{\bibfnamefont{K.~I.} \bibnamefont{Ishikawa}}
  \bibnamefont{et~al.} (\bibinfo{collaboration}{PACS-CS})
  (\bibinfo{year}{2009}), \eprint{0905.0962}.

\bibitem[{\citenamefont{Collaboration}(2009)}]{r0_ETMC_Scaling_Paper}
\bibinfo{author}{\bibfnamefont{E.}~\bibnamefont{Collaboration}},
  \bibinfo{howpublished}{to be published} (\bibinfo{year}{2009}).

\bibitem[{\citenamefont{Alexandrou et~al.}(2007)}]{Alexandrou:2007qq}
\bibinfo{author}{\bibfnamefont{C.}~\bibnamefont{Alexandrou}}
  \bibnamefont{et~al.} (\bibinfo{collaboration}{ETM Collaboration}),
  \bibinfo{journal}{PoS} \textbf{\bibinfo{volume}{LAT2007}},
  \bibinfo{pages}{087} (\bibinfo{year}{2007}), \eprint{arXiv:0710.1173
  [hep-lat]}.

\bibitem[{\citenamefont{Tiburzi and Walker-Loud}(2008)}]{Tiburzi:2008bk}
\bibinfo{author}{\bibfnamefont{B.~C.} \bibnamefont{Tiburzi}} \bibnamefont{and}
  \bibinfo{author}{\bibfnamefont{A.}~\bibnamefont{Walker-Loud}},
  \bibinfo{journal}{Phys. Lett.} \textbf{\bibinfo{volume}{B669}},
  \bibinfo{pages}{246} (\bibinfo{year}{2008}), \eprint{0808.0482}.

\bibitem[{\citenamefont{Frink et~al.}(2005)\citenamefont{Frink, Meissner, and
  Scheller}}]{Frink:2005ru}
\bibinfo{author}{\bibfnamefont{M.}~\bibnamefont{Frink}},
  \bibinfo{author}{\bibfnamefont{U.-G.} \bibnamefont{Meissner}},
  \bibnamefont{and} \bibinfo{author}{\bibfnamefont{I.}~\bibnamefont{Scheller}},
  \bibinfo{journal}{Eur. Phys. J.} \textbf{\bibinfo{volume}{A24}},
  \bibinfo{pages}{395} (\bibinfo{year}{2005}), \eprint{hep-lat/0501024}.

\bibitem[{\citenamefont{Frink and Meissner}(2004)}]{Frink:2004ic}
\bibinfo{author}{\bibfnamefont{M.}~\bibnamefont{Frink}} \bibnamefont{and}
  \bibinfo{author}{\bibfnamefont{U.-G.} \bibnamefont{Meissner}},
  \bibinfo{journal}{JHEP} \textbf{\bibinfo{volume}{07}}, \bibinfo{pages}{028}
  (\bibinfo{year}{2004}), \eprint{hep-lat/0404018}.

\bibitem[{\citenamefont{Gasser et~al.}(1988)\citenamefont{Gasser, Sainio, and
  Svarc}}]{Gas88}
\bibinfo{author}{\bibfnamefont{J.}~\bibnamefont{Gasser}},
  \bibinfo{author}{\bibfnamefont{M.}~\bibnamefont{Sainio}}, \bibnamefont{and}
  \bibinfo{author}{\bibfnamefont{A.}~\bibnamefont{Svarc}},
  \bibinfo{journal}{Nucl. Phys. B} \textbf{\bibinfo{volume}{307}},
  \bibinfo{pages}{779} (\bibinfo{year}{1988}).

\bibitem[{\citenamefont{Nagels et~al.}(1979{\natexlab{a}})}]{Nagels:1979xh}
\bibinfo{author}{\bibfnamefont{M.~M.} \bibnamefont{Nagels}}
  \bibnamefont{et~al.}, \bibinfo{journal}{Nucl. Phys.}
  \textbf{\bibinfo{volume}{B147}}, \bibinfo{pages}{189}
  (\bibinfo{year}{1979}{\natexlab{a}}).

\bibitem[{\citenamefont{Nagels et~al.}(1979{\natexlab{b}})\citenamefont{Nagels,
  Rijken, and de~Swart}}]{Nagels:1978sc}
\bibinfo{author}{\bibfnamefont{M.~M.} \bibnamefont{Nagels}},
  \bibinfo{author}{\bibfnamefont{T.~A.} \bibnamefont{Rijken}},
  \bibnamefont{and} \bibinfo{author}{\bibfnamefont{J.~J.}
  \bibnamefont{de~Swart}}, \bibinfo{journal}{Phys. Rev.}
  \textbf{\bibinfo{volume}{D20}}, \bibinfo{pages}{1633}
  (\bibinfo{year}{1979}{\natexlab{b}}).

\bibitem[{\citenamefont{Tiburzi and Walker-Loud}(2006)}]{Tiburzi:2005na}
\bibinfo{author}{\bibfnamefont{B.~C.} \bibnamefont{Tiburzi}} \bibnamefont{and}
  \bibinfo{author}{\bibfnamefont{A.}~\bibnamefont{Walker-Loud}},
  \bibinfo{journal}{Nucl. Phys.} \textbf{\bibinfo{volume}{A764}},
  \bibinfo{pages}{274} (\bibinfo{year}{2006}), \eprint{hep-lat/0501018}.

\bibitem[{\citenamefont{Baron et~al.}(2008)}]{Baron:2008xa}
\bibinfo{author}{\bibfnamefont{R.}~\bibnamefont{Baron}} \bibnamefont{et~al.}
  (\bibinfo{collaboration}{ETM}), \bibinfo{journal}{PoS}
  \textbf{\bibinfo{volume}{LATTICE2008}}, \bibinfo{pages}{094}
  (\bibinfo{year}{2008}), \eprint{0810.3807}.

\end{thebibliography}

\end{document}